\documentclass[aps,pra,superscriptaddress,twocolumn,showpacs,floatfix,10pt]{revtex4-1}

\usepackage{amsmath}
\usepackage{amssymb}
\usepackage{bm}
\usepackage{graphicx}
\usepackage{appendix}
\usepackage{longtable}
\usepackage{color}

\definecolor{darkgreen}{rgb}{0.0, 0.3, 0.0}
\definecolor{armygreen}{rgb}{0.29, 0.33, 0.13}

\begin{document}


\title{Nonlinear optical response of a two-dimensional quantum-dot supercrystal:\\
Emerging multistability, periodic and aperiodic self-oscillations,\\ chaos, and transient chaos}

\author{Igor V. Ryzhov}
\affiliation{Herzen State Pedagogical University, 191186 St. Petersburg, Russia}
\author{Ramil F. Malikov}
\affiliation{Akmullah State Pedagogical University of Bashkortostan, 450000 Ufa, Russia}

\author{Andrey V. Malyshev}
\email[Corresponding author:]{a.malyshev@fis.ucm.es}
\affiliation{GISC, Departamento de F\'{\i}sica de Materiales,
Universidad Complutense, E-28040 Madrid, Spain} \affiliation{Ioffe
Physical-Technical Institute, 26 Politechnicheskaya str., 194021
St. Petersburg, Russia}

\author{Victor\ A.\ Malyshev}
\email[Corresponding author:]{v.malyshev@rug.nl}
\affiliation{Zernike Institute for Advanced Materials, University of Groningen, Nijenborgh 4, 9747
AG Groningen, The Netherlands}

\date{\today}

\begin{abstract}
We conduct a theoretical study of the nonlinear optical response of a two-dimensional semiconductor quantum dot supercrystal subjected to a quasi-resonant continuous wave excitation. A constituent quantum dot is modeled as a three-level ladder-like system (comprising the ground, the one-exciton, and the bi-exction states).
To study the stationary response of the supercrystal, we propose an exact linear parametric method of solving the nonlinear steady-state problem, while to address the supercrystal optical dynamics qualitatively, we put forward a novel method to calculate the bifurcation diagram of the system.
Analyzing the dynamics, we demonstrate that the supercrystal can exhibit multistability, periodic and aperiodic self-oscillations, and chaotic behavior, depending on parameters of the supercrystal and excitation conditions.
 The effects originate from the interplay of the intrinsic nonlinearity of quantum dots and the retarded inter-dot dipole-dipole interaction. The latter provides a positive feedback which results in the exotic supercrystal optical dynamics.
These peculiarities of the supercrystal optical response open up a possibility for all-optical applications and devices. In particular, an all-optical switch, a tunable generator of THz pulses (in self-oscillating regime), a noise generator (in chaotic regime), and a tunable bistable mirror can be designed.
\end{abstract}

\pacs{ 78.67.-n  
       73.20.Mf  
       85.35.-p  
}
\maketitle

\section{Introduction}
\label{Intro}
\begin{figure}[ht!]
\begin{center}
\includegraphics[width=\columnwidth]{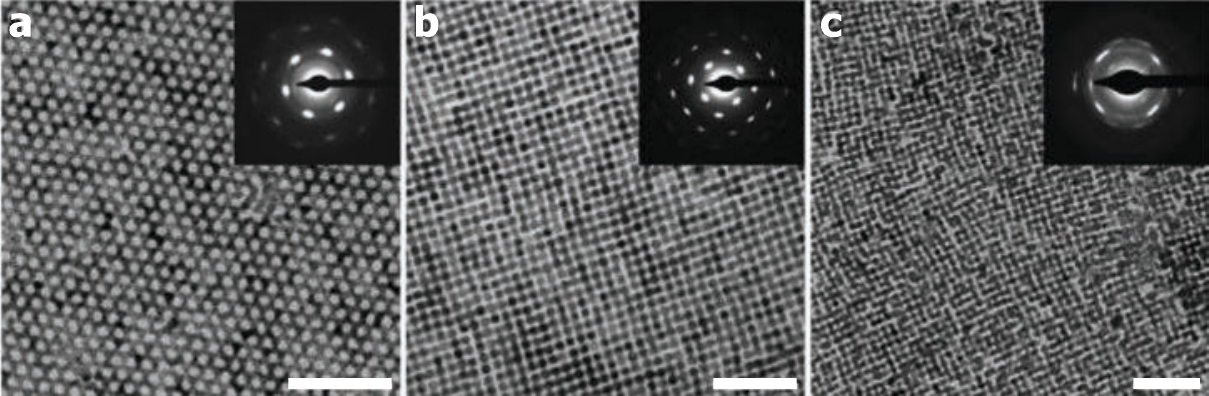}
\caption{\label{fig:Supercrystal} 
PbSe rocksalt 2D nanostructures with (a) honeycomb and (b) square lattice symmetry, (c) - CdSe nanostructure with a compressed zincblende and slightly distorted square lattices (scale bars, 50 nm). Insets show the electro-diffractograms in the [111] (a) and [100] (b,c) projections. The figure is from Ref.~\cite{Evers2013}.}
\end{center}
\end{figure}

In the last decade, the so-called metamaterials, a class of new materials not existing in nature, received a great deal of attention (see for recent reviews Refs.~\cite{Zheludev2010,Zheludev2012,Soukoulis2010,Liu2011,Alu2016}). Super-crystals comprising regularly spaced quantum emitters represent one of the examples of metamaterials with tunable optical properties which can be controlled by the geometry and chemical composition of components.~\cite{Baimuratov2013} Modern nanotechnology has at its disposal a variety of methods to fabricate such systems~\cite{Evers2013,Boneschanscher2014,Baranov2015,Ushakova2016,Liu2017}. In Fig.~\ref{fig:Supercrystal}, a few examples of  ultrathin sheets of regularly spaced semiconductor nanocrystals grown by the method of oriented attachment (see for details Ref.~\cite{Evers2013}) are present.
\begin{figure}
\begin{center}
\includegraphics[width=0.6\columnwidth]{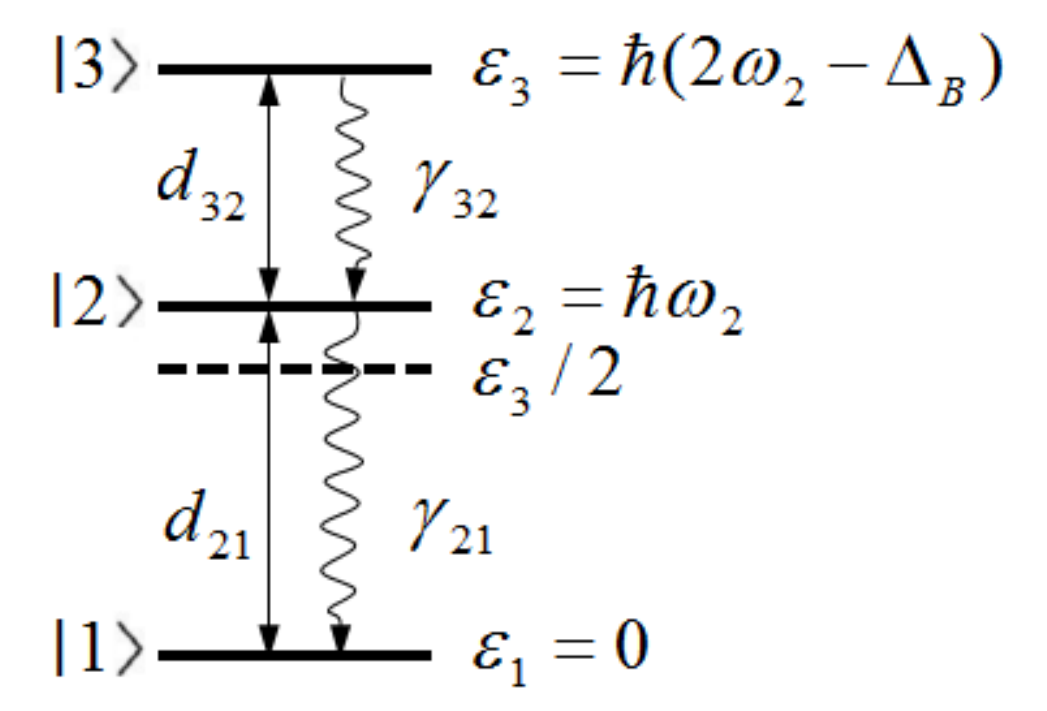}
\caption{\label{fig:Schematics} 
Energy level diagram of the ladder-type three-level SQD: $|1 \rangle$, $|2 \rangle$, and $|3 \rangle$ are the ground, one-exciton, and bi-exciton  states, respectively. The energies of corresponding states are $\varepsilon_1 =0$, $\varepsilon_2 =\hbar \omega_2$ and $\varepsilon_3 = \hbar(2\omega_2 -\Delta_B)$, where, $\hbar\Delta_B$ is the bi-exciton binding energy. Allowed transitions with the corresponding transition dipole moments $\bf{d}_{21}$ and $\bf{d}_{32}$ are indicated by solid arrows. Wavy arrows denote the allowed spontaneous transitions with rates $\gamma_{32}$ and $\gamma_{21}$. The dashed horizontal line shows the energy of the coherent two-photon resonance (the corresponding $1\leftarrow 3$ transition occurs with the simultaneous absorption of two $\varepsilon_3/2$ photons).}
\end{center}
\end{figure}

As is well known, a thin layer of two-level emitters (atoms, molecules, $J$-aggregates), the thickness $L$ of which is much smaller than the radiation wavelength $\lambda$ in the layer, can act as an all-optical bistable element~\cite{Bowden1986a,Bowden1986b,Zakharov1988,Basharov1988,Benedict1990,Benedict1991,Oraevsky1994,Malyshev2000,Glaeske2000,Klugkist2007, Malikov2017a}. For bistability to occur, two factors are required: nonlinearity of the material and a positive feedback. Interplay of these two factors leads to a situation when the system has two stable states; switching between them is governed again by an external optical signal. The nonlinearity of the layer is ensured by the fact that two-level emitters are nonlinear systems. The positive feedback originates from the secondary field, which is generated by the emitters themselves; this is the so-called “intrinsic feedback”, i.e., here a cavity (external feedback) is not required.

A two-dimensional (2D) semiconductor quantum dot (SQD) supercrystal represents a limiting case of a thin layer. In this paper, we conduct a theoretical study of the nonlinear optical response of such a system. A single SQD is considered as a point-like system with three consecutive levels of the ground, one-exciton, and bi-exction states (corresponding to the so-called ladder or $\Theta$ level scheme). Due to the high density of SQDs and high oscillator strengths of the SQD's transitions, the total (retarded) dipole-dipole SQD-SQD interactions have to be taken into account, which is finally done in the mean-field approximation for the point-like dipoles in a homogeneous host for simplicity. The real part of the dipole-dipole interaction results in the dynamic shift of the SQD's energy levels, whereas the imaginary part describes the collective radiative decay of SQDs, both depending on the population differences between the levels. These two effects are crucial for the nonlinear dynamics of the SQD supercrystal. As a result, in addition to bistability, analogous to that for a thin layer of two-level emitters, we found multistability, periodic and aperiodic self-oscillation, and chaotic regimes in the optical response of the SQD supercrystal~\footnote{\label{Note0}It should be noticed that a thin layer of three-level $V$-emitters also shows a similar behavior (see Refs.~\onlinecite{Vlasov2013a,Vlasov2013b}).}. 
To the best of our knowledge, a detailed study of the SQD supercrystal optical response has not been performed so far~\cite{footnote1}.. 
To uncover the character of the instabilities, we use the standard methods of nonlinear dynamics, such as the analysis of the Lyapunov exponents, bifurcation diagrams, phase space maps, and Fourier spectra~\cite{AndronovBook1966,EckmannRevModPhys1985,GuckenheimerBook1986,NeimarkLandaBook1992,OttBook1993,Arnol'dBook1994,AlligoodBook1996,KatokBook1997,KuznetsovBook2004,WieczorekPhysRep2005}.
Important technical results of our study are a new simple parametric method of finding the exact solution of the nonlinear steady-state multilevel Maxwell-Bloch equations and a new method of bifurcation diagram calculation that can be used for a wide class of systems.

The arrangement of the paper is as follows. In the next section, we describe the model of a 2D supercrystal comprised of SQDs and the mathematical formalism to treat its optical response. We use for that the one-particle density matrix formalism within the rotating wave approximation (RWA), where the total retarded dipole-dipole interactions between point-like SQDs are taken into account. In Sec.\ref{Mean-field approximation}, the general formalism is simplified making use of the mean-field approximation, and the mean-field parameters (the collective energy level shift and radiation damping) are calculated. In Sec.~\ref{Numerics}, we present the results of numerical calculations of the supersrystal optical response, including the steady-state solution (Sec.~\ref{steady-state} ), an analysis of bifurcations occurring in the system (Sec.~\ref{Bifurcation diagram}), and the system's dynamics (Sec.~\ref{Time-domain}) for two conditions of excitation: (i) the external field is tuned into the one-exciton transition and (ii) it is in resonance with the coherent two-photon transition (with the simultaneous absorption of two photons). A rationale for the physical mechanism of the effects found is provided in  Sec.~\ref{Discussion}. In Sec.~\ref{Reflectance}  we show that the 2D SQD supercrystal can operate as a bistable nanoscale mirror. Section~\ref{Summary} concludes the paper.

\section{Model and theoretical background}
\label{Model}
We consider a 2D supercrystal comprising identical semiconductor quantum dots (SQDs).
The optical excitations in an SQD are confined excitons. In such a system, the degeneracy of the one-exciton state is lifted due to the anisotropic electron-hole exchange, leading to two split linearly polarized one-exciton states (see, e.g., Refs.~\cite{Stufler2006,Jundt2008,Gerardot2009}). In this case, the ground state is coupled to the bi-exciton state via the linearly polarized one-exciton transitions. By choosing the appropriate polarization of the applied field, i.e., selecting one of the single-exciton states, the system effectively acquires a three-level ladder-like structure with a ground state $|1\rangle$, one exciton state $|2\rangle$, and bi-exciton state $|3\rangle$ with corresponding energies $\varepsilon_1 = 0$, $\varepsilon_2 = \hbar\omega_{2}$, and $\varepsilon_3 = \hbar\omega_3 = \hbar(2\omega_2 - \Delta_B)$, where $\hbar\Delta_B$ is bi-exciton binding energy (see Fig.~\ref{fig:Schematics}). Within this model, the allowed transitions, induced by the external field, are $|1\rangle \leftrightarrow |2\rangle$ and $|2\rangle \leftrightarrow |3\rangle$, which are characterized by the transition dipole moments $\bf{d}_{21}$ and $\bf{d}_{32}$, respectively. For the sake of simplicity, we assume that they are real. The states $|3\rangle$ and $|2\rangle$ spontaneously decay with rates $\gamma_{32}$ and $\gamma_{21}$, accordingly. Note that the bi-exciton  state $|3\rangle$, having no allowed transition dipole moment from the ground state $|1\rangle$, can be populated either via consecutive $|1\rangle \rightarrow |2\rangle \rightarrow |3\rangle$ transitions or via the simultaneous absorption of two photons of frequency $\omega_3/2$. In what follows, we will consider both options.

The optical dynamics of SQDs is described by means of the Lindblad quantum master equation for the density operator $\rho(t)$~\cite{Lindblad1976,Blum2012}, which in the rotating frame (with the frequency $\omega_0$ of the external field) reads as
\begin{widetext}
\begin{subequations}
\label{MasterEqAndHamiltonian}
\begin{equation}
\dot{\rho}(t) = -\frac{i}{\hbar} \left[{H^{RWA}}(t),\rho(t)\right ] + {\cal L}\{\rho(t)\}~,
\label{LindbladEq}
\end{equation}
\begin{equation}
H^{RWA}(t) =  \hbar \sum_{\bf n} \left( \Delta_{21}\sigma^{\bf n}_{22} + \Delta_{31}\sigma^{\bf n}_{33} \right)
        - i\hbar \sum_{\bf n} \left[ {\Omega}^{\bf n}_{21}(t) \sigma^{\bf n}_{21} + \Omega^{\bf n}_{32}(t) \sigma^{\bf n}_{32} \right] + \mathrm{H.c.}\ ,
\label{HamiltonianRWA}
\end{equation}
\begin{equation}
{\cal L}\{\rho(t)\} =
           \frac{\gamma_{21}}{2} \sum_{\bf n} \left( \left[ \sigma^{\bf n}_{12} \rho(t),
           \sigma^{\bf n}_{21}\right]  + \left[ \sigma^{\bf n}_{12},
           \rho(t)\, \sigma^{\bf n}_{21}\right]\right)
           + \frac{\gamma_{32}}{2} \sum_{\bf n} \left( \left[
           \sigma^{\bf n}_{23}\rho(t),\sigma^{\bf n}_{32}\right]
           + \left[ \sigma^{\bf n}_{23},\rho(t) \, \sigma^{\bf n}_{32}\right]\right)\ ,
\label{LindbladOperator}
\end{equation}
\begin{equation}
\sigma^{\bf n}_{ij} = |{\bf n}i\rangle \langle j{\bf n}|~, \quad i,j = 1,2,3~.
\end{equation}
\end{subequations}
\end{widetext}
In Eq.~(\ref{LindbladEq}), $\hbar$ is the reduced Plank constant, $H^{RWA}$ is the SQD Hamiltonian in the RWA, $[A,B]$ denotes the commutator, ${\cal L}$ is the Lindblad relaxation operator, given by Eq.~(\ref{LindbladOperator})~\cite{Lindblad1976,Blum2012}.
In Eq.~(\ref{HamiltonianRWA}), $ \Delta_{21} = \omega_2 - \omega_0$ and $\Delta_{31} = \omega_3 - 2\omega_0$ are the energies of states $|2 \rangle$ and $|3 \rangle$ in the rotating frame, accordingly. Alternatively, these quantities can be interpreted as
the detunings away from the one-photon resonance and the coherent two-photon resonance, respectively. $\Omega^{\bf n}_{21}(t) = \bf{d}_{21} \bf{E_{n}}(t)/\hbar$ and $\Omega^{\bf n}_{32}(t) = \bf{d}_{32}\bf{E_n}(t)/\hbar$, where $\bf{E_n}(t)$ is the slowly-varying amplitude of the total field driving the optical transitions in the ${\bf n}$-th SQD, $\bm{\mathcal E}_{\bf n}(t) = {\bf E_n}(t)\exp(-i\omega_0 t) + c.c.$. The latter is the sum of the applied field $\bm{\mathcal E}^0_{\bf n}(t) = {\bf E}^0_{\bf n}(t)\exp(-i\omega_0 t) + c.c.$ and the field produced by all others SQDs in place of the ${\bf n}$-th SQD, $\bm{\mathcal E}^{loc}_{\bf n}(t) = \sum_{\bf m} \bm{\mathcal E}_{\bf mn}(t) = \sum_{\bf m} {\bf E_{mn}}(t)\exp(-i\omega_0 t) + c.c.$, where the amplitude ${\bf E_{mn}}(t)$ is given by (see, e.g., Refs.~\onlinecite{Zaitsev1983,Benedict1996})
\begin{widetext}
\begin{eqnarray}
\label{Emn}
{\bf E_{mn}}(t) = \left\{\left[ \frac{3}{r^3_{\bf mn}} - \frac{3ik_0}{r^2_{\bf mn}}  - \frac{k_0^2}{r_{\bf mn}}\right]({\bf d_{21}u_{mn}}){\bf u_{mn}}
- \left[ \frac{1}{r^3_{\bf mn}} - \frac{ik_0}{r^2_{\bf mn}} - \frac{k_0^2}{r_{\bf mn}}\right]{\bf d_{21}}  \right\}\rho^{\bf m}_{21}(t^\prime)e^{ik_0r_{\bf mn}} \nonumber
\\
+ \left\{\left[ \frac{3}{r^3_{\bf mn}} - \frac{3ik_0}{r^2_{\bf mn}}  - \frac{k_0^2}{r_{\bf mn}}\right]({\bf d_{32}u_{mn}}){\bf u_{mn}}
- \left[ \frac{1}{r^3_{\bf mn}} - \frac{ik_0}{r^2_{\bf mn}} - \frac{k_0^2}{r_{\bf mn}}\right]{\bf d}_{32}  \right\}\rho^{\bf m}_{32}(t^\prime) e^{ik_0r_{\bf mn}}~,
\end{eqnarray}
\end{widetext}
where $r_{\bf mn}$ is the distance between sites ${\bf m}$ and ${\bf n}$, $k_0 = \omega_0/c$ ($c$ is the speed of light in vacuum), ${\bf u_{mn}} = {\bf r_{mn}}/r_{\bf mn}$ is the unit vector along ${\bf r_{mn}}$, and $t^\prime = t - r_{\bf mn}/c$. Equation~(\ref{Emn}) represents the field (amplitude) produced by an oscillating dipole ${\bf d_{21}}R^{\bf m}_{21}(t^\prime) + {\bf{d_{32}}}R^{\bf m}_{32}(t^\prime)$ situated at a point ${\bf r_m}$ in another point ${\bf r_n}$ at an instant $t$, accounting for retardation: $t - t^\prime = r_{\bf mn}/c$~\footnote{Strictly speaking, this field should be the field acting {\it inside} the SQD, the latter differs from the field acting {\it on} the SQD by a screening factor which depends on the system geometry and material parameters. In the simplest case of a spherical dot in a homogeneous environment this factor can be obtained analytically (see, e. g., Ref.~\onlinecite{Malyshev2011}). A realistic SQD array is a considerably more complicated system involving a non-homogeneous host, at least three different materials, and a number of geometrical parameters. We believe that explicit calculation of the screening factors in this case would introduce unnecessary level of detail and obscure further analysis. Therefore, for the sake of simplicity, we consider a SQD as a point-like system in a homogeneous host; all the fields entering the Lindblad equations should be interpreted as those rescaled by appropriate screening factors.}.
Using Eq.~(\ref{Emn}), the fields $\Omega^{\bf n}_{\alpha\beta}(t)$ ($\alpha\beta = 21,32$) can be written in the form
\begin{widetext}
\begin{subequations}
\begin{equation}
\label{Omega21n}
\Omega^{\bf n}_{\alpha\beta}(t) = \Omega^{0{\bf n}}_{\alpha\beta}(t)
    + \sum_{\bf m} (\gamma_{\alpha\beta}^{\bf mn} + i\Delta_{\alpha\beta}^{\bf mn}) \rho_{\alpha\beta}^{\bf m}(t^\prime)~,
\end{equation}
\begin{equation}
\label{gamma21n}
\gamma_{\alpha\beta}^{\bf mn} = \frac{3\gamma_{\alpha\beta}}{4(k_0a)^3} \left\{ \left[(k_0a)^2 \frac{\kappa_{\alpha\beta}^{\mathbf mn}}{|{\bf m-n}|}
- \frac{\chi_{\alpha\beta}^{\mathbf mn}}{|{\bf m-n}|^3} \right] \sin(k_0a|{\bf m-n}|) + k_0a \frac{\chi_{\alpha\beta}^{\mathbf mn}}{|{\bf m-n}|^2}\cos(k_0a|{\bf m-n}|) \right\}
\end{equation}
\begin{equation}
\label{Delta21n}
\Delta_{\alpha\beta}^{\bf mn} = \frac{3\gamma_{\alpha\beta}}{4(k_0a)^3} \left\{ \left[ \frac{\chi_{\alpha\beta}^{\mathbf mn}}{|{\bf m-n}|^3}
- (k_0a)^2 \frac{\kappa_{\alpha\beta}^{\mathbf mn}}{|{\bf m-n}|} \right] \cos(k_0a|{\bf m-n}|) + k_0a \frac{\chi_{\alpha\beta}^{\mathbf mn}}{|{\bf m-n}|^2}\sin(k_0a|{\bf m-n}|) \right\}
\end{equation}
\begin{equation}
\label{kappa21mn}
\kappa_{\alpha\beta}^{\mathbf mn} = 1 - ({\bf e}_{\alpha\beta}{\bf u_{mn}})^2~,
    \quad \chi_{\alpha\beta}^{\mathbf mn} = 1 - 3({\bf e}_{\alpha\beta}{\bf u_{mn}})^2~.
\end{equation}
\end{subequations}
\end{widetext}
We used in Eqs.~(\ref{gamma21n}) and~(\ref{Delta21n}) the expression $\gamma_{\alpha\beta} =
4|{\bf d}_{\alpha\beta}|^2 \omega_{\alpha\beta}^3/(3\hbar c^3)$.
In Eq.~(\ref{kappa21mn}), ${\bf e}_{\alpha\beta} = {\bf d}_{\alpha\beta}/d_{\alpha\beta}$ is the unit vector along ${\bf d}_{\alpha\beta}$.
The matrices $\gamma_{\alpha\beta}^{\bf mn}$ and $\Delta_{\alpha\beta}^{\bf mn}$ represent the real and imaginary parts of the retarded dipole-dipole interaction of $\bf n$-th and $\bf m$-th SQDs.

Equation~(\ref{LindbladEq}), written in the site basis $|{\bf n}i \rangle$ ($i = 1,2,3$), reads as
\begin{widetext}
\begin{subequations}
\label{allrhon}
\begin{equation}
\label{rho11n}
\dot{\rho}^{\bf n}_{11} = \gamma_{21} \rho^{\bf n}_{22} + \Omega^{\bf n}_{21} {\rho^{\bf n *}_{21}} + {\Omega^{\bf n *}_{21}} \rho^{\bf n}_{21}~,
\end{equation}
\begin{equation}
\label{rho22n}
\dot{\rho}^{\bf n}_{22} =  -\gamma_{21} \rho^{\bf n}_{22} + \gamma_{32} \rho^{\bf n}_{33} - \Omega^{\bf n}_{21} \rho^{\bf n*}_{21}
    - \Omega^{\bf n*}_{21} \rho^{\bf n}_{21} + \Omega^{\bf n}_{32} \rho^{\bf n*}_{32} + \Omega^{\bf n*}_{32} \rho^{\bf n}_{32}~,
\end{equation}
\begin{equation}
\label{rho33n}
\dot{\rho}^{\bf n}_{33} = -\gamma_{32} \rho^{\bf n}_{33} - \Omega^{\bf n}_{32} \rho^{\bf n *}_{32} - \Omega^{\bf n*}_{32} \rho_{32}~,
\end{equation}
\begin{equation}
\label{R21n}
\dot{\rho}^{\bf n}_{21} = - \left( i\Delta_{21} + \frac{1}{2}\gamma_{21} \right) \rho^{\bf n}_{21} + \Omega^{\bf n}_{21}(\rho^{\bf n}_{22}
    - \rho^{\bf n}_{11}) + \Omega^{\bf n*}_{32} \rho_{31}~,
\end{equation}
\begin{equation}
\label{R32n}
\dot{\rho}^{\bf n}_{32} = - \left[ i\Delta_{32} + \frac{1}{2} (\gamma_{32} + \gamma_{21}) \right] \rho^{\bf n}_{32}
    +\Omega^{\bf n}_{32}(\rho^{\bf n}_{33} - \rho^{\bf n}_{22}) - \Omega^{\bf n *}_{21} \rho^{\bf n}_{31}~,
\end{equation}
\begin{equation}
\label{R31n}
\dot{\rho}^{\bf n}_{31} = - \left( i\Delta_{31} + \frac{1}{2}\gamma_{32} \right) \rho^{\bf n}_{31} - \Omega^{\bf n}_{32} \rho^{\bf n}_{21}
    + \Omega^{\bf n}_{21} \rho^{\bf n}_{32}~,
\end{equation}
\end{subequations}
\end{widetext}
where $\Omega^{\bf n}_{21}$ and $\Omega^{\bf n}_{32}$ are given by Eqs.~(\ref{Omega21n}) -~(\ref{kappa21mn}). The time dependence of all relevant quantities is dropped here.

It is worth to noting that Eqs.~(\ref{rho11n}) -~(\ref{R31n}) represent a set of equations for the one-particle density matrix, where the quantum correlations of the dipole operators of different SQDs are neglected that implies that $\langle \hat{d}_{\bf n} \hat{d}_{\bf m}\rangle = \langle \hat{d}_{\bf n}\rangle\ldots\langle \hat{d}_{\bf m}\rangle$, where $\langle ... \rangle$ denotes the quantum mechanical average. A proof of this assumption is a stand-alone problem to be solved, which is beyond the scope of this paper.

\section{Mean-field approximation}
\label{Mean-field approximation}
The set of Eqs.~(\ref{rho11n}) - (\ref{R31n}) allows one to study the optical response of a SQD monolayer, without any limitation to the layer's size, lattice geometry, and the spatial profile of the external field amplitude ${\bf E}^{\bf n}_0$. Here, we restrict our consideration to a spatially homogeneous case, when all relevant quantities entering Eqs.~(\ref{rho11n}) - (\ref{R31n}) do not depend on the SQD's position $\bf n$. In fact, this approximation is equivalent to taking into account the Lorentz local field correction to the field acting on an emitter, which has been widely used when analyzing the optical response of dense bulk media, both linear~\cite{BornAndWolf,Friedberg1973} and nonlinear~\cite{Hopf1984,Bowden1986b,Benedict1990,Malyshev1997a,Malyshev1997b,Malikov2017a}. This approximation intuitively seems to be appropriate for an infinite layer, however, for a finite sample, its validity should be examined. Nevertheless, as we show below, even this simplest model predicts a variety of fascinating effects. We consider a simple square lattice of SQDs in order to avoid unnecessary computational complications.

Thus, we neglect the spatial dependence of all functions in Eqs.~(\ref{rho11n})-(\ref{R31n}). Additionally, we assume for the sake of simplicity that the transition dipoles ${\bf d}_{21}$ and ${\bf d}_{32}$ are parallel to each other, ${\bf d}_{32} = \mu {\bf d}_{21} \equiv \mu{\bf d}$ (not a principal limitation). Accordingly, $\gamma_{32} = \mu^2 \gamma_{21} \equiv \mu^2 \gamma$ and $\Omega_{32} = \mu \Omega_{21} \equiv \mu \Omega$. Then the system of equations~(\ref{rho11n}) - (\ref{R31n}) takes the form~\footnote{Note that Eqs.~(\ref{rho11})-~(\ref{Omega21}) are algebraically equivalent to the equations for a heterodimer comprising a metallic nanoparticle and a semiconductor quantum dot subjected to a quasi-resonant irradiation (see Refs.~\onlinecite{Artuso2013} and~\onlinecite{Nugroho2017})}
\begin{widetext}
\begin{subequations}
\label{allrho}
\begin{equation}
\label{rho11}
\dot{\rho}_{11} = \gamma \rho_{22} + \Omega \rho_{21}^* + \Omega^* \rho_{21}~,
\end{equation}
\begin{equation}
\label{rho22}
\dot{\rho}_{22} =  - \gamma \rho_{22} + \mu^2 \gamma \rho_{33} - \Omega\rho_{21}^* - \Omega^* \rho_{21} + \mu (\Omega\rho_{32}^* + \Omega^*\rho_{32})~,
\end{equation}
\begin{equation}
\label{rho33}
\dot{\rho}_{33} = -\mu^2 \gamma \rho_{33} - \mu (\Omega \rho_{32}^* + \Omega^* \rho_{32})~,
\end{equation}
\begin{equation}
\label{R21}
\dot{\rho}_{21} = - \left( i\Delta_{21} + \frac{1}{2}\gamma \right) \rho_{21} + \Omega(\rho_{22} - \rho_{11}) + \mu \Omega^* \rho_{31}~,
\end{equation}
\begin{equation}
\label{R32}
\dot{\rho}_{32} = - \left[ i\Delta_{32} + \frac{1}{2} (1 + \mu^2)\gamma \right] \rho_{32} + \mu \Omega(\rho_{33} - \rho_{22}) - \Omega^* \rho_{31}~,
\end{equation}
\begin{equation}
\label{R31}
\dot{\rho}_{31} = - \left( i\Delta_{31} + \frac{1}{2}\mu^2 \gamma \right) \rho_{31} - \mu \Omega \rho_{21} + \Omega \rho_{32}~,
\end{equation}
\begin{equation}
\label{Omega21}
\Omega = \Omega_0 + (\gamma_R + i\Delta_L) (\rho_{21} + \mu \rho_{32})~,
\end{equation}
\end{subequations}
\end{widetext}
where the constants $\gamma_R$ and $\Delta_L$ are given by
\begin{widetext}
\begin{subequations}
\label{gammaRDeltaL}
\begin{equation}
\label{gammaR}
\gamma_R = \sum_{\bf m(\neq n)} \gamma_{21}^{\bf mn} =
\frac{3\gamma}{4(k_0a)^3} \sum_{\bf n \neq 0} \left\{ \left[(k_0a)^2 \frac{\kappa_\mathbf n}{|{\bf n}|}
- \frac{\chi_{\mathbf n}}{|{\bf n}|^3} \right] \sin(k_0a|{\bf n}|) + k_0a \frac{\chi_{\mathbf n}}{|{\bf n}|^2}\cos(k_0a|{\bf n}|) \right\}~,
\end{equation}
\begin{equation}
\label{DeltaL}
\Delta_L = \sum_{\bf m(\neq n)} \Delta_{21}^{\bf mn} = \frac{3\gamma}{4(k_0a)^3} \sum_{\bf n \neq 0} \left\{ \left[ \frac{\chi_{\mathbf n}}{|{\bf n}|^3}
- (k_0a)^2 \frac{\kappa_\mathbf n}{|{\bf n}|} \right] \cos(k_0a|{\bf n}|) + k_0a \frac{\chi_{\mathbf n}}{|{\bf n}|^2}\sin(k_0a|{\bf n}|) \right\}~,
\end{equation}
\begin{equation}
\label{kappa}
\kappa_{\mathbf n} = 1 - ({\bf eu_n})^2~, \quad\quad \chi_{\mathbf n} = 1 - 3({\bf eu_n})^2~.
\end{equation}
\end{subequations}
\end{widetext}
Recall that the summation in Eqs.~(\ref{gammaR}) and~(\ref{DeltaL}) runs over sites ${\bf n} = (n_x,n_y)$ of a simple square lattice, where $n_x = 0, \pm 1, \pm 2, \pm 3 ...\pm N_x$, $n_y = 0, \pm 1, \pm 2, \pm 3 ...\pm N_y$, and ${\bf e} = {\bf d}/d$ is the unit vector along the transition dipole moment ${\bf d}_{21}$.

Next, we are interested in ($k_0a$)-scaling of the constants $\gamma_R$ and $\Delta_L$. First, consider a point-like system, when the lateral lattice sizes $N_xa$ and $N_ya$ are much smaller that the reduced wavelength $\lambdabar = k_0^{-1}$. Then, making the expansion of sine- and cosine-functions in Eqs.~(\ref{gammaR}) and~(\ref{DeltaL}) to the lowest order with respect to $k_0a \ll 1$, one finds
\begin{subequations}
\label{gammaRDeltaLpoint-like}
\begin{equation}
\label{gammaRpoint-like}
\gamma_R =
\frac{3\gamma}{4} \sum_{\bf n \neq 0} \kappa_{\mathbf n} = \frac{3}{8}\gamma N~,
\end{equation}
\begin{eqnarray}
\label{DeltaLpoint-like}
\Delta_L &=&
\frac{3\gamma}{4(k_0a)^3} \sum_{\bf n \neq 0} \frac{\chi_{\mathbf n}}{|{\bf n}|^3}
= -\frac{3\gamma}{2(k_0a)^3}\zeta(3/2)\beta(3/2)\nonumber \\
&\simeq& -3.39 \frac{\gamma}{(k_0a)^3} = -3.39 \gamma \left(\frac{\lambdabar}{a}\right)^3~,
\end{eqnarray}
\end{subequations}
where $N = 4 N_xN_y$ is the total number of sites in the lattice, $\zeta(x)$ is the Riman $z$-function and $\beta(x) = \sum_{n=0}^\infty (-1)^n (2n + 1)^{-x}$ is the analytical continuation of the Dirichlet series~\cite{Glasser1972}. When deriving Eq.~(\ref{gammaRpoint-like}) we used the fact that $\sum_{\bf n \neq 0} \kappa_{\bf n} = N/2$. Furthermore, the formula~(\ref{DeltaLpoint-like}) follows from Eq.~(A20) of Ref.~\onlinecite{Christiansen1998} at $\theta = \pi/2$. As is seen from Eq.~(\ref{gammaRpoint-like}), $\gamma_R$ does not depend on $k_0a$; it is determined by the total number of SQDs in the lattice and describes the collective (Dicke) radiative relaxation of SQDs as all the SQD's dipoles are in phase for a point-like system~\cite{Zaitsev1983,Benedict1996,Dicke1954}. Oppositely, $\Delta_L$ shows ($k_0a$)-scaling, corresponding to the near-zone dipole-dipole interaction of a given SQD with all others.

For a large system ($N_xa, N_ya \gg \lambdabar$), one has to use Eqs.~(\ref{gammaR}) and~(\ref{DeltaL}) to calculate $\gamma_R$ and $\Delta_L$, keeping all terms when performing summation. It turns out that the sums in Eqs.~(\ref{gammaRpoint-like}) and~(\ref{DeltaLpoint-like}), which contain summands proportional to $|{\bf n}|^{-1}$, converge very slowly as the lattice size increases, which results in diminishing oscillations of $\gamma_R$ and $\Delta_L$ around their asymptotic values given by (see Appendix A)
\begin{subequations}
\label{gammaRDeltaLextended}
\begin{equation}
\label{gammaRextended}
\gamma_R \simeq 4.51 \frac{\gamma}{(k_0a)^2} = 4.51 \gamma \left(\frac{\lambdabar}{a}\right)^2~.
\end{equation}
\begin{equation}
\label{DeltaLextended}
\Delta_L \simeq -3.35 \frac{\gamma}{(k_0a)^3} = -3.35 \gamma \left(\frac{\lambdabar}{a}\right)^3~,
\end{equation}
\end{subequations}
As follows from Eq.~(\ref{gammaRextended}), for a large system,  the collective radiation rate $\gamma_R$ is determined by a number of SQDs within an area on the order of $\lambdabar^2$: all SQD's dipoles are in phase there. Recall that for a linear chain of emitters, $\gamma_R \sim \lambdabar/a$.~\cite{Zaitsev1983} On the contrary, the near-zone dipole-dipole interaction $\Delta_L$ changes insignificantly compared with that for a point-like system [compare Eq.(\ref{DeltaLextended}) with Eq.~(\ref{DeltaLpoint-like})].

It should be noticed that irrespectively of the system size, the inequality $|\Delta_L| \gg \gamma_R$ is always fulfilled for a dense system, $\lambdabar \gg a$. We will use this relationship in our analysis of the supercrystal's optical response.

\section{Numerics}
\label{Numerics}
We performed calculations of the system dynamics
for two resonance conditions: (i) — the applied  field $\Omega_0$ is in resonance with the one-exciton transition $\omega_0 = \omega_2$ ($\Delta_{21} = 0, \Delta_{32} = - \Delta_B$) of a single emitter (conventionally called in what follows as one-photon resonance) and (ii)— it is tuned to the two-photon resonance, $\omega_0 = \omega_3/2$ ($\Delta_{21} = \Delta_B/2, \Delta_{32} = - \Delta_B/2$). In reality, however, the single emitter resonance $\Delta_{21} = 0$ is redshifted due to the near-zone SQD-SQD interactions by $|\Delta_L|$, so that the resonance in the linear low field intensity regime is defined by the condition $\Delta_{21} = |\Delta_L|$ (see Sec.~\ref{Discussion} for detail).

In our numerical calculations we use the typical values of optical parameters of the SQDs (emitting in the visible) and SQD supercrystals (see, e.g., Fig.~\ref{fig:Supercrystal}). More specifically, the spontaneous decay rate $\gamma \approx 3\cdot 10^9$~s$^{-1}$ and the ratio $\mu = d_{32}/d_{21}  = \sqrt{\gamma_{32}/\gamma_{21}} = \sqrt{2/3}$. The magnitudes of $\gamma_R$ and $\Delta_L$ depend on the ratio $\lambdabar/a$. Taking $\lambdabar \sim 100 - 200$ nm and $a \sim 10 - 20$ nm, one obtains the following estimates for these two constants: $\gamma_R \sim 10^{12}$ s$^{-1}$  and $|\Delta_L| \sim 10^{13}$ s$^{-1}$. The typical values of the biexciton binding energy $\Delta_B$ are on the order of several meV, $\Delta_B \sim 2.5 - 10~\mathrm{meV} \sim 10^{12}$~s$^{-1}$, although for some 2D systems, like transition metal dichalcogenides~\cite{Mai2014,Mak2016}, it can be one order of magnitude larger. Therefore, the biexciton binding energy $\hbar\Delta_B$ is considered as a variable parameter. In what follows, the spontaneous emission rate $\gamma$ is used as the unit of all frequency-dimensional quantities, whereas $\gamma^{-1}$ as the time unit. According to our estimates, we set in $\gamma_R = 100\gamma$ and $|\Delta_L| = 1000\gamma$.

The equations~(\ref{rho11})-~(\ref{Omega21}) belongs to a class of so-called stiff differential equations, characterized by several significantly different time scales. In our case, these are defined by $\gamma^{-1}\gg \gamma_R^{-1} \gg |\Delta_L|^{-1}$. 
We therefore use specialized integration routines adapted for systems of such stiff equations, in particular, the ODE23tb of MATLAB and some implementations of methods based on the backward differentiation formulas.

\subsection{Steady-state analysis}
\label{steady-state}
\begin{figure}[ht!]
\begin{center}
\includegraphics[width=\columnwidth]{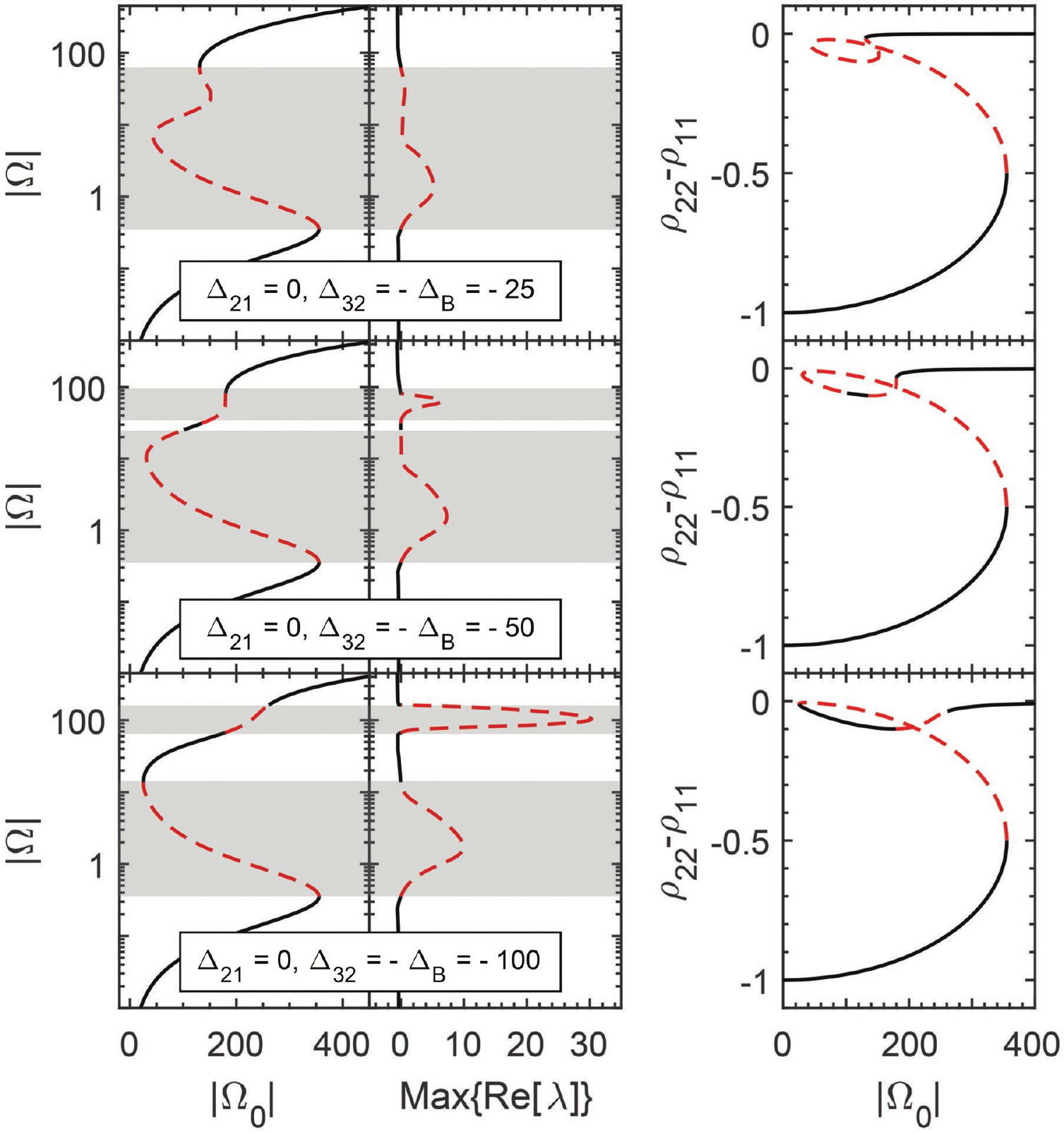}
\caption{\label{fig:one-photon_steady-state} 
Steady-state solutions to Eqs.~(\ref{rho11})-(\ref{Omega21}) for the case of one-photon resonance ($\Delta_{21} = 0, \Delta_{32} = -\Delta_B$) and for different values of the bi-exciton binding energy $\Delta_B$ (shown in the plots). The left column shows dependencies of the total field magnitude $|\Omega|$ on the excitation field $|\Omega_0|$, right column -- dependencies of the population difference $\rho_{22} - \rho_{11}$ on the $|\Omega_0|$. Unstable regions of the stationary solutions are indicated by gray shading; the maximum values of the real parts of Lyapunov exponents $\max_k{\mathrm{Re}\lambda_k}$ are shown in the middle column (see text for details). All frequency-dimension quantities are given in units of $\gamma$.}
\end{center}
\end{figure}
\begin{figure}[ht!]
\begin{center}
\includegraphics[width=\columnwidth]{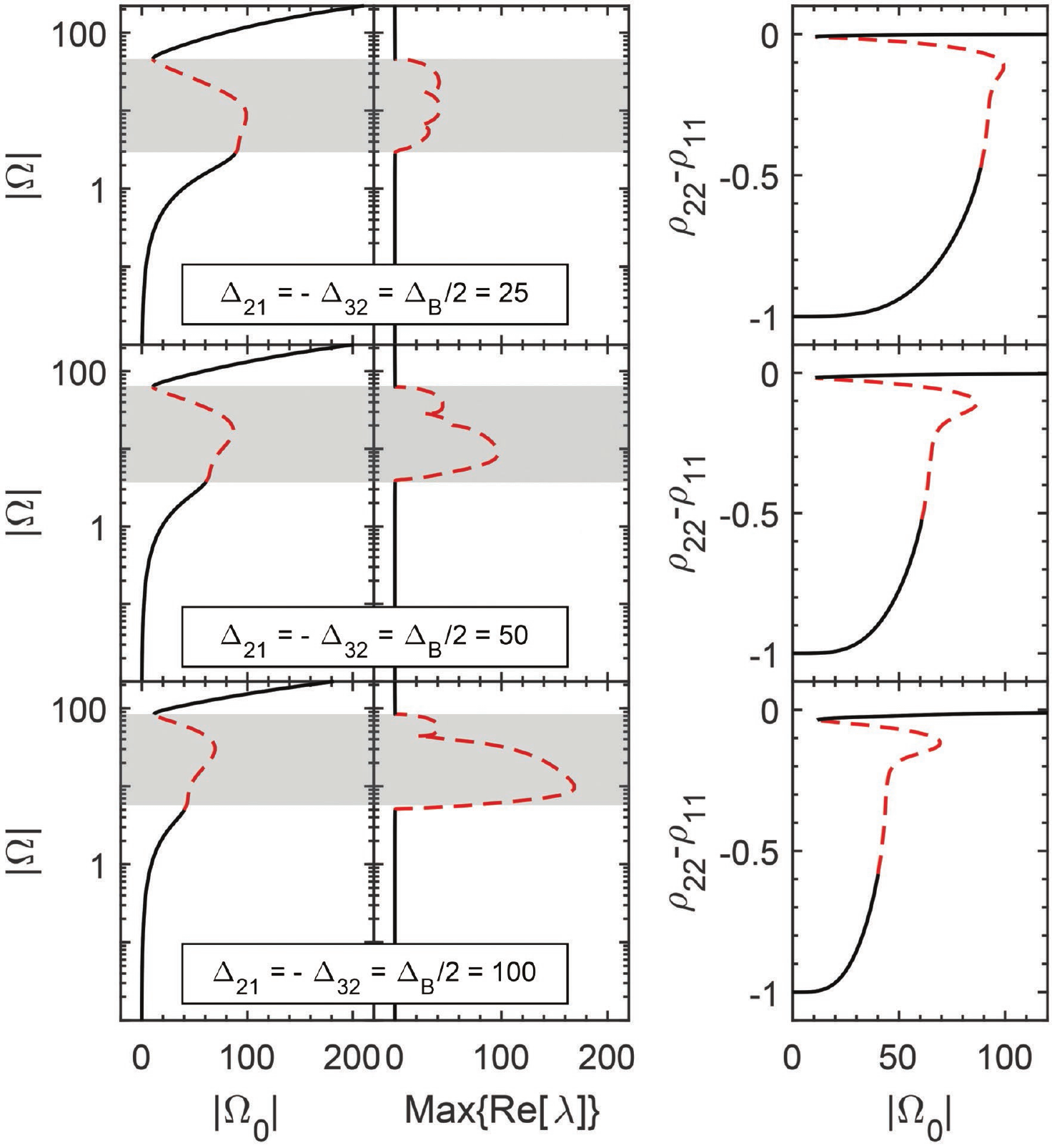}
\caption{\label{fig:two-photon_steady-state}  Same as in Fig.~\ref{fig:one-photon_steady-state}, but for the case of the two-photon resonance ($\Delta_{21} = -\Delta_{32} = \Delta_B/2$).}
\end{center}
\end{figure}

As the first step of studying the system optical response, we turn to the steady-state regime. By setting the time derivatives in Eqs.~(\ref{rho11})-(\ref{Omega21}) to zero, we obtain the system of stationary nonlinear equations which we solve by our new exact parametric method (detailed in Appendix B). The results for different values of the biexciton binding energy $\Delta_B$ are presented below in the series of figures.

Figures~\ref{fig:one-photon_steady-state} and~\ref{fig:two-photon_steady-state} show the dependence of the total field magnitude $|\Omega|$ (leftmost column) and the population difference $Z_{21} = \rho_{22} - \rho_{11}$ (rightmost column) on the external field magnitude $|\Omega_0|$ calculated for the one-photon ($\omega_0 = \omega_2,\, \Delta_{21} = 0,\, \Delta_{32} = -\Delta_B$) and two-photon ($\omega_0 = \omega_3/2,\, \Delta_{21} = -\Delta_{32} = \Delta_B/2$ resonance, respectively. As is seen from the figures, the total field magnitude $|\Omega|$ can have several solutions (up to five for $\Delta_{32} = -50$) for a given value of the external field magnitude $|\Omega_0|$, which can give rise to multistability and hysteresis phenomenon (see Sec.~\ref{Optical hysteresis}). We analyzed the stability of different branches by the standard Lyapunov exponents analysis~\cite{EckmannRevModPhys1985,KatokBook1997}. To this end, we calculated the eigenvalues $\lambda_k$ ($k=1\ldots 8$) of the Jacobian matrix of the right hand side of Eqs.~(\ref{rho11})-(\ref{Omega21}) as a function of $|\Omega|$. The exponent with the maximal real part, $\max_k{\mathrm{Re}\lambda_k}$, determines the stability of the steady-state solution: if $\max_k{\mathrm{Re}\lambda_k} \leq 0$, the solution is stable and unstable otherwise. The values of $\max_k{\mathrm{Re}\lambda_k}$ are plotted in the middle panels of Figs.~\ref{fig:one-photon_steady-state} and~\ref{fig:two-photon_steady-state}. The shaded regions show the unstable parts of the steady-state solutions (with $\max_k{\mathrm{Re}\lambda_k} > 0$).

We stress that not only the branches with the negative slope are unstable, which is always the case, but some parts of the branches with the positive slopes as well. This occurs for both the one- and two-photon resonance conditions. Quite remarkably, in the case of the one-photon resonance with $\Delta_B = 100$, a part of the upper branch of the steady-state solution is unstable. Moreover, for $\Delta_B = 50$, two unstable regions of the upper branch are separated by a stable one. The nature of these instabilities is discussed in Secs.~\ref{Bifurcation diagram} and~\ref{Time-domain}.

\subsection{Bifurcation diagram}
\label{Bifurcation diagram}
\begin{figure*}[ht!]
\begin{center}
\includegraphics[width=0.66\columnwidth]{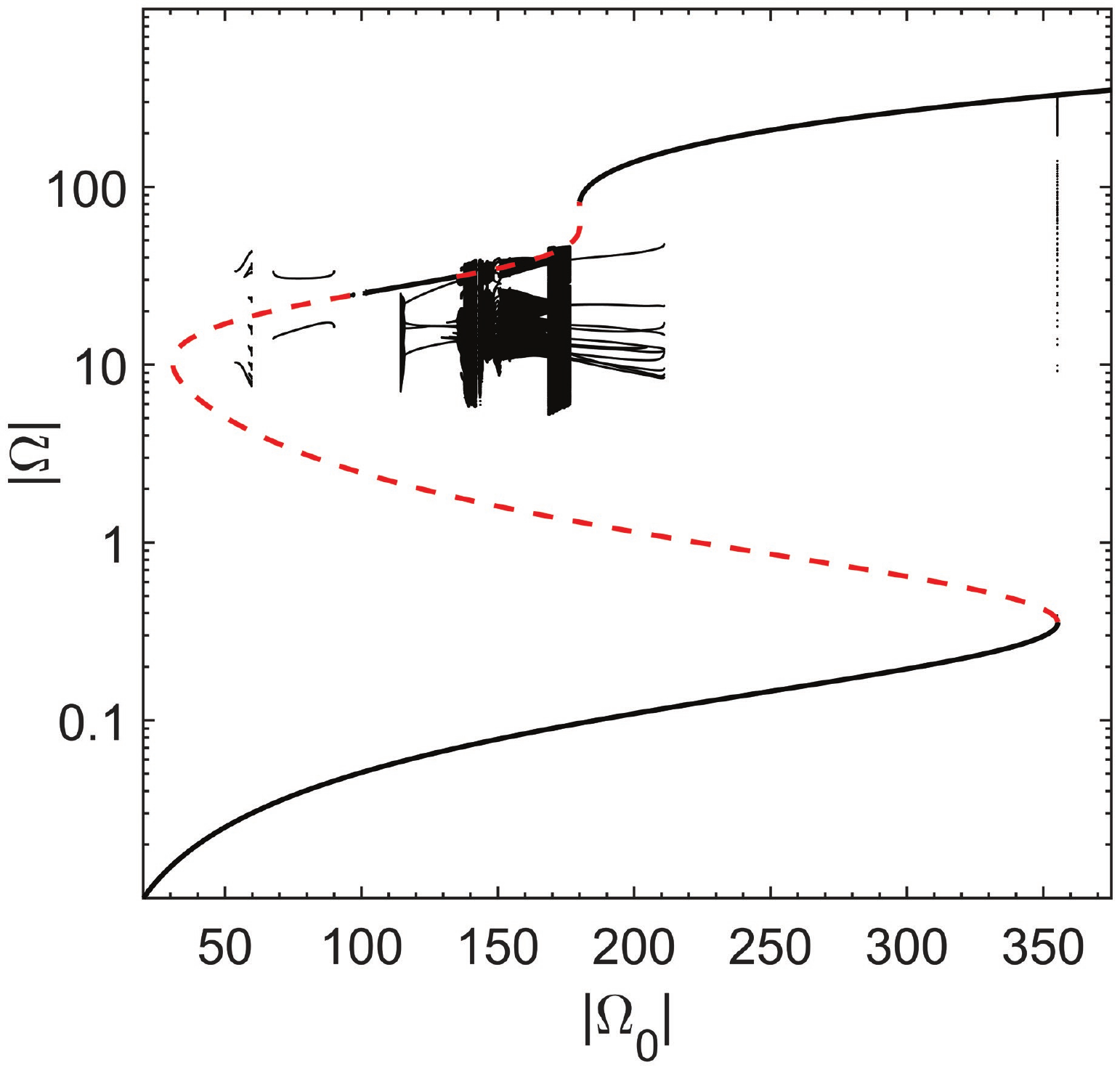}
\includegraphics[width=0.66\columnwidth]{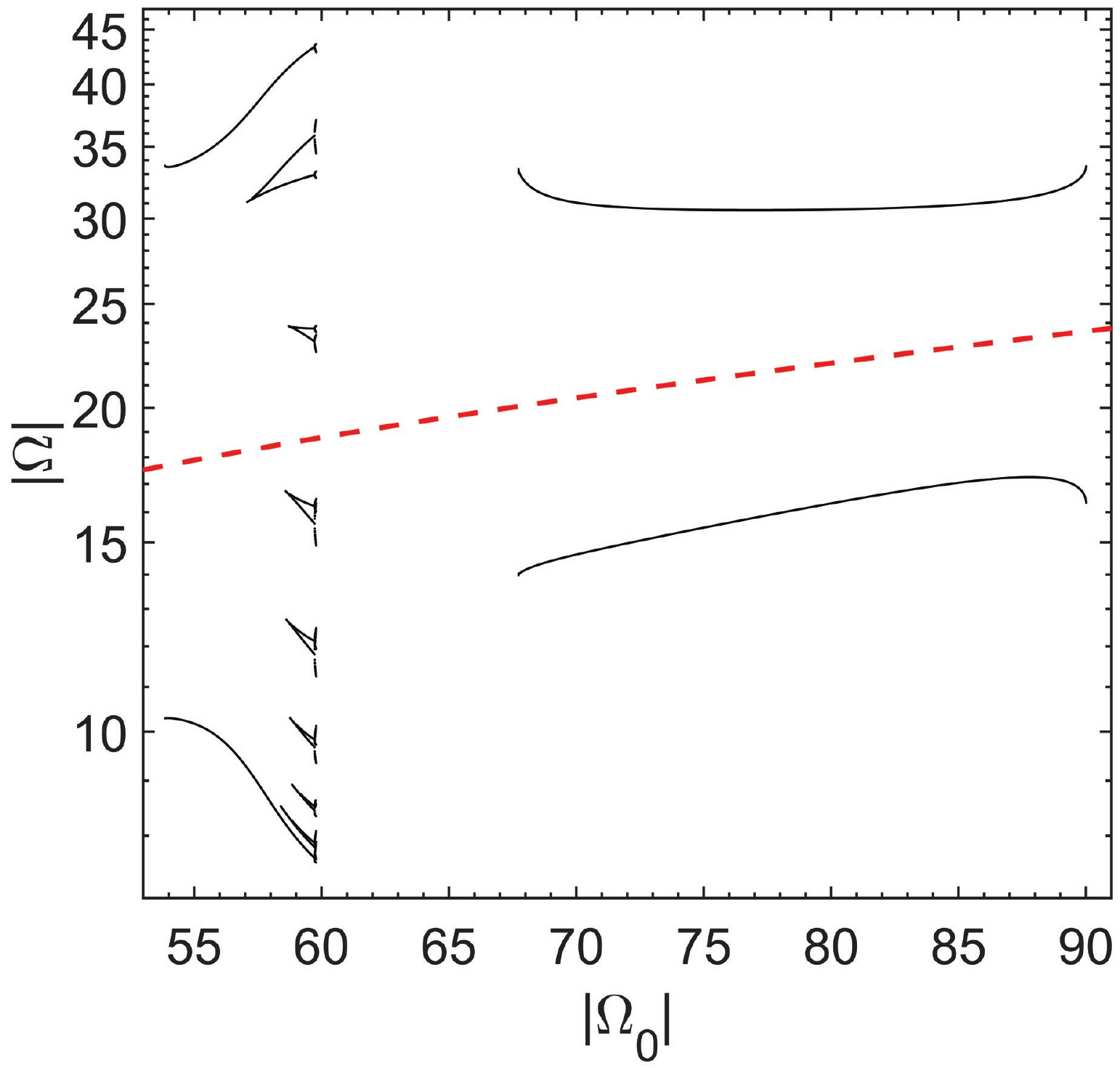}
\includegraphics[width=0.66\columnwidth]{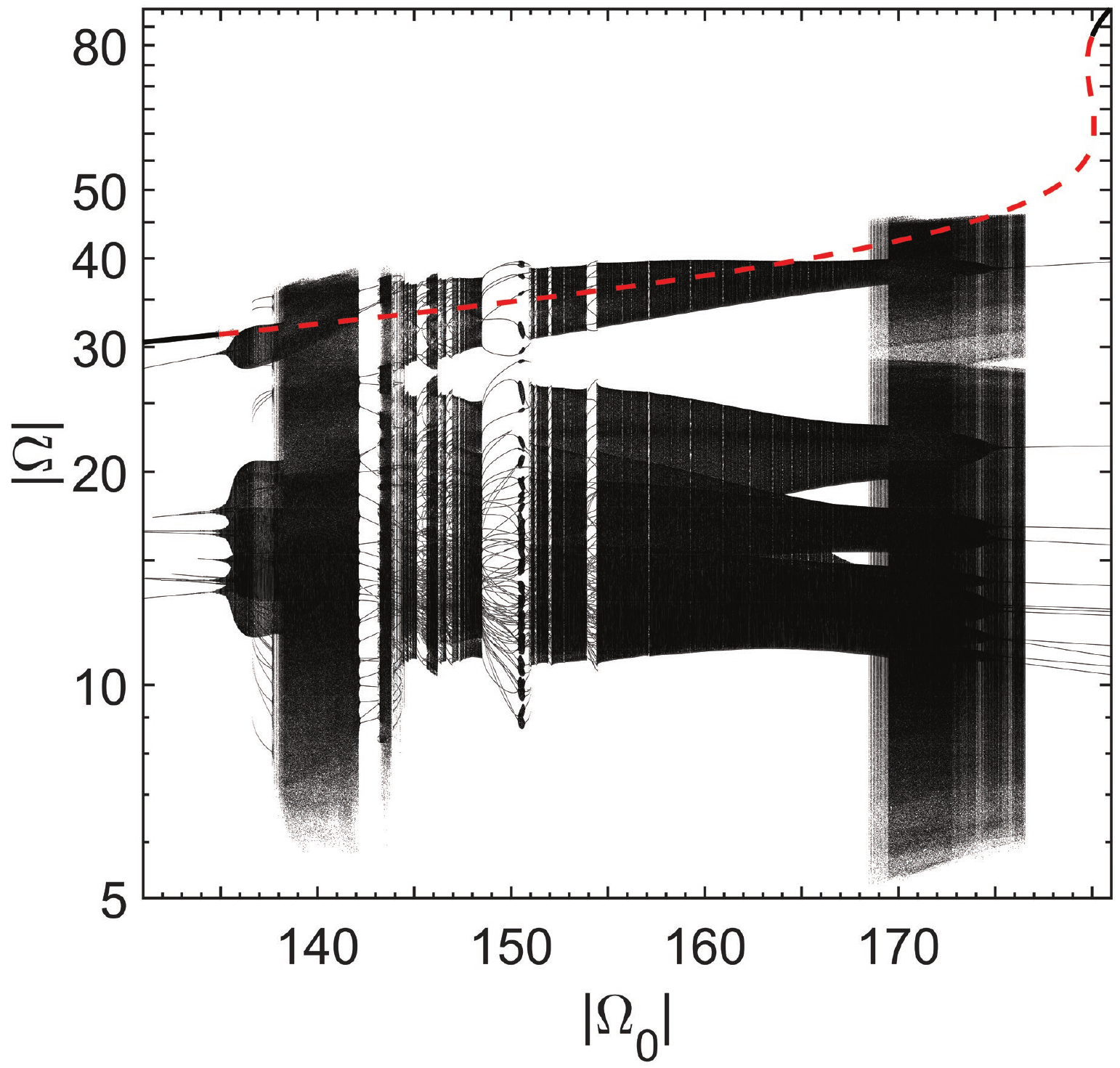}
\end{center}
\caption{\label{fig:One-photon bifurcation diagram DeltaB = 50}
Left panel: the overall bifurcation diagram [extrema of the total field magnitude $|\Omega(t)|$ on attractors as a function of the external field magnitude $|\Omega_0|$] calculated for the case of one-photon resonance $\Delta_{21} = 0, \Delta_{32} = -\Delta_B = -50$ (see text for details of the calculation method). Dashed red line shows the unstable part of the steady-state solution from  Fig.~\ref{fig:one-photon_steady-state}, which is given for reference. Middle and right panels: blow-ups of the regions of the bifurcation diagram with nontrivial dynamics (see also Sec.~\ref{Time-domain}).}
\end{figure*}
\begin{figure}[ht!]
\begin{center}
\includegraphics[width=0.7\columnwidth]{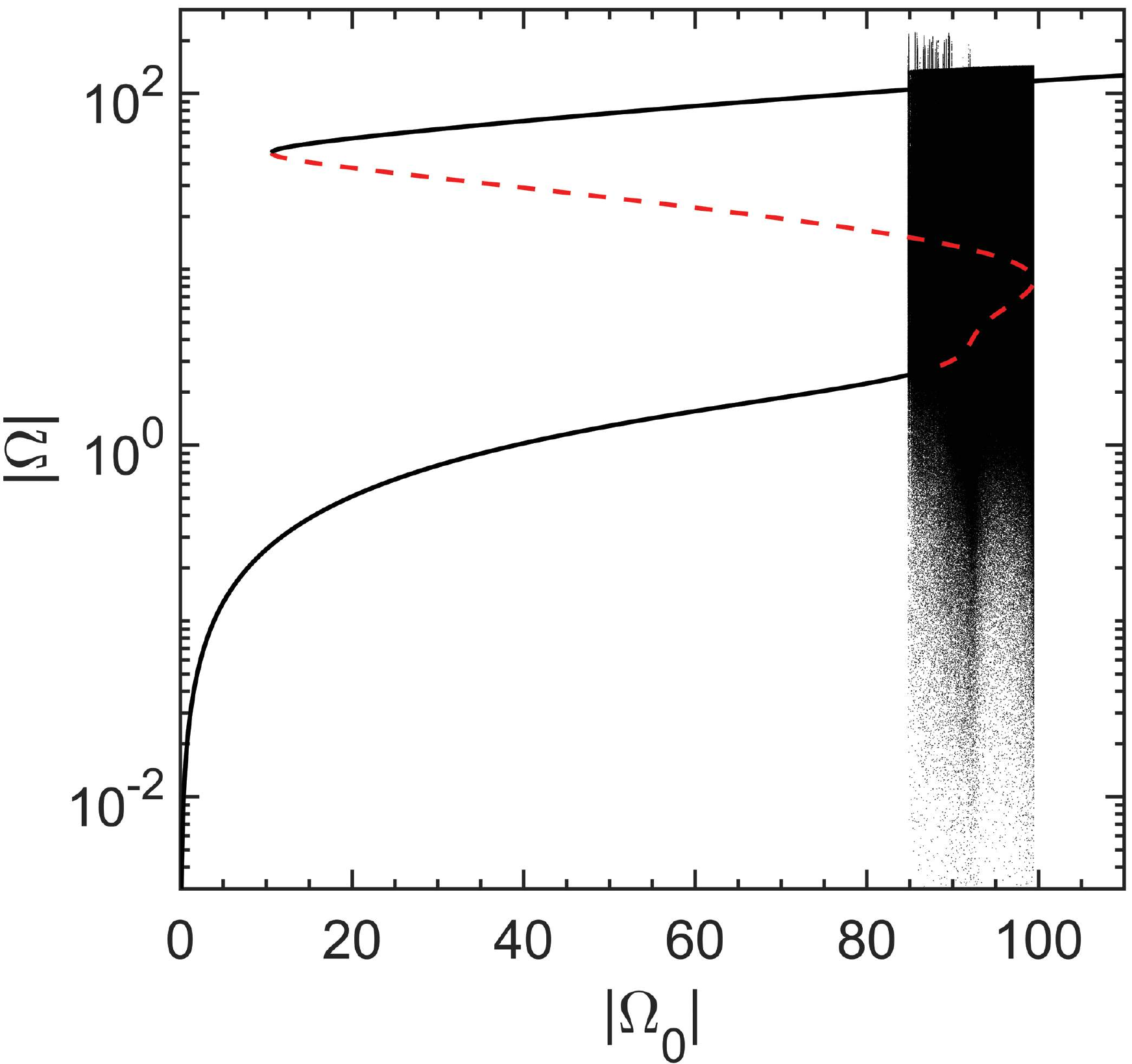}
\includegraphics[width=0.7\columnwidth]{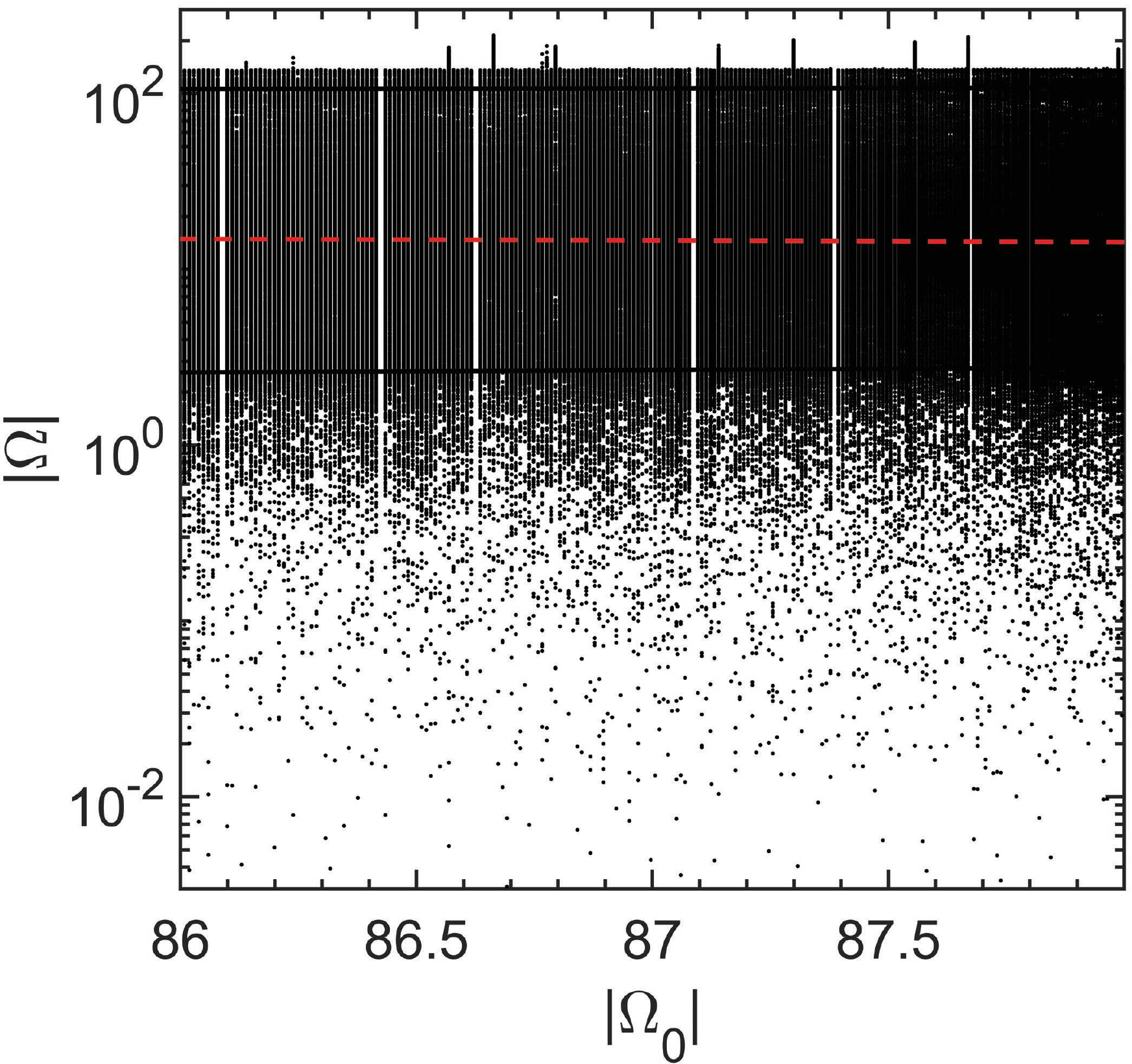}
\end{center}
\caption{\label{fig:Two-photon bifurcation diagram DeltaB = 50}
Same as in Fig.~\ref{fig:One-photon bifurcation diagram DeltaB = 50} but for the case of two-photon resonance ($\Delta_{21} = -\Delta_{32} = \Delta_B/2$) with $\Delta_B = 50$.  Upper panel -- the overall diagram, lower panel -- a blow-up of a fragment of the above diagram showing its fine structure. All quantities are given in units of $\gamma$.}
\end{figure}

The bifurcation diagram is a very useful tool providing an insight into possible scenarios of the system behavior in a graphical way~\cite{GuckenheimerBook1986,NeimarkLandaBook1992,Arnol'dBook1994,KuznetsovBook2004} by portraying the system dynamics qualitatively as a function of a control (bifurcation) parameter.  In order to construct the bifurcation diagram, we address the dynamics of the total field magnitude $|\Omega(t)|$ (which is one of the possible measurable outputs) as a function of the external field magnitude $|\Omega_0|$, which is the most natural bifurcation parameter for the system under consideration. To this end, for each $|\Omega_0|$ we plot a set of characteristic points of $|\Omega(t)|$, namely, all the extrema of the latter on an attractor. 

Our proposed method of bifurcation diagram calculation is as follows. First, we note that the steady-state characteristics ($|\Omega|$ vs $|\Omega_0|$ dependence in Figs.~\ref{fig:one-photon_steady-state} and~\ref{fig:two-photon_steady-state}) are multivalued, which can complicate numerical procedures considerably if the external field amplitude $\Omega_0$ is swept. Therefore, we sweep the total field amplitude $\Omega$ instead. 
As we argue in Appendix B, 
the latter can be considered to be real without loss of generality [if appropriate phase transformations are performed, see Eq.~(\ref{Omegaphasetrans})]. Thus, for each real valued $\Omega$, we use Eq.~(\ref{Omega21}) to obtain the unique stationary (complex valued) $\Omega_0$, whose {\it absolute value} is used as the external field amplitude in Eqs.~(\ref{rho11})-(\ref{Omega21}) to calculate the system dynamics. After going through a transient phase, the dynamics converges to an attractor. Then, we obtain all the extrema of the absolute value of the total field amplitude $|\Omega(t)|$ on the trajectory over a sufficiently long time interval. All such extrema are plotted as points for the current value of $|\Omega_0|$, forming the bifurcation diagram. 

The distribution of the extrema provides qualitative information on possible types of the system dynamics. For example, if the dynamics converges to a stable fixed point, all the extrema collapse onto a single point (within the precision of the numerical method). The point coincides with the stable stationary value of the field (the extrema exist because the solution is typically still oscillating about the stationary value due to finite precision of numerical methods). If the dynamics converges to a periodic orbit, all the extrema collapse onto a finite set of points separated by gaps (see Fig.~\ref{fig:One-photon bifurcation diagram DeltaB = 50}, middle panel). Quasi-periodic oscillations can turn up as vertical bars separated by gaps (see Fig.~\ref{fig:One-photon bifurcation diagram DeltaB = 50}, right panel), while chaos would probably display itself as a continuous vertical line. The proposed representation of the system dynamics is somewhat similar to the Lorenz map~\cite{Lorenz1963} (in the sense that it uses extrema), but it contains considerably more information. On the other hand, it is also resembling the Poincare map~\cite{EckmannRevModPhys1985} (in the way it represents different types of dynamics), but it is less complicated to calculate than the latter while providing almost equivalent qualitative information. We believe therefore that the proposed method of bifurcation diagram calculation is quite advantageous.

The choice of the initial conditions becomes very important when scanning for attractors with nontrivial dynamics (those different from a stable fixed point). Ideally, one should try out all possible initial conditions for each value of the bifurcation parameter, which is hardly feasible. Hereafter, we assume that the system can manifest interesting dynamics when it is "not too far" in the phase space from the unstable branches of the steady-state characteristics; we therefore use the following procedure to choose the initial conditions. At each step, i.e., for each value of $|\Omega|$, we are inspecting the solution from the previous step. If the previous solution appears to be on a nontrivial attractor, we take the previous solution at the final time instant as the initial condition for the current step. Thus, we try to keep the system in the basin of attraction of the nontrivial attractor. Otherwise, if the systems is converging to a stable fixed point at the previous step, we take the steady state solution corresponding to the current value of $|\Omega|$ as the initial condition. Such a solution can be on an unstable part of the stationary curve and yield some interesting dynamics. Besides, we are sweeping the parameter $|\Omega|$ across the window of interest back and forth, intending to discover the most complete set of attractors.

Finally, to ascertain that the dynamics has converged to an attractor, in other words, to make sure that the transient phase of the dynamics has passed, we apply the following procedure. We integrate the system over consecutive time intervals $\Delta T$ and calculate the range $R_n$ of the function $|\Omega(t)|$ at each interval, i.e., $$R_n=\max_{\Delta T}{|\Omega(t)|}-\min_{\Delta T}{|\Omega(t)|}\ ,$$ where $n$ is the step number. The dynamics is considered to converge to an attractor when the range $R_n$ stops growing and its change from one step to the next becomes sufficiently small in the following sense: $|R_{n+1} - R_n| < \epsilon_a$ and $|R_{n+1} - R_n|/R_n < \epsilon_r$~\footnote{Additionally, one can apply the same criteria to the average $A_n=(\min_{\Delta T}{|\Omega(t)|}+\max_{\Delta T}{|\Omega(t)|})/2$ to account for and get rid of the drift of the average towards an attractor}. In any case, the integration was stopped when the integration time reached $T_{max}$. Unfortunately, we can not propose any general method to estimate the parameters $\Delta T$, $\epsilon_a$, $\epsilon_r$, and $T_{max}$. To determine their appropriate values, we analyzed the system dynamics on different types of attractors and found out that the set $\Delta T = 50$, $\epsilon_a = 0.001$, $\epsilon_r = 0.001$ and $T_{max}=10^4$ was working well in all cases we considered. Finally, when the dynamics converges to an attractor, the extrema of $|\Omega(t)|$ are calculated on the last time interval; their values are used to construct the bifurcation diagram as explained above.

Figure~\ref{fig:One-photon bifurcation diagram DeltaB = 50} presents the bifurcation diagram calculated for the case of the one-photon resonance $\Delta_{21} = 0, \Delta_{32} = - \Delta_B$, and the biexciton binding energy $\Delta_B = 50$. The steady-state characteristics from Fig.~\ref{fig:one-photon_steady-state} is also plotted; the stable stationary branches (black lines) form the trivial part of the bifurcation diagram, while the unstable branches (red dashed line) are given for reference. The middle and right panels of the figure show blow-ups of the parts of the diagram with nontrivial dynamics. As expected, these parts are located in proximity to the unstable branches of the steady state. The figure shows that, although there are stable stationary solutions for all values of the external field magnitude $|\Omega_0|$, the system dynamics can be highly nontrivial, manifesting a wide range of attractor types. In particular, fingerprints of stable fixed points, periodic and aperiodic orbits, and chaotic trajectories can be seen. 

As observed from Fig.~\ref{fig:One-photon bifurcation diagram DeltaB = 50}, the system undergoes multiple bifurcations.
Consider, for example, the bifurcation of limit cycles existing within the range of the external field magnitude $67 \lesssim |\Omega_0| \lesssim 90$ (middle panel of Fig.~\ref{fig:One-photon bifurcation diagram DeltaB = 50}).
When $|\Omega_0|$ crosses the left boundary of the interval, being swept down, the limit cycle disappears and the system is attracted to a stable fixed point which resides in the lower stable branch of the steady-state characteristics.
This scenario resembles a subcritical Andronov-Hopf bifurcation~\cite{GuckenheimerBook1986,Arnol'dBook1994,KuznetsovBook2004}.
Once at the stable branch, the system remains at this trivial attractor even if the field magnitude is swept up to fall again within the interval, where self-oscillations can exist. Here, we deal with hysteresis of the bifurcation diagram.

We turn now to the case when the system is in a stable fixed point belonging to the intermediate positive-slope branch of the steady-state characteristics, surrounded by the unstable parts, i.e., within the interval $100 \lesssim |\Omega_0| \lesssim 135$ (see the panel of Fig.~\ref{fig:one-photon_steady-state} for $\Delta_B = 50$ and the right panel of Fig.~\ref{fig:One-photon bifurcation diagram DeltaB = 50}). If the external field magnitude $|\Omega_0|$ starts to increase and crosses the right boundary of the interval, a limit cycle is created from a stable fixed point at $|\Omega_0|\approx 137$. This change of the character of dynamics resembles a supercritical Andronov-Hopf bifurcation~\cite{GuckenheimerBook1986,Arnol'dBook1994,KuznetsovBook2004}. Further, if the external field magnitude is swept back (starts to decrease), the system would follow the non-trivial attractor until its lower field extreme (at $|\Omega_0|\approx 115$), where the auto-oscillation disappears, and the system is attracted back to the stable fixed point at the upper steady-state branch.

Figure~\ref{fig:Two-photon bifurcation diagram DeltaB = 50} shows the extrema diagram calculated for the case of the two-photon resonance ($\Delta_{21} = - \Delta_{32} = \Delta_B/2 = 25$). The black vertical feature at $85 \lesssim |\Omega_0| \lesssim 100$ represents the most interesting part of the diagram with nontrivial dynamics. The feature consists of very densely packed points forming practically continuous vertical lines, which indicates that the extrema of the total field magnitude $|\Omega(t)|$ might be distributed randomly and that the signal is presumably of a chaotic nature. We confirmed the latter by calculating the Lyapunov spectra using the standard method based on the QR factorization (decomposition of a matrix into a product of an orthogonal matrix Q and an upper triangular one R)~\cite{Shimada1979,Benettin1980a,Benettin1980b,Wolf1985,Dieci1994,Dieci1997,Scales1997} and found that a typical spectrum contains one positive exponent, a zero one, and negative remaining exponents. The latter $(+,0,-,\ldots,-)$ pattern of the signs of Lyapunov exponents is known to be a fingerprint of a chaotic trajectory. The typical value of the corresponding Lyapunov dimension, estimated using the Kaplan and Yorke's conjecture~\cite{Frederickson1983,Kaplan2006}, is $d_L \geq 4$.

However, this chaos may turns up to be transient, in the sense that, if the system is let to evolve for sufficiently long time, it will finally be attracted to one of the stable steady-state points. Such events can be seen in the blow-up shown in the lower panel of Fig.~\ref{fig:Two-photon bifurcation diagram DeltaB = 50}: the white gaps in the feature correspond to solutions that converged toward the stable stationary curve for $t\leq T_{max}$. Our calculations showed, that the time during which the transient chaotic dynamics exists is hardly predictable, besides this time seem to be very sensitive to initial conditions and the integration method, which is a typical feature of a transient chaos (see Refs.~\onlinecite{Lai2011,Tel2015} and references therein).

Regarding bifurcations occurring in the present case, we can state with definiteness only about those which arise at the edges of the black feature: at the left edge, a stable fixed point loses its stability and bifurcates into a chaotic trajectory, while at the right one, the back bifurcation takes place.  

Thus, the system dynamics can be very complex demonstrating a large variety of attractor types and bifurcations, some of which can manifest hysteresis. A detailed study of all possible bifurcations goes far beyond the scope of this paper; in what follows, we restrict ourselves to addressing some of the most prominent system dynamics scenarios in more detail.

\subsection{Time-domain analysis}
\label{Time-domain}
\begin{figure*}[ht!]
\begin{center}
\includegraphics[width=0.7\textwidth]{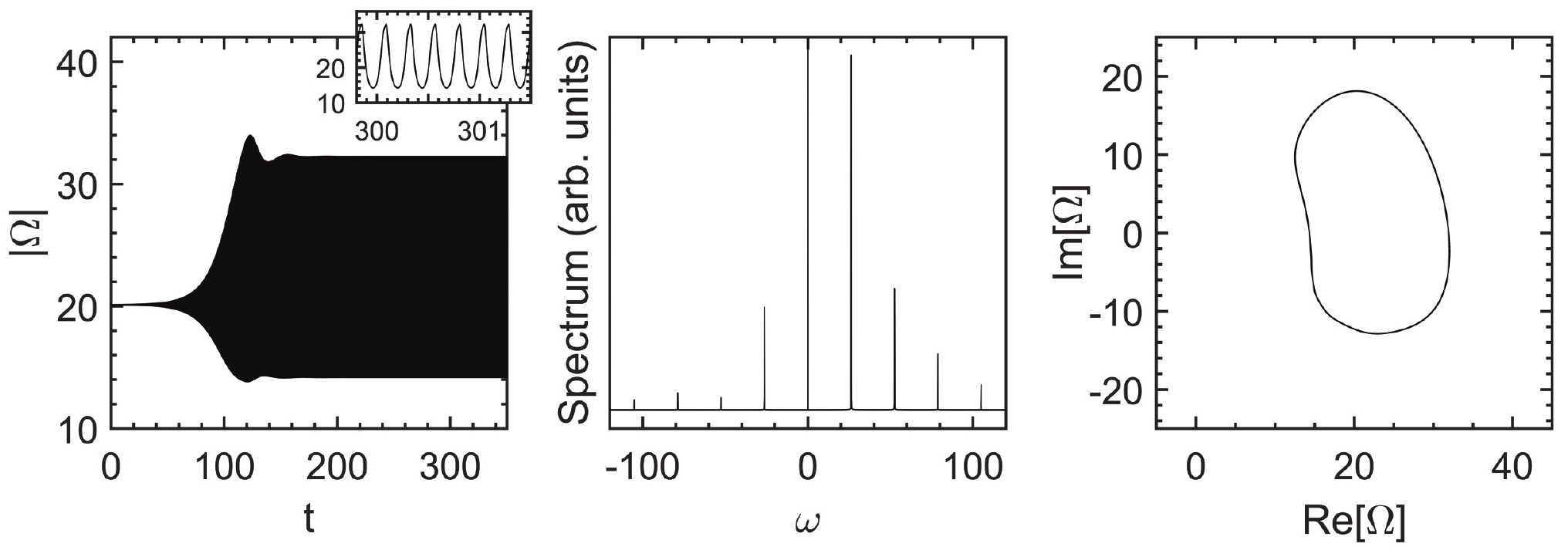}
\includegraphics[width=0.7\textwidth]{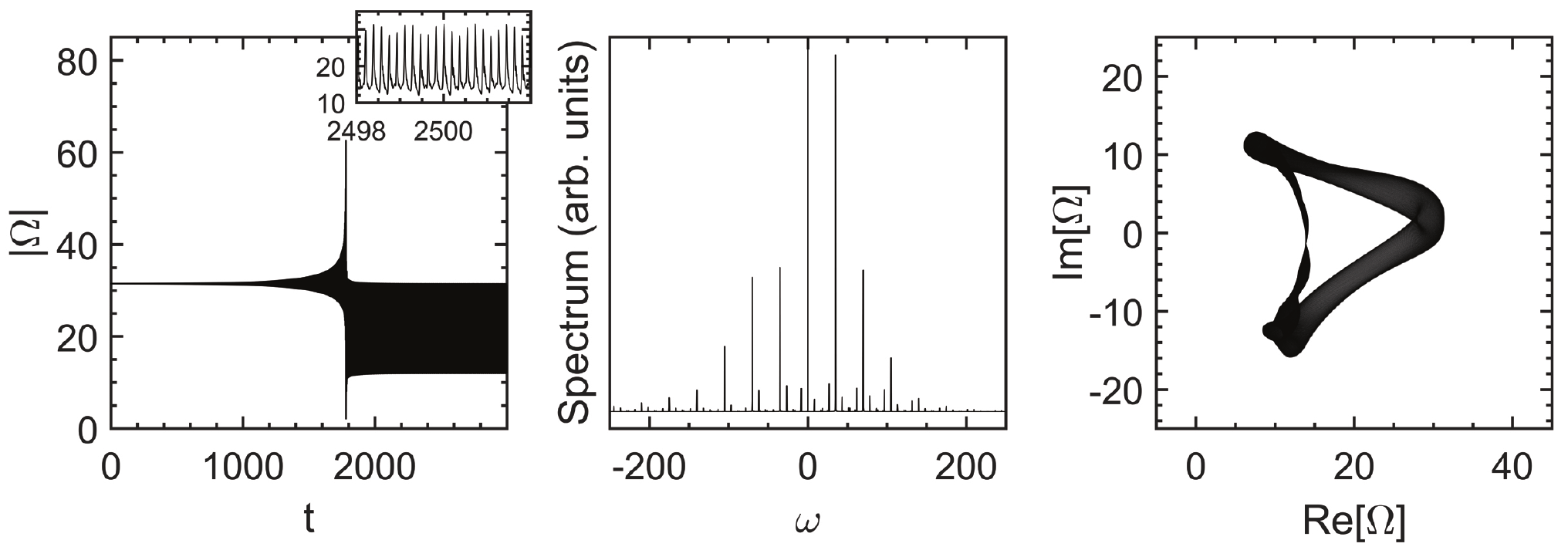}
\includegraphics[width=0.7\textwidth]{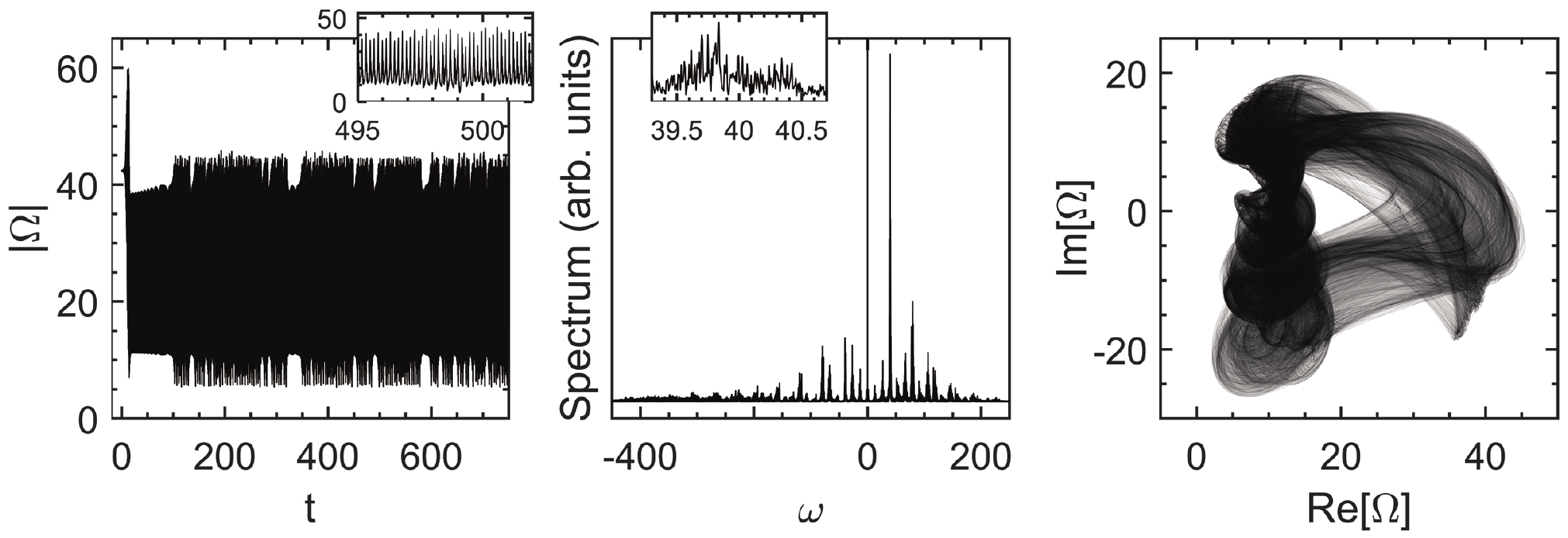}
\end{center}
\caption{\label{fig:DeltaB = 50-a}
The dynamics of the total field magnitude $|\Omega(t)|$ (left column, insets show blow-ups of the dynamics), the Fourier spectrum (middle column), and two-dimensional map of  ($\mathrm{Re}[\Omega], \mathrm{Im}[\Omega]$) on the attractor (right column). The results are calculated for the case of the one-photon resonance ($\Delta_{21} = 0$, $\Delta_{32} = - \Delta_B$) with $\Delta_B = 50$. The initial conditions were taken to be on the steady-state characteristics at $|\Omega_0| = 68.1, |\Omega| = 20.1257$ (upper row), $|\Omega_0| = 136, |\Omega| = 31.4674$ (middle row), and $|\Omega_0| = 170, |\Omega| = 42.3836$ (bottom row). All frequency-dimension quantities are given in units of $\gamma$, while time is in units of $\gamma^{-1}$.}
\end{figure*}

In this section, we present and discuss system dynamics on a variety of nontrivial attractors for either the one-photon ($\omega = \omega_2$) or two-photon ($\omega = \omega_3/2$) resonant excitation. We solve Eqs.~(\ref{rho11})--(\ref{Omega21}) for two types of initial conditions: the system is initially in the ground state ($\rho_{11} = 1$, while all other density matrix elements are equal to zero) or in a given steady-state\footnote{When the system is exactly in a stationary state it remains there forever. However, due to the final precision of numerical methods and the initial state itself, the system is in very small vicinity of the exact stationary state. Therefore it is either attracted to the steady state (if it is a stable fixed point) or drifts away from it if the stationary state is unstable.} corresponding to the external field magnitude $|\Omega_0|$. 

\subsubsection{One-photon resonance $(\Delta_{21} = 0$, $\Delta_{32} = -\Delta_B)$}
\label{One-photon resonance}

Figure~\ref{fig:DeltaB = 50-a} shows the results of time-domain calculations performed for the case of the one-photon resonance ($\Delta_{21} = 0$) with $\Delta_{32} = -\Delta_B = -50$. Three points on the unstable parts of the steady-state solution were used as initial conditions: ($|\Omega_0| = 68.1, |\Omega| = 20.1257$) - upper row, ($|\Omega_0| = 136, |\Omega| = 31.4674$) - middle row, and ($|\Omega_0| = 170, |\Omega| = 42.3836$) - bottom row.

The left panels in Fig.~\ref{fig:DeltaB = 50-a} show the time evolution of the total field magnitude $|\Omega|$. As is seen, after some delay which correlates well with the values of inverse Lyapunov exponents for the corresponding points of the steady-state solution (see Fig.~\ref{fig:one-photon_steady-state}, the middle panel for $\Delta_B = 50$) an instability starts to develop. At longer times, the latter acquires a sustained form, indicating that the system is on an attractor. In the middle panels, the Fourier spectra ($|\int dt \exp(i\omega t)\Omega(t)|$) on the attractor are plotted. The right panels shows the trajectories on attractors in the reduced phase space $(\mathrm{Re}\,\Omega,\mathrm{Im}\,\Omega)$.

The figure shows that the character of motion on the attractor depends on the initial point. For example, for $|\Omega_0| = 68.1, |\Omega| = 20.1257$ (the upper row in Fig.~\ref{fig:DeltaB = 50-a}), the system dynamics looks like a simple self-oscillations [see the left panel and also the inset for a blow up of the dynamics of $|\Omega(t)|$]. Accordingly, the Fourier spectrum (middle panel) contains a few well-defined harmonics of the base frequency while the phase space map (right panel) represents a closed curve, commonly called a limit cycle~\cite{EckmannRevModPhys1985,KatokBook1997}. The pattern of the Lyapunov exponents signs ($0,-,\ldots,-$) is also typical for a limit cycle.

For $|\Omega_0| = 136, |\Omega| = 31.4674$ (the middle row in Fig.~\ref{fig:DeltaB = 50-a}), the dynamics manifests signature of aperiodic oscillations. In this case the Fourier spectrum is also discrete, but now together with the equidistant peaks there are also satellites with incommensurate frequencies. The phase space map represents a stripe-like trajectory, densely filling a finite area in the phase space. This is a signature of aperiodic motion on a hypertorus.

Finally, for $|\Omega_0| = 170, |\Omega| = 42.3836$ (the bottom row), the dynamics is more complicated (see the inset in the left panel). The Fourier spectrum consists of a set of broadened peaks at a noisy background (see also the inset in the middle panel). This regime is chaotic; our calculations of the Lyapunov exponents spectrum confirm that: the signs of the exponents have the typical ($+,0,-,\ldots,-$) pattern and the Lyapunov dimension is $d_L\approx 3.42$.

\subsubsection{Two-photon resonance $(\Delta_{21} = -\Delta_{32} = \Delta_B/2)$}
\label{Two-photon resonance}
\begin{figure*}[ht!]
\begin{center}
\includegraphics[width=0.7\textwidth]{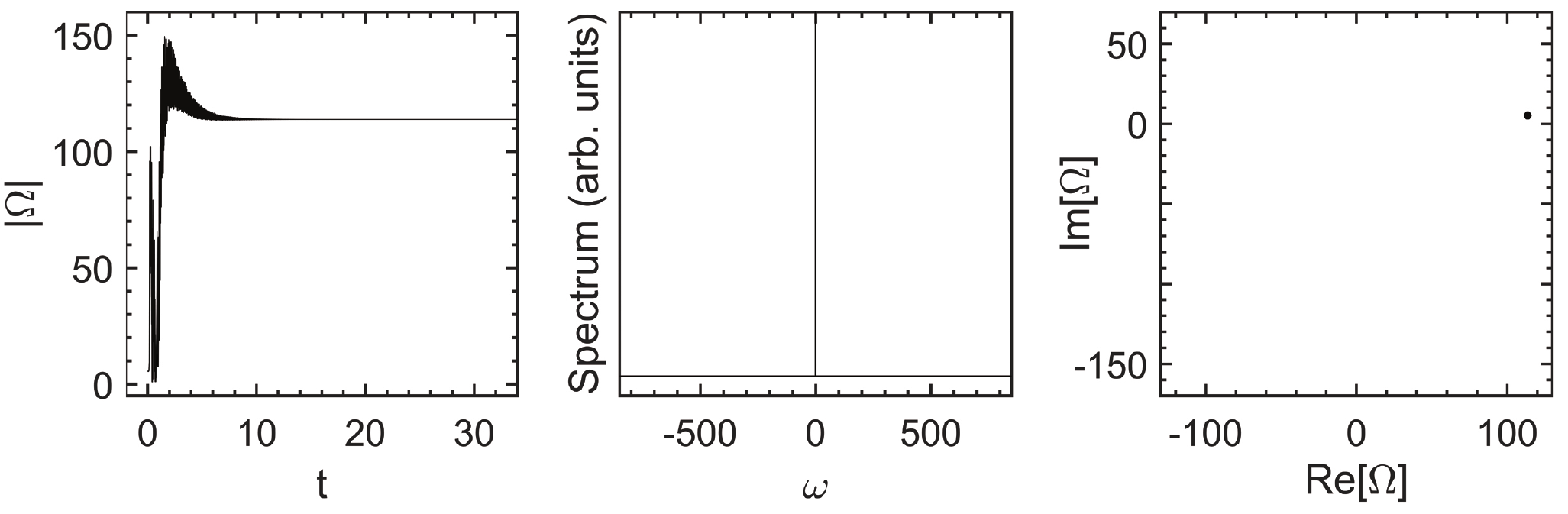}
\includegraphics[width=0.7\textwidth]{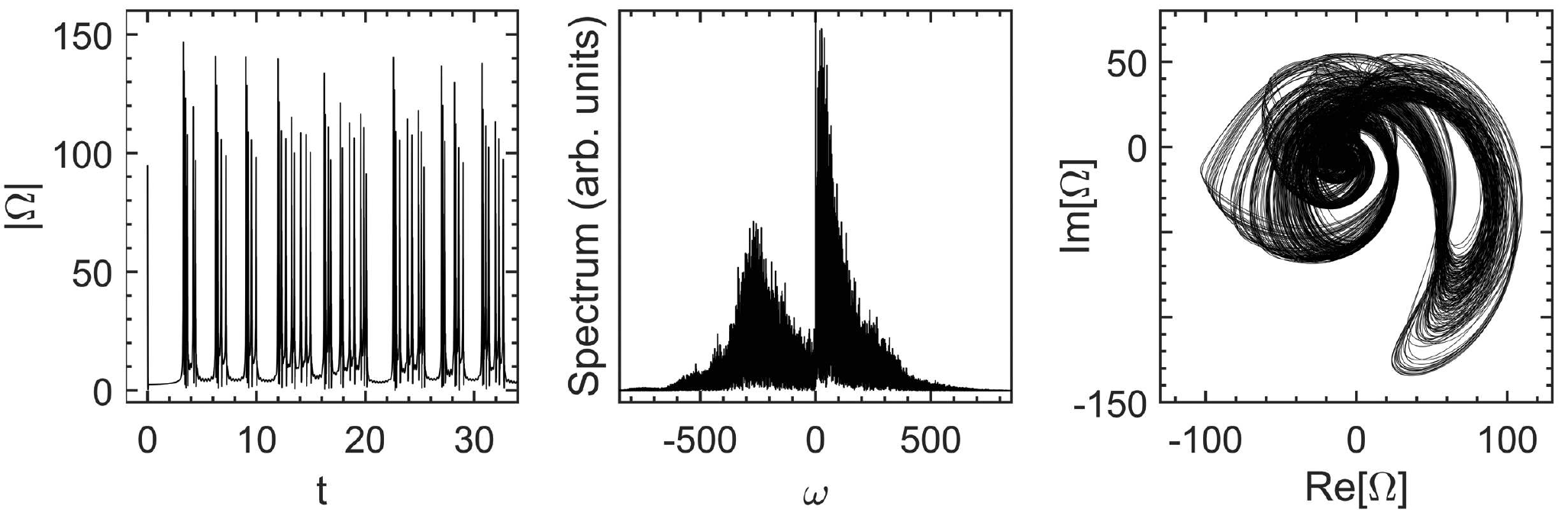}
\includegraphics[width=0.7\textwidth]{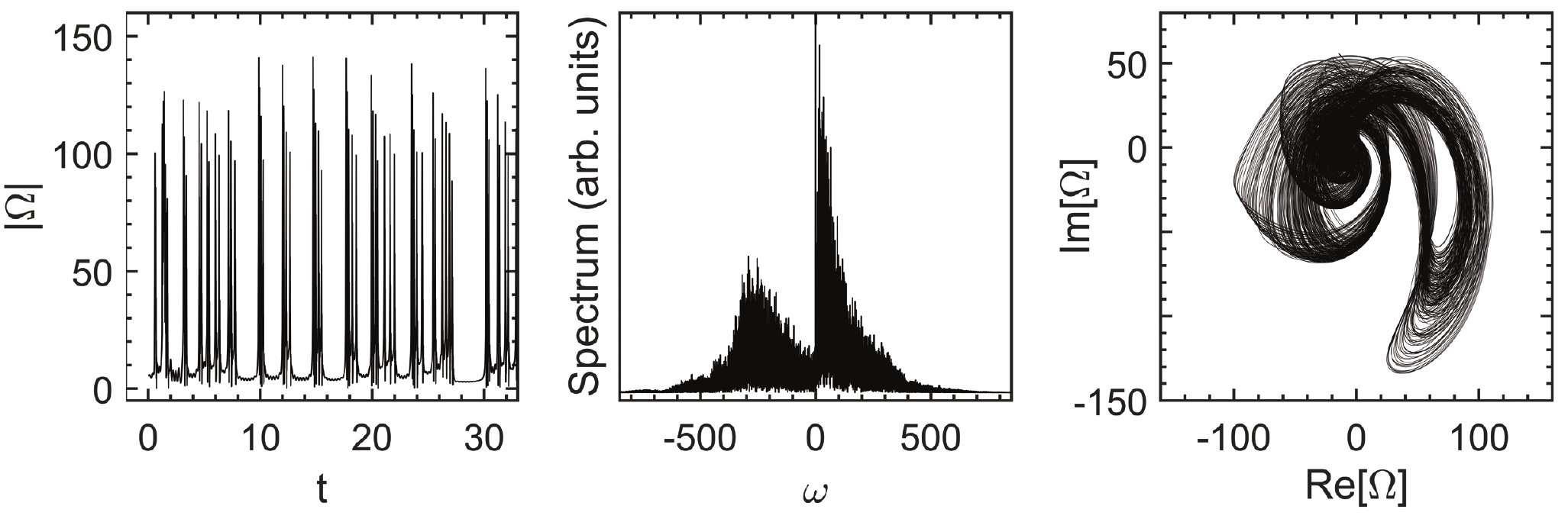}
\end{center}
\caption{\label{fig:Delta21 = -Delta32 = 25a}
Same as in Fig.~\ref{fig:DeltaB = 50-a}, but for the case of the two-photon resonance ($\Delta_{21} = -\Delta_{32} = \Delta_B/2 = 25$) and two values of of the external field magnitude $|\Omega_0|$. Top and middle rows - the system resides initially in the steady-state points $(|\Omega_0| = 95, |\Omega| = 5.5214)$ and $(|\Omega_0| = 95.2, |\Omega| = 5.6121)$, respectively. Bottom row - the system initially is in the ground state [$\rho_{11}(0) = 1$] and $|\Omega_0| = 95$.}
\end{figure*}

In the case of the two-photon resonance ($\omega_0 = \omega_3/2$), a part of the lower branch of the steady state solution with a positive slope is unstable (see Fig.~\ref{fig:two-photon_steady-state}). As a result, the dynamics can be nontrivial even if the system is initially in the ground state, in contrast to the case of the one-photon resonance discussed in the preceding section. We therefore consider both the steady-state and the ground-state initial conditions; the corresponding results are presented in Fig.~\ref{fig:Delta21 = -Delta32 = 25a}.

The top row in Fig.~\ref{fig:Delta21 = -Delta32 = 25a} shows the system dynamics for $\Delta_B = 50$ and the steady-state initial condition at ($|\Omega_0| = 95, |\Omega| = 5.5$). As is seen, after a short transient phase, the system evolves towards the fixed point on the upper stable branch of the stationary curve that corresponds to $|\Omega_0| = 95$. Accordingly, the Fourier spectrum on the attractor consists of a single peak at zero frequency and the phase space map is a point.

On the contrary, if the ground-state initial condition is used for the same external field magnitude $|\Omega_0| = 95$ (middle row), the dynamics is seemingly chaotic, manifesting a very irregular train of pulses. The Fourier spectrum is practically continuous in this case, while the reduced phase space map of the trajectory seem to have a completely filled volume. The sign pattern of the Lyapunov exponents is ($+,0,-,\ldots,-$) indicating that the trajectory is indeed chaotic; the Lyapunov dimension is $d_L\approx 4.7$. 

The results of calculations, performed for another value of the external field magnitude $|\Omega_0| = 95.2$, turned out to be essentially independent on the initial conditions. The output for the unstable steady-state point ($|\Omega_0| = 95.2, |\Omega| = 5.6$) is shown in the bottom row of Fig.~\ref{fig:Delta21 = -Delta32 = 25a} and reveals a chaotic behavior of the system, in contrast with the unstable steady-state point ($|\Omega_0| = 95, |\Omega| = 5.5$).

\subsubsection{Optical hysteresis}
\label{Optical hysteresis}
\begin{figure}[ht!]
\begin{center}
\includegraphics[width=0.7\columnwidth]{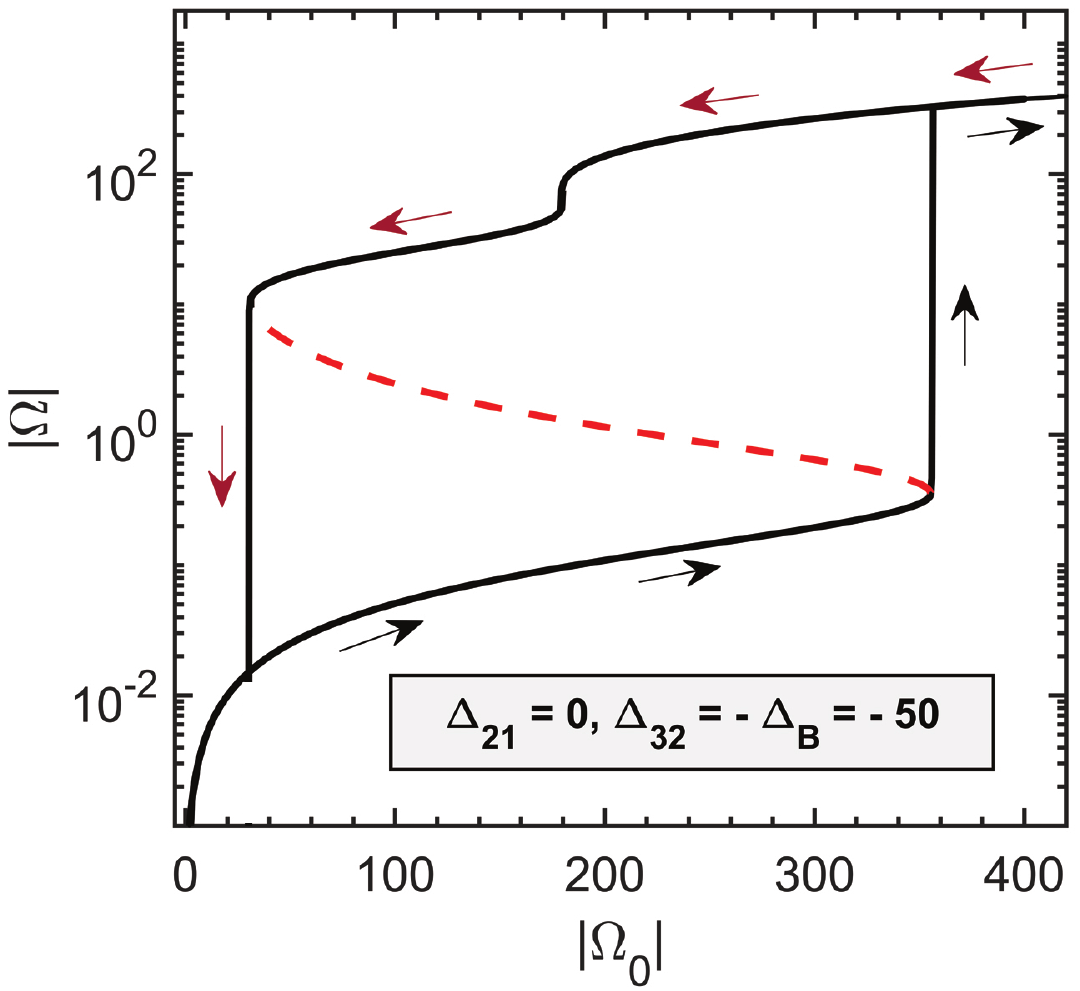}
\end{center}
\caption{\label{fig:Hysteresis one-photon DeltaB = 50} 
Optical hysteresis of the total field magnitude $|\Omega|$ calculated by solving Eqs.~(\ref{rho11})-(\ref{Omega21}) for the case of the one-photon resonance ($\Delta_{21} = 0$, $\Delta_{32} = - \Delta_B$) for $\Delta_B = 50$. The external field magnitude $|\Omega_0|$ was slowly swept back and forth across the multivalued part of the steady-state curve from Fig.~\ref{fig:one-photon_steady-state}, which is also shown for reference. The arrows indicate the sweep direction.}
\end{figure}
\begin{figure}[hb]
\begin{center}
\includegraphics[width=0.7\columnwidth]{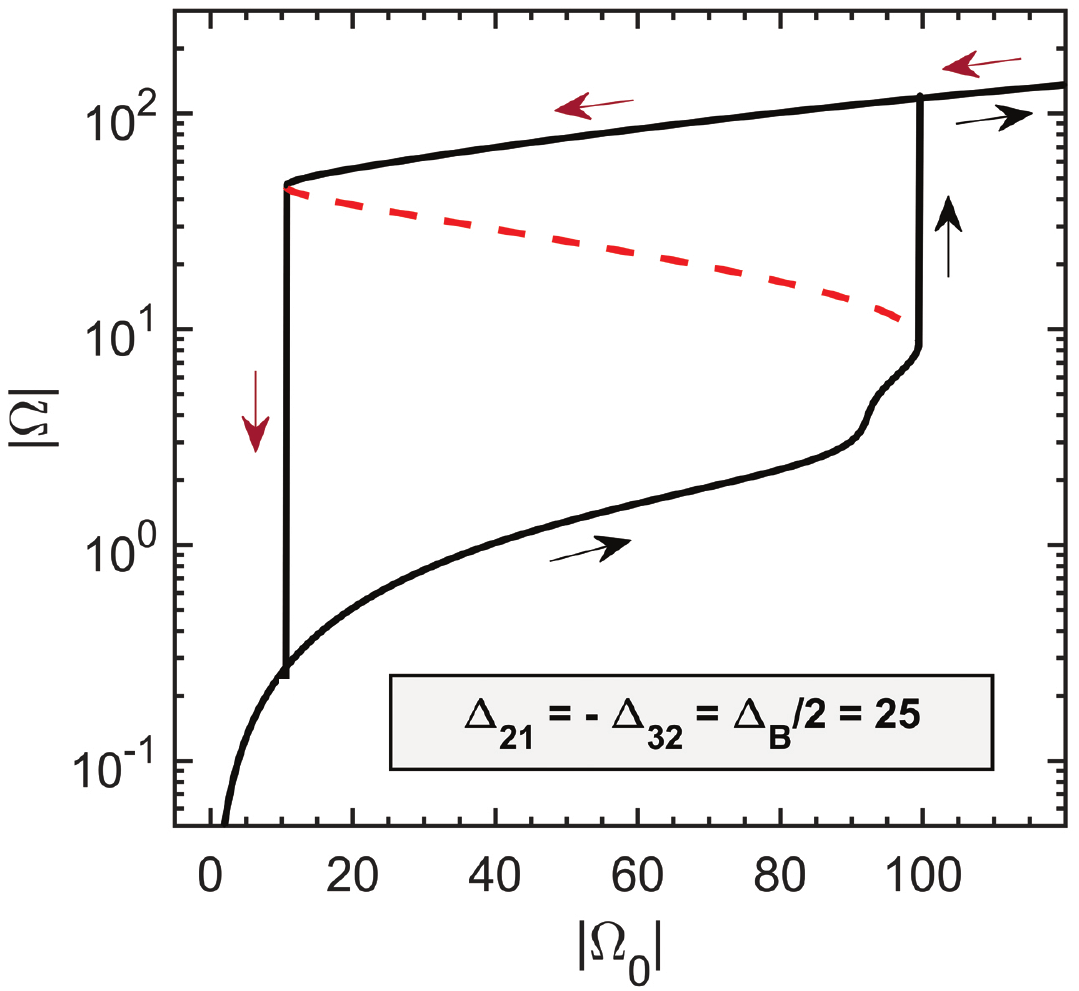}
\end{center}
\caption{\label{fig:Hysteresis two-photon DeltaB = 50} Same as in Fig.~\ref{fig:Hysteresis one-photon DeltaB = 50}, but for the case of the two-photon resonance ($\Delta_{21} = -\Delta_{32} = -\Delta_B/2$) with $\Delta_B = 50$.}
\end{figure}

The multivalued character of the steady-state (see Figs.~\ref{fig:one-photon_steady-state} and~\ref{fig:two-photon_steady-state}), can give rise to a hysteresis of the system response, when the external field magnitude $|\Omega_0|$ is slowly swept back and forth. It is unclear, however, whether the hysteresis loops are stable, because some parts of the steady-state solutions, through which the system is driven by the field, are unstable. Figures~\ref{fig:Hysteresis one-photon DeltaB = 50} and~\ref{fig:Hysteresis two-photon DeltaB = 50} show the corresponding results for the one- and two-photon resonance excitation, respectively. In both cases, the hysteresis loops appear to be stable. 

In the hysteresis loop calculations, the external field magnitude $|\Omega_0|$ was swept linearly in the following way $|\Omega_0| = 0.002 t$ for $0\leq t \leq T$ and $|\Omega_0| = 0.002 (2T - t)$ for $T\leq t \leq 2T$, where the time $T$ is chosen in such a way that the whole multivalued part of the steady-state characteristics is scanned.

From Figs.~\ref{fig:Hysteresis one-photon DeltaB = 50} and~\ref{fig:Hysteresis two-photon DeltaB = 50} it follows that in both cases, the optical response is bistable within a window of external field amplitudes. As the external field magnitude $|\Omega_0|$ is increased from zero, the total field magnitude $|\Omega|$ follows the lower branch of the steady state characteristics until it reaches the right critical point at which $|\Omega|$ abruptly jumps up to the upper stable branch where the system is saturated. On decreasing $|\Omega_0|$ the system remains on the upper branch until $|\Omega_0|$ reaches the left critical point, where the system abruptly jumps down to the lower branch, completing the hysteresis loop. Branches with the negative slope are not accessible in the adiabatic numerical experiment.

\subsection{Discussion}
\label{Discussion}
As we argue above, the considered system demonstrates a very rich optical dynamics: multistability, periodic and aperiodic self-oscillations, and dynamical chaos. The origin of such a behavior is derived from the secondary field produced by the SQDs, which depends on the current state of SQDs. This can provide a strong enough positive feedback resulting finally in instabilities. If the secondary field is neglected all above mentioned effects disappear.

Below, we discuss the underlying nonlinearities giving rise to the exotic SQD supercrystal optical response. To this end, let us consider Eqs.~(\ref{R21}) and~(\ref{R32}). Substituting  into Eqs.~(\ref{R21}) and~(\ref{R32}) the expression (\ref{Omega21}) for the field $\Omega$, one gets
\begin{widetext}
\begin{subequations}
\begin{eqnarray}
\label{R21 extended}
    \dot{\rho}_{21} = &-& \left[ i(\Delta_{21} - \Delta_L Z_{21}) + \frac{1}{2}\gamma - \gamma_R Z_{21} \right] \rho_{21}
    \nonumber \\
    &+& \mu(\gamma_R + i\Delta_L)Z_{21}\rho_{32}  + \mu(\gamma_R - i\Delta_L)(\rho_{21}^* + \mu\rho_{32}^*) \rho_{31}
    + \Omega_0 (Z_{21} + \mu\rho_{31})~,
\end{eqnarray}
\begin{eqnarray}
\label{R32 extended}
    \dot{\rho}_{32} = &-& \left[ i(\Delta_{32} - \mu^2\Delta_L Z_{32}) + \frac{1}{2} (1 + \mu^2)\gamma -\mu^2\gamma_R Z_{32} \right] \rho_{32}
    \nonumber \\
    &+& \mu(\gamma_R + i\Delta_L)Z_{32}\rho_{21} - (\gamma_R - i\Delta_L)(\rho_{21}^* + \mu\rho_{32}^*) \rho_{31}
    + \Omega_0 (\mu Z_{32} - \rho_{31})~,
\end{eqnarray}
\end{subequations}
\end{widetext}
As is seen, these equations contain a number of nonlinear terms, however a special attention should be paid to the first terms in the right-hand sides, which describe oscillations and decay of the off-diagonal density matrix elements $\rho_{21}$ and $\rho_{32}$. Note that the secondary field results in an additional frequency detuning $\Delta_L Z_{21}$ and damping $\gamma_R Z_{21}$ for $\rho_{21}$ and, respectively, $\mu^2\Delta_L Z_{32}$ and $\mu^2\gamma_R Z_{32}$ for $\rho_{32}$. These additional quantities depend on the corresponding population differences $Z_{21}$ and $Z_{32}$. Thus, the following renormalizations are evident: $\Delta_{21} \mapsto \Delta_{21} - \Delta_L Z_{21}$ and $\gamma/2 \mapsto \gamma/2 - \gamma_R Z_{21}$ for the transition $2 \leftrightarrow 1$, and $\Delta_{32} \mapsto \Delta_{32} - \mu^2\Delta_L Z_{32}$ and $(1 + \mu^2)\gamma/2 \mapsto (1 + \mu^2)\gamma/2 - \mu^2\gamma_R Z_{21}$ for the transition $3 \leftrightarrow 2$. 

Before the external field is switched on and the system is in the ground state, the population difference $Z_{21} = -1$, whereas $Z_{32} = 0$, because the states $|2 \rangle$ and $|3 \rangle$ are not populated. Accordingly, only the ($1 \leftrightarrow 2$) transition experiences the above mentioned renormalization, whereas the ($2 \leftrightarrow 3$) transition does not. Thus, the initial values of the parameters of the $1 \leftrightarrow 2$ transition detuning and decay rate are $\Delta_{21} - |\Delta_L|\approx - |\Delta_L|$ and $\gamma/2 + \gamma_R \approx \gamma_R$, respectively (here we took into account that $|\Delta_L| \gg \Delta_{21}$ and $\gamma_R \gg \gamma/2$). All other resonance detunings and decay rates keep their bare values.

When the external field is switched on, the system starts to evolve reaching finally the strong excitation regime.  Alongside the dynamic shift $\Delta_L Z_{21}$ is increasing whereas the shift $\mu^2\Delta_L Z_{32}$ is decreasing, which is driving the initially off-resonance situation towards a better resonance condition for both transitions. As a result, the redistribution of the level populations and the competition between transitions come into play creating necessary conditions for emerging instabilities (see Ref.~\cite{Nugroho2017} for more details).

The system manifests the bistability and hysteresis because the values of parameters $\Delta_L$, $\gamma_R$ are
far above the bistability threshold~\cite{Friedberg1989,Malyshev2012}.

Finally, as far as parameters are concerned, we would like to note that the model has so many of them that a complete study of the whole parameter space is a hardly feasible. However, for particular systems, such as supercrystals comprising semiconductor quantum dots, some parameters are well known. In particular, the relaxation rates $\gamma_{21}$ and $\gamma_{32}$ and the relationship between them, while the biexciton binding energy $\Delta_B$ can vary by a factor of about $4-5$. To demonstrate the possible impact of variations of the latter parameter, we presented results for a range of values of $\Delta_B$ (see Figs.~\ref{fig:one-photon_steady-state} and \ref{fig:two-photon_steady-state}).

The parameters $\Delta_L$ and $\gamma_R$ (that are related to the secondary field) were kept fixed throughout the study. They have been estimated on the basis of experimental data presented in Fig.~\ref{fig:Supercrystal}. In principle, both $\Delta_L$ and $\gamma_R$ vary if the lattice constant of the supercrystal is different. We performed additional calculations (not presented here) for the values of these parameters twice as small as the ones used in this paper. As can be expected, the results were quantitatively different but the system was manifesting the same wide range of nontrivial dynamics. The robustness of the dynamics is related to the fact that $\Delta_L$ is the largest parameter in the problem and it therefore determines the optical response. Only when the value of $\Delta_L$ becomes comparable to that of $\Delta_B$, the system becomes stable and all nontrivial dynamics scenarios disappear.

\section{Reflectance}
\label{Reflectance}

In our analysis of the system's nonlinear response, we addressed the total field $\Omega$ acting on an emitter. Although this field can be measured by near-field techniques, it is less demanding to measure the reflected or transmitted fields. These are determined by the far-zone part of $\Omega$ and are given by the following expressions:
\begin{subequations}
\begin{equation}
\label{Reflected field}
\Omega_\mathrm{refl} = \gamma_R (\rho_{21} + \mu\rho_{32})~.
\end{equation}
\begin{equation}
\label{Transmitted field}
\Omega_\mathrm{tr} = \Omega_0 + \gamma_R (\rho_{21} + \mu\rho_{32})~.
\end{equation}
\end{subequations}
The reflectance $R$ and transmittance $T$ are then defined as
\begin{equation}
\label{Reflection and Transmission}
R = \left|\frac{\Omega_\mathrm{refl}}{\Omega_0}\right|^2, \quad T = \left|\frac{\Omega_\mathrm{tr}}{\Omega_0}\right|^2~.
\end{equation}

Let us first consider the linear regime of excitation and restrict ourselves to analyzing the steady-state reflectance. In this case, the major contribution to the secondary field comes from $\rho_{21}$ which is given by
\begin {equation}
\label{eq:rho21 linear}
    \rho_{21} = - \frac{\Omega_0}{\frac{1}{2}\gamma + \gamma_R +i(\Delta_{21} + \Delta_L)}~.
\end{equation}
Substituting Eq.~(\ref{eq:rho21 linear}) into (\ref{Reflection and Transmission}), one obtains the following approximate expression for the reflectance $R$:
\begin{equation}
\label{Reflectance linear}
    R = \left| \frac{\gamma_R}{\frac{1}{2}\gamma + \gamma_R +i(\Delta_{21} + \Delta_L)}\right|^2~.
\end{equation}
It follows from the latter expression that for the range of relatively small detunings used so far in our calculations ($\Delta_{21}<100$), the reflectance 
$$R \approx\left|\frac{\gamma_R}{\Delta_L}\right|^2\ll 1\ ,$$ 
because $|\Delta_L| \gg \Delta_{21}, \gamma_R$. 
Remarkably, if the excitation frequency is in the vicinity of the resonance renormalized by the near field, i.e., $|\Delta_{21}+\Delta_L|\ll\gamma_R$, the reflectance of the system is close to unity, $R \approx 1$. Thus, in this region of frequencies, the SQD supercrystal operates as a {\it perfect mirror}. It has been reported recently that an atomically thin mirror can be realized based on a monolayer of MoSe$_2$~\cite{Back2018,Scuri2018}. SQD supercrystals represent yet another class of nanoscopically thin reflectors. The advantage of the latter, however, is that the properties of the SQD-based mirror can be controlled by the geometry and materials of the nanostructure.

\begin{figure}[ht!]
\begin{center}
\includegraphics[width=0.7\columnwidth]{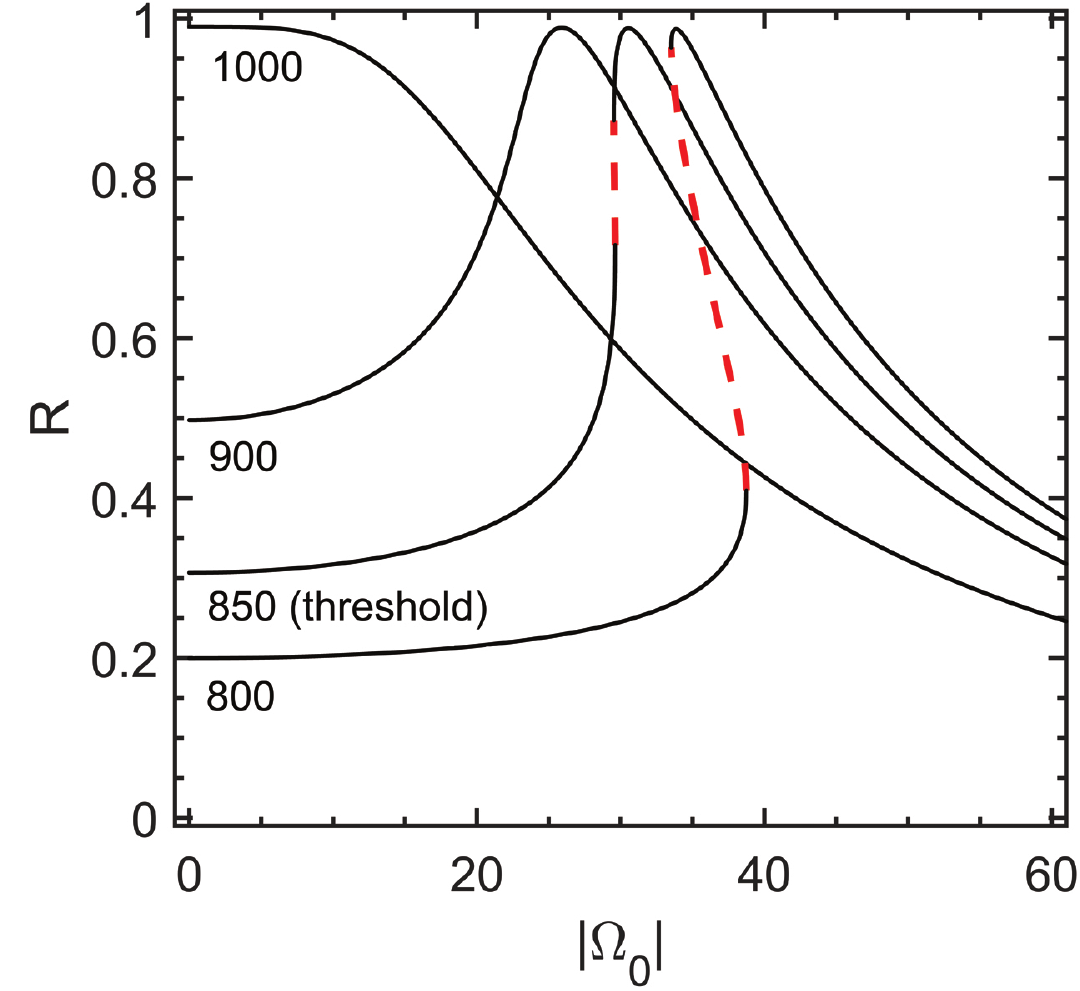}
\end{center}
\caption{\label{fig:Reflection DeltaB = 50} 
The dependence of the steady-state reflectance $R$ on the external field magnitude $|\Omega_0|$ calculated for different values of the detuning $\Delta_{21}$ in the vicinity of the renormalized resonance, $\Delta_{21} \leq - \Delta_L$. The results are calculated for the biexciton binding energy $\Delta_B = 50$, the values of $\Delta_{21}$ are given in the plot, herewith $\Delta^{th}_{21} = 850$ is the threshold detuning below which the reflectance become bistable. Red parts of the curves are unstable branches of the reflectance.
}
\end{figure}

Now, we turn to the nonlinear regime of reflectance in the vicinity of the renormalized resonance $\Delta_{21} \approx -\Delta_L$. We calculated the $|\Omega_0|$-dependence of the reflectance $R$ for a set of detunings above the renormalized resonance, $\Delta_{21} \leq -\Delta_L$. The results are presented in Fig.~\ref{fig:Reflection DeltaB = 50}. The figure shows that at the exact resonance ($\Delta_{21} = -\Delta_L = 1000$), the reflectance decreases monotonously as the external field magnitude $|\Omega_0|$ increases. This behavior is explained by the dependence of the current detuning $\Delta_{21}^\prime = \Delta_{21} - \Delta_L Z_{21}$ on the population difference [see Eq.(\ref{R21 extended})]: as the system is being excited, it is driven away from the renormalized resonance and, consequently, reflects less.

If the system is initially out of the renormalized resonance ($\Delta_{21} \leq -\Delta_L$), the low-field reflectance is relatively small according to Eq.~(\ref{Reflectance linear}). As the system is being excited, it is driven towards the resonance ($\Delta_{21} - \Delta_LZ_{21} \to 0$) and, at some $|\Omega_0|$, manifests again almost unity reflectance (Fig.~\ref{fig:Reflection DeltaB = 50}). Furthermore, starting some critical value of $\Delta_{21}$, namely, $\Delta^{th}_{21} = 850$ for the set of parameters used, the reflectance becomes three-valued within some window of external field amplitudes, manifesting the optical bistability. The critical value $\Delta^{th}_{21} = 850$ is in a good agreement with the theoretical estimate made within the framework of an effective two-level model, $\Delta_{21} = -\Delta_L - \sqrt{3}\gamma_R \approx 827$~\cite{Benedict1990}. A small deviation from the calculated value is probably due to the third biexciton level, a small admixture of which affects slightly the threshold value.

Finally, we note that the discussed reflectance properties are almost independent on the biexciton binding energy $\Delta_B$, so our results should apply to a wide range of SDQ supercrystals.

\section{Summary 
}
\label{Summary}

We conducted a theoretical study of the optical response of a two-dimensional semiconductor quantum dot supercrystal subjected to a monochromatic quasiresonant excitation. A constituent SQD was modeled as a three-level ladder-like system with the ground, one-exciton and biexciton states. The set of parameters used in our study is typical for SQDs emitting in the visible range, such as, CdSe and CdSe/ZnSe.
We took into account the SQD dipole-dipole interaction within the framework of the mean field approximation. 

To address the stationary response of the system, we developed a novel exact linear parametric method of solving the nonlinear steady-state problem which has multivalued solutions in all considered cases. Analyzing the Lyapunov exponents at the stationary characteristics, we found stable and unstable branches of the steady-state solutions. We provided a physical insight into the nature of the instabilities which have their origin in the competition between the ground-to-one exciton and one exciton-to-biexciton transitions, driven by the near-field SQD-SQD interactions. The stability analysis provided us with a solid starting point for further study of the system dynamics, which we first addressed qualitatively. To this end we put forward a novel method to calculate the bifurcation diagram of the system which gives a general overview of possible system dynamics. It turned out that the 2D supercrystal optical response can manifest very different dynamics under a continuous wave excitation: periodic or aperiodic self-oscillations and probably chaotic behavior. The frequency of self-oscillations depends on the external field magnitude and, for the set of parameters used, falls in the THz region.

Our results suggest various applications of the 2D SQD supercrystals, such as: an all-optical bistable switch, an ultra thin tunable bistable mirror, a tunable generator of trains of THz pulses (in self-oscillation regime), and as a noise generator (in chaotic regime). The intrinsic sensitivity of the optical response to the initial conditions in the chaotic regime could be of interest for information encryption~\cite{Gao2008}. All these findings make the considered system a promising candidate for practical applications in all-optical information processing and computing.


\acknowledgments
A. V. M. acknowledge support from Spanish MINECO grants MAT2013-46308 and MAT2016-75955. I. V. R. acknowledges support from the Russian Foundation for Basic Research, project no. 15-02-08369. A. V. M. is grateful to R. Noskov for useful discussions on bifurcation diagrams and P. \'A. Zapatero for prior collaboration.

\begin{appendix}
\section{Numerical evaluation of $\gamma_R$ and $\Delta_L$}
\label{ApendixA}
\begin{figure}[ht!]
\begin{center}
\includegraphics[width=0.8\columnwidth]{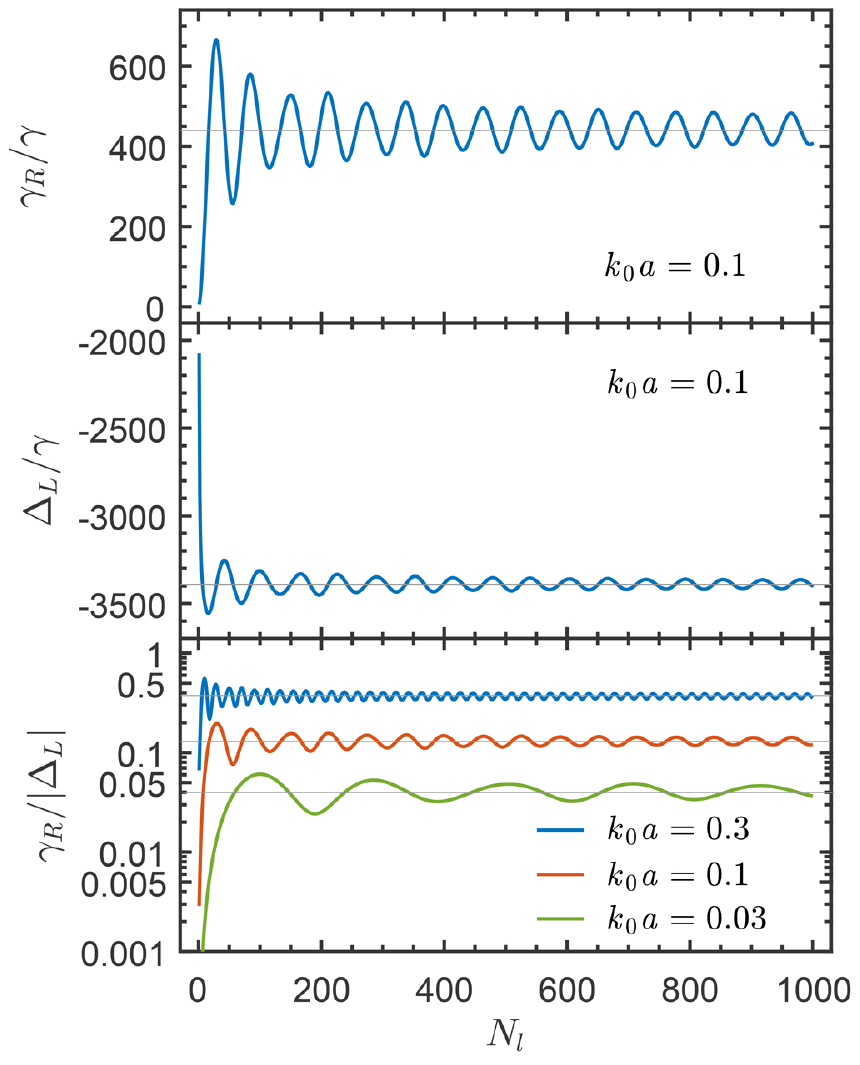}
\caption{\label{fig:gammaR and DeltaL} 
The lateral size dependence of the collective radiation rate $\gamma_R$ (upper plot), the near-zone dipole-dipole interaction of SQDs, $\Delta_L$ (middle plot), and the ratio $\gamma_R/|\Delta_L|$ (lower plot) calculated from Eqs.~(\ref{gammaR}) and (\ref{DeltaL}) for different values of $k_0a$ (indicated in the plots). Thin horizontal lines are guides for the eye showing the asymptotic values of the oscillating functions.
}
\end{center}
\end{figure}

Here, we evaluate numerically $\gamma_R$ and $\Delta_L$, given by Eqs.~(\ref{gammaR}) and~(\ref{DeltaL}), for a large square system ($N_xa, N_ya \gg \lambdabar$, $N_x = N_y = N_l$). In Fig.~\ref{fig:gammaR and DeltaL}, we plotted $\gamma_R$, $\Delta_L$, and the ratio $\gamma_R/|\Delta_L|$ against the system lateral size $N_l$ for different values of $k_0a$. As can be seen from the figure, these quantities manifest decaying oscillations around their asymptotic values, which reflect slow convergence of the sums that contain terms proportional to $|{\bf n}|^{-1}$. Comparing these data with the expected $(k_0a)^{-2}$-scaling of $\gamma_R$ and $(k_0a)^{-3}$-scaling of $\Delta_L$ [which follow from (\ref{gammaR}) and~(\ref{DeltaL})] we obtained the approximate numerical formulas Eqs.~(\ref{gammaRextended}) and~(\ref{DeltaLextended}) which describe excellently all numerical data presented in Fig.~\ref{fig:gammaR and DeltaL}.

\section{Solution of the steady-state problem}
\label{AppendixB}
The steady-state problem is governed by the following set of equations:
%
%
\begin{subequations}
\begin{equation}
\label{a:rho22}
\gamma\rho_{22} + \Omega \rho_{21}^* + \Omega^* \rho_{21} = 0~,
\end{equation}
\begin{equation}
\label{a:rho33}
\mu\gamma \rho_{33} + \Omega \rho_{32}^* + \Omega^* \rho_{32} = 0~,
\end{equation}
\begin{equation}
\label{a:R21}
\Omega(\rho_{22} - \rho_{11}) - \left( i\Delta_{21} + \frac{\gamma}{2} \right) \rho_{21}  + \mu \Omega^* \rho_{31} = 0~,
\end{equation}
\begin{equation}
\label{a:R32}
\mu \Omega(\rho_{33} - \rho_{22}) - \left[ i\Delta_{32} + \frac{\gamma}{2} (1 + \mu^2) \right] \rho_{32}  - \Omega^* \rho_{31} = 0~,
\end{equation}
\begin{equation}
\label{a:R31}
- \left( i\Delta_{31} + \frac{\gamma}{2}\mu^2 \right) \rho_{31} - \mu \Omega \rho_{21} + \Omega \rho_{32} = 0~,
\end{equation}
\begin{equation}
\label{a:Normalization}
\rho_{11} + \rho_{22} + \rho_{33} = 1~.
\end{equation}
\end{subequations}
%
%
Thus, originally the system of nine nonlinear coupled equations for the density matrix elements should be solved to find the dependence of these elements and the total field $\Omega$ on the external field $\Omega_0$. The two fields are related by Eq.~(\ref{Omega21}) which we rewrite for convenience in the following form:
\begin{equation}
\label{Omega0}
\Omega_0 = \Omega - (\gamma_R + i\Delta_L)(\rho_{21} + \mu\rho_{32})\ .
\end{equation}

Traditionally, one or another numerical method of direct solution of the nonlinear system (\ref{a:rho22})-(\ref{a:Normalization}) is used. Below we propose a much more efficient and essentially linear parametric method to solve this nonlinear problem.

First, we note that Eqs.~(\ref{a:rho22})-(\ref{a:Normalization}) and (\ref{Omega0}) are invariant under the following phase transformation:
\begin{subequations}
\begin{equation}
\rho_{21} \mapsto \rho_{21}\,e^{i\,\varphi},\;\;
\rho_{32} \mapsto \rho_{32}\,e^{i\,\varphi},\;\;
\rho_{31} \mapsto \rho_{31}\,e^{2i\,\varphi}
\label{rhophasetrans}
\end{equation}
\begin{equation}
\Omega \mapsto \Omega\,e^{i\,\varphi},\quad
\Omega_0 \mapsto \Omega_0\,e^{i\,\varphi}\ ,
\label{Omegaphasetrans}
\end{equation}
\label{phasetrans}
\end{subequations}
where $\varphi$ is an arbitrary phase.
Second, the system of Eqs.~(\ref{a:rho22})-(\ref{a:Normalization}) is linear in the density matrix elements if $\Omega$ is considered to be a parameter. Furthermore, Eq.~(\ref{Omegaphasetrans}) suggests that instead of (naturally) treating the external field amplitude $\Omega_0$ as a real quantity, one can consider the total field amplitude $\Omega$ to be real (the phase of $\Omega$ can be chosen arbitrarily; the zero phase is just the most conventional choice). 

Importantly, the system of Eqs.~(\ref{a:rho22})-(\ref{a:Normalization}), as being a system of linear equations, can be solved analytically and the {\it unique} parametric dependence of all density matrix elements on $\Omega$ can be obtained. Then Eq.~(\ref{Omega0}) provides the {\it unique} parametric dependence of the external field $\Omega_0$ on the real total field $\Omega$.
The sought dependencies of the density matrix elements on the external field $\Omega_0$ can then be obtained in the parametric way, varying the real $\Omega$ within an appropriate interval of values. Finally, to recover the ``traditional'' case, in which the external field amplitude $\Omega_0$ is real, the transformations (\ref{phasetrans}) can be used with the phase $\varphi = -\mathrm{arg}\,\Omega_0$ given by:
\begin{equation}
\varphi = -\mathrm{arg}\left[
\, \Omega - (\gamma_R + i\Delta_L)(\rho_{21} + \mu\rho_{32}) \,
\right]\ .
\end{equation}

To conclude, we note that our method of solving the nonlinear mean-field steady-state equations for the density matrix elements is quite general and, therefore, can probably be applied to a broad class of similar systems.
\end{appendix}

\bibliographystyle{apsrev4-1}
\bibliography{bibliography-3levels}

\begin{thebibliography}{79}%
\makeatletter
\providecommand \@ifxundefined [1]{%
 \@ifx{#1\undefined}
}%
\providecommand \@ifnum [1]{%
 \ifnum #1\expandafter \@firstoftwo
 \else \expandafter \@secondoftwo
 \fi
}%
\providecommand \@ifx [1]{%
 \ifx #1\expandafter \@firstoftwo
 \else \expandafter \@secondoftwo
 \fi
}%
\providecommand \natexlab [1]{#1}%
\providecommand \enquote  [1]{``#1''}%
\providecommand \bibnamefont  [1]{#1}%
\providecommand \bibfnamefont [1]{#1}%
\providecommand \citenamefont [1]{#1}%
\providecommand \href@noop [0]{\@secondoftwo}%
\providecommand \href [0]{\begingroup \@sanitize@url \@href}%
\providecommand \@href[1]{\@@startlink{#1}\@@href}%
\providecommand \@@href[1]{\endgroup#1\@@endlink}%
\providecommand \@sanitize@url [0]{\catcode `\\12\catcode `\$12\catcode
  `\&12\catcode `\#12\catcode `\^12\catcode `\_12\catcode `\%12\relax}%
\providecommand \@@startlink[1]{}%
\providecommand \@@endlink[0]{}%
\providecommand \url  [0]{\begingroup\@sanitize@url \@url }%
\providecommand \@url [1]{\endgroup\@href {#1}{\urlprefix }}%
\providecommand \urlprefix  [0]{URL }%
\providecommand \Eprint [0]{\href }%
\providecommand \doibase [0]{http://dx.doi.org/}%
\providecommand \selectlanguage [0]{\@gobble}%
\providecommand \bibinfo  [0]{\@secondoftwo}%
\providecommand \bibfield  [0]{\@secondoftwo}%
\providecommand \translation [1]{[#1]}%
\providecommand \BibitemOpen [0]{}%
\providecommand \bibitemStop [0]{}%
\providecommand \bibitemNoStop [0]{.\EOS\space}%
\providecommand \EOS [0]{\spacefactor3000\relax}%
\providecommand \BibitemShut  [1]{\csname bibitem#1\endcsname}%
\let\auto@bib@innerbib\@empty
\bibitem [{\citenamefont {Zheludev}(2010)}]{Zheludev2010}%
  \BibitemOpen
  \bibfield  {author} {\bibinfo {author} {\bibfnamefont {N.~I.}\ \bibnamefont
  {Zheludev}},\ }\href@noop {} {\bibfield  {journal} {\bibinfo  {journal}
  {Science}\ }\textbf {\bibinfo {volume} {328}},\ \bibinfo {pages} {582}
  (\bibinfo {year} {2010})}\BibitemShut {NoStop}%
\bibitem [{\citenamefont {Zheludev}\ and\ \citenamefont
  {Kivshar}(2012)}]{Zheludev2012}%
  \BibitemOpen
  \bibfield  {author} {\bibinfo {author} {\bibfnamefont {N.~I.}\ \bibnamefont
  {Zheludev}}\ and\ \bibinfo {author} {\bibfnamefont {Y.~S.}\ \bibnamefont
  {Kivshar}},\ }\href {\doibase 10.1038/nmat3431} {\bibfield  {journal}
  {\bibinfo  {journal} {Nature Materials}\ }\textbf {\bibinfo {volume} {11}},\
  \bibinfo {pages} {917} (\bibinfo {year} {2012})}\BibitemShut {NoStop}%
\bibitem [{\citenamefont {Soukoulis}\ and\ \citenamefont
  {Wegener}(2010)}]{Soukoulis2010}%
  \BibitemOpen
  \bibfield  {author} {\bibinfo {author} {\bibfnamefont {C.~M.}\ \bibnamefont
  {Soukoulis}}\ and\ \bibinfo {author} {\bibfnamefont {M.}~\bibnamefont
  {Wegener}},\ }\href@noop {} {\bibfield  {journal} {\bibinfo  {journal}
  {Science}\ }\textbf {\bibinfo {volume} {330}},\ \bibinfo {pages} {1633}
  (\bibinfo {year} {2010})}\BibitemShut {NoStop}%
\bibitem [{\citenamefont {Liu}\ and\ \citenamefont {Zhang}(2011)}]{Liu2011}%
  \BibitemOpen
  \bibfield  {author} {\bibinfo {author} {\bibfnamefont {Y.}~\bibnamefont
  {Liu}}\ and\ \bibinfo {author} {\bibfnamefont {X.}~\bibnamefont {Zhang}},\
  }\href@noop {} {\bibfield  {journal} {\bibinfo  {journal} {Chem. Soc. Rev.}\
  }\textbf {\bibinfo {volume} {40}},\ \bibinfo {pages} {2494} (\bibinfo {year}
  {2011})}\BibitemShut {NoStop}%
\bibitem [{\citenamefont {Al\`u}(2016)}]{Alu2016}%
  \BibitemOpen
  \bibfield  {author} {\bibinfo {author} {\bibfnamefont {A.}~\bibnamefont
  {Al\`u}},\ }\href@noop {} {\bibfield  {journal} {\bibinfo  {journal} {Nat.
  Materials}\ }\textbf {\bibinfo {volume} {15}},\ \bibinfo {pages} {1229}
  (\bibinfo {year} {2016})}\BibitemShut {NoStop}%
\bibitem [{\citenamefont {Baimuratov}\ \emph {et~al.}(2013)\citenamefont
  {Baimuratov}, \citenamefont {Rukhlenko}, \citenamefont {Turkov},
  \citenamefont {Baranov},\ and\ \citenamefont {Fedorov}}]{Baimuratov2013}%
  \BibitemOpen
  \bibfield  {author} {\bibinfo {author} {\bibfnamefont {A.~S.}\ \bibnamefont
  {Baimuratov}}, \bibinfo {author} {\bibfnamefont {I.~D.}\ \bibnamefont
  {Rukhlenko}}, \bibinfo {author} {\bibfnamefont {V.~K.}\ \bibnamefont
  {Turkov}}, \bibinfo {author} {\bibfnamefont {A.~V.}\ \bibnamefont {Baranov}},
  \ and\ \bibinfo {author} {\bibfnamefont {A.~V.}\ \bibnamefont {Fedorov}},\
  }\href@noop {} {\bibfield  {journal} {\bibinfo  {journal} {Sci. Rep.}\
  }\textbf {\bibinfo {volume} {3}},\ \bibinfo {pages} {1727} (\bibinfo {year}
  {2013})}\BibitemShut {NoStop}%
\bibitem [{\citenamefont {Evers}\ \emph {et~al.}(2013)\citenamefont {Evers},
  \citenamefont {Goris}, \citenamefont {Bals}, \citenamefont {Casavola},
  \citenamefont {de~Graaf}, \citenamefont {van Roij}, \citenamefont
  {Dijkstra},\ and\ \citenamefont {Vanmaekelbergh}}]{Evers2013}%
  \BibitemOpen
  \bibfield  {author} {\bibinfo {author} {\bibfnamefont {W.~H.}\ \bibnamefont
  {Evers}}, \bibinfo {author} {\bibfnamefont {B.}~\bibnamefont {Goris}},
  \bibinfo {author} {\bibfnamefont {S.}~\bibnamefont {Bals}}, \bibinfo {author}
  {\bibfnamefont {M.}~\bibnamefont {Casavola}}, \bibinfo {author}
  {\bibfnamefont {J.}~\bibnamefont {de~Graaf}}, \bibinfo {author}
  {\bibfnamefont {R.}~\bibnamefont {van Roij}}, \bibinfo {author}
  {\bibfnamefont {M.}~\bibnamefont {Dijkstra}}, \ and\ \bibinfo {author}
  {\bibfnamefont {D.}~\bibnamefont {Vanmaekelbergh}},\ }\href@noop {}
  {\bibfield  {journal} {\bibinfo  {journal} {Nano Lett.}\ }\textbf {\bibinfo
  {volume} {13}},\ \bibinfo {pages} {2317} (\bibinfo {year}
  {2013})}\BibitemShut {NoStop}%
\bibitem [{\citenamefont {Boneschanscher}\ \emph {et~al.}(2014)\citenamefont
  {Boneschanscher}, \citenamefont {Evers}, \citenamefont {Geuchies},
  \citenamefont {Altantzis}, \citenamefont {Goris}, \citenamefont {Rabouw},
  \citenamefont {van Rossum}, \citenamefont {van~der Zant}, \citenamefont
  {Siebbeles}, \citenamefont {Tendeloo}, \citenamefont {Swart}, \citenamefont
  {Hilhorst}, \citenamefont {Petukhov}, \citenamefont {Bals},\ and\
  \citenamefont {Vanmaekelbergh}}]{Boneschanscher2014}%
  \BibitemOpen
  \bibfield  {author} {\bibinfo {author} {\bibfnamefont {M.~P.}\ \bibnamefont
  {Boneschanscher}}, \bibinfo {author} {\bibfnamefont {W.~H.}\ \bibnamefont
  {Evers}}, \bibinfo {author} {\bibfnamefont {J.~J.}\ \bibnamefont {Geuchies}},
  \bibinfo {author} {\bibfnamefont {T.}~\bibnamefont {Altantzis}}, \bibinfo
  {author} {\bibfnamefont {B.}~\bibnamefont {Goris}}, \bibinfo {author}
  {\bibfnamefont {F.~T.}\ \bibnamefont {Rabouw}}, \bibinfo {author}
  {\bibfnamefont {S.~A.~P.}\ \bibnamefont {van Rossum}}, \bibinfo {author}
  {\bibfnamefont {H.~S.~J.}\ \bibnamefont {van~der Zant}}, \bibinfo {author}
  {\bibfnamefont {L.~D.~A.}\ \bibnamefont {Siebbeles}}, \bibinfo {author}
  {\bibfnamefont {G.~V.}\ \bibnamefont {Tendeloo}}, \bibinfo {author}
  {\bibfnamefont {I.}~\bibnamefont {Swart}}, \bibinfo {author} {\bibfnamefont
  {J.}~\bibnamefont {Hilhorst}}, \bibinfo {author} {\bibfnamefont {A.~V.}\
  \bibnamefont {Petukhov}}, \bibinfo {author} {\bibfnamefont {S.}~\bibnamefont
  {Bals}}, \ and\ \bibinfo {author} {\bibfnamefont {D.}~\bibnamefont
  {Vanmaekelbergh}},\ }\href@noop {} {\bibfield  {journal} {\bibinfo  {journal}
  {Science}\ }\textbf {\bibinfo {volume} {344}},\ \bibinfo {pages} {1377}
  (\bibinfo {year} {2014})}\BibitemShut {NoStop}%
\bibitem [{\citenamefont {Baranov}\ \emph {et~al.}(2015)\citenamefont
  {Baranov}, \citenamefont {Ushakova}, \citenamefont {Golubkov}, \citenamefont
  {Litvin}, \citenamefont {Parfenov}, \citenamefont {Fedorov},\ and\
  \citenamefont {Berwick}}]{Baranov2015}%
  \BibitemOpen
  \bibfield  {author} {\bibinfo {author} {\bibfnamefont {A.~V.}\ \bibnamefont
  {Baranov}}, \bibinfo {author} {\bibfnamefont {E.~V.}\ \bibnamefont
  {Ushakova}}, \bibinfo {author} {\bibfnamefont {V.~V.}\ \bibnamefont
  {Golubkov}}, \bibinfo {author} {\bibfnamefont {A.~P.}\ \bibnamefont
  {Litvin}}, \bibinfo {author} {\bibfnamefont {P.~S.}\ \bibnamefont
  {Parfenov}}, \bibinfo {author} {\bibfnamefont {A.~V.}\ \bibnamefont
  {Fedorov}}, \ and\ \bibinfo {author} {\bibfnamefont {K.}~\bibnamefont
  {Berwick}},\ }\href@noop {} {\bibfield  {journal} {\bibinfo  {journal}
  {Langmuir}\ }\textbf {\bibinfo {volume} {31}},\ \bibinfo {pages} {506}
  (\bibinfo {year} {2015})}\BibitemShut {NoStop}%
\bibitem [{\citenamefont {Ushakova}\ \emph {et~al.}(2016)\citenamefont
  {Ushakova}, \citenamefont {Cherevkov}, \citenamefont {Litvin}, \citenamefont
  {Parfenov}, \citenamefont {Volgina}, \citenamefont {Kasatkin}, \citenamefont
  {Fedorov},\ and\ \citenamefont {Baranov}}]{Ushakova2016}%
  \BibitemOpen
  \bibfield  {author} {\bibinfo {author} {\bibfnamefont {E.~V.}\ \bibnamefont
  {Ushakova}}, \bibinfo {author} {\bibfnamefont {S.~A.}\ \bibnamefont
  {Cherevkov}}, \bibinfo {author} {\bibfnamefont {A.~P.}\ \bibnamefont
  {Litvin}}, \bibinfo {author} {\bibfnamefont {P.~S.}\ \bibnamefont
  {Parfenov}}, \bibinfo {author} {\bibfnamefont {D.-O.~A.}\ \bibnamefont
  {Volgina}}, \bibinfo {author} {\bibfnamefont {I.~A.}\ \bibnamefont
  {Kasatkin}}, \bibinfo {author} {\bibfnamefont {A.~V.}\ \bibnamefont
  {Fedorov}}, \ and\ \bibinfo {author} {\bibfnamefont {A.~V.}\ \bibnamefont
  {Baranov}},\ }\href@noop {} {\bibfield  {journal} {\bibinfo  {journal} {J.
  Phys. Chem. C}\ }\textbf {\bibinfo {volume} {120}},\ \bibinfo {pages} {25061}
  (\bibinfo {year} {2016})}\BibitemShut {NoStop}%
\bibitem [{\citenamefont {Liu}\ \emph {et~al.}(2017)\citenamefont {Liu},
  \citenamefont {Luo}, \citenamefont {Bao}, \citenamefont {Liu}, \citenamefont
  {Ning}, \citenamefont {Abdelwahab}, \citenamefont {Li}, \citenamefont {Nai},
  \citenamefont {Hu}, \citenamefont {Zhao}, \citenamefont {Liu}, \citenamefont
  {Quek},\ and\ \citenamefont {Loh}}]{Liu2017}%
  \BibitemOpen
  \bibfield  {author} {\bibinfo {author} {\bibfnamefont {W.}~\bibnamefont
  {Liu}}, \bibinfo {author} {\bibfnamefont {X.}~\bibnamefont {Luo}}, \bibinfo
  {author} {\bibfnamefont {Y.}~\bibnamefont {Bao}}, \bibinfo {author}
  {\bibfnamefont {Y.~P.}\ \bibnamefont {Liu}}, \bibinfo {author} {\bibfnamefont
  {G.-H.}\ \bibnamefont {Ning}}, \bibinfo {author} {\bibfnamefont
  {I.}~\bibnamefont {Abdelwahab}}, \bibinfo {author} {\bibfnamefont
  {L.}~\bibnamefont {Li}}, \bibinfo {author} {\bibfnamefont {C.~T.}\
  \bibnamefont {Nai}}, \bibinfo {author} {\bibfnamefont {Z.~G.}\ \bibnamefont
  {Hu}}, \bibinfo {author} {\bibfnamefont {D.}~\bibnamefont {Zhao}}, \bibinfo
  {author} {\bibfnamefont {B.}~\bibnamefont {Liu}}, \bibinfo {author}
  {\bibfnamefont {S.~Y.}\ \bibnamefont {Quek}}, \ and\ \bibinfo {author}
  {\bibfnamefont {K.~P.}\ \bibnamefont {Loh}},\ }\href@noop {} {\bibfield
  {journal} {\bibinfo  {journal} {Nat. Chemistry}\ }\textbf {\bibinfo {volume}
  {9}},\ \bibinfo {pages} {563} (\bibinfo {year} {2017})}\BibitemShut {NoStop}%
\bibitem [{\citenamefont {Ben-Aryeh}\ \emph
  {et~al.}(1986{\natexlab{a}})\citenamefont {Ben-Aryeh}, \citenamefont
  {Bowden},\ and\ \citenamefont {Englund}}]{Bowden1986a}%
  \BibitemOpen
  \bibfield  {author} {\bibinfo {author} {\bibfnamefont {Y.}~\bibnamefont
  {Ben-Aryeh}}, \bibinfo {author} {\bibfnamefont {C.~M.}\ \bibnamefont
  {Bowden}}, \ and\ \bibinfo {author} {\bibfnamefont {J.~C.}\ \bibnamefont
  {Englund}},\ }\href@noop {} {\bibfield  {journal} {\bibinfo  {journal} {Opt.
  Commun.}\ }\textbf {\bibinfo {volume} {59}},\ \bibinfo {pages} {224}
  (\bibinfo {year} {1986}{\natexlab{a}})}\BibitemShut {NoStop}%
\bibitem [{\citenamefont {Ben-Aryeh}\ \emph
  {et~al.}(1986{\natexlab{b}})\citenamefont {Ben-Aryeh}, \citenamefont
  {Bowden},\ and\ \citenamefont {Englund}}]{Bowden1986b}%
  \BibitemOpen
  \bibfield  {author} {\bibinfo {author} {\bibfnamefont {Y.}~\bibnamefont
  {Ben-Aryeh}}, \bibinfo {author} {\bibfnamefont {C.~M.}\ \bibnamefont
  {Bowden}}, \ and\ \bibinfo {author} {\bibfnamefont {J.~C.}\ \bibnamefont
  {Englund}},\ }\href@noop {} {\bibfield  {journal} {\bibinfo  {journal} {Phys.
  Rev. A}\ }\textbf {\bibinfo {volume} {34}},\ \bibinfo {pages} {3917}
  (\bibinfo {year} {1986}{\natexlab{b}})}\BibitemShut {NoStop}%
\bibitem [{\citenamefont {Zakharov}\ and\ \citenamefont
  {Manykin}(1988)}]{Zakharov1988}%
  \BibitemOpen
  \bibfield  {author} {\bibinfo {author} {\bibfnamefont {S.~M.}\ \bibnamefont
  {Zakharov}}\ and\ \bibinfo {author} {\bibfnamefont {E.~A.}\ \bibnamefont
  {Manykin}},\ }\href@noop {} {\bibfield  {journal} {\bibinfo  {journal}
  {Poverkhnost'}\ }\textbf {\bibinfo {volume} {2}},\ \bibinfo {pages} {137}
  (\bibinfo {year} {1988})}\BibitemShut {NoStop}%
\bibitem [{\citenamefont {Basharov}(1988)}]{Basharov1988}%
  \BibitemOpen
  \bibfield  {author} {\bibinfo {author} {\bibfnamefont {A.~M.}\ \bibnamefont
  {Basharov}},\ }\href@noop {} {\bibfield  {journal} {\bibinfo  {journal} {Sov.
  Phys. JETP}\ }\textbf {\bibinfo {volume} {67}},\ \bibinfo {pages} {1741}
  (\bibinfo {year} {1988})}\BibitemShut {NoStop}%
\bibitem [{\citenamefont {Benedict}\ \emph {et~al.}(1990)\citenamefont
  {Benedict}, \citenamefont {Zaitsev}, \citenamefont {Malyshev},\ and\
  \citenamefont {Trifonov}}]{Benedict1990}%
  \BibitemOpen
  \bibfield  {author} {\bibinfo {author} {\bibfnamefont {M.~G.}\ \bibnamefont
  {Benedict}}, \bibinfo {author} {\bibfnamefont {A.~I.}\ \bibnamefont
  {Zaitsev}}, \bibinfo {author} {\bibfnamefont {V.~A.}\ \bibnamefont
  {Malyshev}}, \ and\ \bibinfo {author} {\bibfnamefont {E.~D.}\ \bibnamefont
  {Trifonov}},\ }\href@noop {} {\bibfield  {journal} {\bibinfo  {journal} {Opt.
  Spectrosc.}\ }\textbf {\bibinfo {volume} {68}},\ \bibinfo {pages} {473}
  (\bibinfo {year} {1990})}\BibitemShut {NoStop}%
\bibitem [{\citenamefont {Benedict}\ \emph {et~al.}(1991)\citenamefont
  {Benedict}, \citenamefont {Malyshev}, \citenamefont {Trifonov},\ and\
  \citenamefont {Zaitsev}}]{Benedict1991}%
  \BibitemOpen
  \bibfield  {author} {\bibinfo {author} {\bibfnamefont {M.~G.}\ \bibnamefont
  {Benedict}}, \bibinfo {author} {\bibfnamefont {V.~A.}\ \bibnamefont
  {Malyshev}}, \bibinfo {author} {\bibfnamefont {E.~D.}\ \bibnamefont
  {Trifonov}}, \ and\ \bibinfo {author} {\bibfnamefont {A.~I.}\ \bibnamefont
  {Zaitsev}},\ }\href@noop {} {\bibfield  {journal} {\bibinfo  {journal} {Phys.
  Rev. A}\ }\textbf {\bibinfo {volume} {43}},\ \bibinfo {pages} {3845}
  (\bibinfo {year} {1991})}\BibitemShut {NoStop}%
\bibitem [{\citenamefont {Oraevsky}\ \emph {et~al.}(1994)\citenamefont
  {Oraevsky}, \citenamefont {Jones},\ and\ \citenamefont
  {Bandy}}]{Oraevsky1994}%
  \BibitemOpen
  \bibfield  {author} {\bibinfo {author} {\bibfnamefont {A.~N.}\ \bibnamefont
  {Oraevsky}}, \bibinfo {author} {\bibfnamefont {D.~J.}\ \bibnamefont {Jones}},
  \ and\ \bibinfo {author} {\bibfnamefont {D.~K.}\ \bibnamefont {Bandy}},\
  }\href@noop {} {\bibfield  {journal} {\bibinfo  {journal} {Opt. Commun.}\
  }\textbf {\bibinfo {volume} {111}},\ \bibinfo {pages} {163} (\bibinfo {year}
  {1994})}\BibitemShut {NoStop}%
\bibitem [{\citenamefont {Malyshev}\ and\ \citenamefont
  {Conejero~Jarque}(2000)}]{Malyshev2000}%
  \BibitemOpen
  \bibfield  {author} {\bibinfo {author} {\bibfnamefont {V.~A.}\ \bibnamefont
  {Malyshev}}\ and\ \bibinfo {author} {\bibfnamefont {E.}~\bibnamefont
  {Conejero~Jarque}},\ }\href@noop {} {\bibfield  {journal} {\bibinfo
  {journal} {Opt. Experess}\ }\textbf {\bibinfo {volume} {6}},\ \bibinfo
  {pages} {227} (\bibinfo {year} {2000})}\BibitemShut {NoStop}%
\bibitem [{\citenamefont {Glaeske}\ \emph {et~al.}(2000)\citenamefont
  {Glaeske}, \citenamefont {Malyshev},\ and\ \citenamefont
  {Feller}}]{Glaeske2000}%
  \BibitemOpen
  \bibfield  {author} {\bibinfo {author} {\bibfnamefont {H.}~\bibnamefont
  {Glaeske}}, \bibinfo {author} {\bibfnamefont {V.~A.}\ \bibnamefont
  {Malyshev}}, \ and\ \bibinfo {author} {\bibfnamefont {K.-H.}\ \bibnamefont
  {Feller}},\ }\href@noop {} {\bibfield  {journal} {\bibinfo  {journal} {J.
  Chem. Phys.}\ }\textbf {\bibinfo {volume} {113}},\ \bibinfo {pages} {1170}
  (\bibinfo {year} {2000})}\BibitemShut {NoStop}%
\bibitem [{\citenamefont {Klugkist}\ \emph {et~al.}(2007)\citenamefont
  {Klugkist}, \citenamefont {Malyshev},\ and\ \citenamefont
  {Knoester}}]{Klugkist2007}%
  \BibitemOpen
  \bibfield  {author} {\bibinfo {author} {\bibfnamefont {J.~A.}\ \bibnamefont
  {Klugkist}}, \bibinfo {author} {\bibfnamefont {V.~A.}\ \bibnamefont
  {Malyshev}}, \ and\ \bibinfo {author} {\bibfnamefont {J.}~\bibnamefont
  {Knoester}},\ }\href@noop {} {\bibfield  {journal} {\bibinfo  {journal} {J.
  Chem. Phys.}\ }\textbf {\bibinfo {volume} {127}},\ \bibinfo {pages} {164705}
  (\bibinfo {year} {2007})}\BibitemShut {NoStop}%
\bibitem [{\citenamefont {Malikov}\ and\ \citenamefont
  {Malyshev}(2017)}]{Malikov2017a}%
  \BibitemOpen
  \bibfield  {author} {\bibinfo {author} {\bibfnamefont {R.~F.}\ \bibnamefont
  {Malikov}}\ and\ \bibinfo {author} {\bibfnamefont {V.~A.}\ \bibnamefont
  {Malyshev}},\ }\href@noop {} {\bibfield  {journal} {\bibinfo  {journal} {Opt.
  Spectrosc.}\ }\textbf {\bibinfo {volume} {122}},\ \bibinfo {pages} {955}
  (\bibinfo {year} {2017})}\BibitemShut {NoStop}%
\bibitem [{Note1()}]{Note1}%
  \BibitemOpen
  \bibinfo {note} {\label {Note0}It should be noticed that a thin layer of
  three-level $V$-emitters also shows a similar behavior (see Refs.~\protect
  \rev@citealp {Vlasov2013a,Vlasov2013b}).}\BibitemShut {Stop}%
\bibitem [{foo()}]{footnote1}%
  \BibitemOpen
  \bibinfo {note} {A preliminary study of the quantum dot supercrystal's
  optical response has been recently reported in
  Refs.~\onlinecite{MalikovEPJWebConf2017,MalyshevJPhysConfSer2019}}\BibitemShut
  {NoStop}%
\bibitem [{\citenamefont {Andronov}\ \emph {et~al.}(1966)\citenamefont
  {Andronov}, \citenamefont {Vitt},\ and\ \citenamefont
  {Khaikin}}]{AndronovBook1966}%
  \BibitemOpen
  \bibfield  {author} {\bibinfo {author} {\bibfnamefont {A.~A.}\ \bibnamefont
  {Andronov}}, \bibinfo {author} {\bibfnamefont {A.~A.}\ \bibnamefont {Vitt}},
  \ and\ \bibinfo {author} {\bibfnamefont {S.~E.}\ \bibnamefont {Khaikin}},\
  }\href@noop {} {\emph {\bibinfo {title} {Theory Of Oscillators}}}\ (\bibinfo
  {publisher} {Pergamon Press, New York},\ \bibinfo {year} {1966})\BibitemShut
  {NoStop}%
\bibitem [{\citenamefont {Eckmann}\ and\ \citenamefont
  {Ruelle}(1985)}]{EckmannRevModPhys1985}%
  \BibitemOpen
  \bibfield  {author} {\bibinfo {author} {\bibfnamefont {J.-P.}\ \bibnamefont
  {Eckmann}}\ and\ \bibinfo {author} {\bibfnamefont {D.}~\bibnamefont
  {Ruelle}},\ }\href@noop {} {\bibfield  {journal} {\bibinfo  {journal} {Rev.
  Mod. Phys.}\ }\textbf {\bibinfo {volume} {57}},\ \bibinfo {pages} {617}
  (\bibinfo {year} {1985})}\BibitemShut {NoStop}%
\bibitem [{\citenamefont {Guckenheimer}\ and\ \citenamefont
  {Holmes}(1986)}]{GuckenheimerBook1986}%
  \BibitemOpen
  \bibfield  {author} {\bibinfo {author} {\bibfnamefont {J.}~\bibnamefont
  {Guckenheimer}}\ and\ \bibinfo {author} {\bibfnamefont {P.}~\bibnamefont
  {Holmes}},\ }\href@noop {} {\emph {\bibinfo {title} {Nonlinear Oscillations,
  Dynamical Systems and Bifurcations of Vector Fields}}}\ (\bibinfo
  {publisher} {Springer, Berlin},\ \bibinfo {year} {1986})\BibitemShut
  {NoStop}%
\bibitem [{\citenamefont {Neimark}\ and\ \citenamefont
  {Landa}(1992)}]{NeimarkLandaBook1992}%
  \BibitemOpen
  \bibfield  {author} {\bibinfo {author} {\bibfnamefont {Y.~I.}\ \bibnamefont
  {Neimark}}\ and\ \bibinfo {author} {\bibfnamefont {P.~S.}\ \bibnamefont
  {Landa}},\ }\href@noop {} {\emph {\bibinfo {title} {Stochastic and Chaotic
  Oscillations}}}\ (\bibinfo  {publisher} {Springer Science \& Business
  Media},\ \bibinfo {year} {1992})\BibitemShut {NoStop}%
\bibitem [{\citenamefont {Ott}(1993)}]{OttBook1993}%
  \BibitemOpen
  \bibfield  {author} {\bibinfo {author} {\bibfnamefont {E.}~\bibnamefont
  {Ott}},\ }\href@noop {} {\emph {\bibinfo {title} {Chaos in Dynamical
  Systems}}}\ (\bibinfo  {publisher} {Cambridge University Press, Cambridge},\
  \bibinfo {year} {1993})\BibitemShut {NoStop}%
\bibitem [{\citenamefont {Arnol'd}\ \emph {et~al.}(1994)\citenamefont
  {Arnol'd}, \citenamefont {(Ed.)}, \citenamefont {Afrajmovich}, \citenamefont
  {Il'yashenko},\ and\ \citenamefont {Shil'nikov}}]{Arnol'dBook1994}%
  \BibitemOpen
  \bibfield  {author} {\bibinfo {author} {\bibfnamefont {V.~I.}\ \bibnamefont
  {Arnol'd}}, \bibinfo {author} {\bibnamefont {(Ed.)}}, \bibinfo {author}
  {\bibfnamefont {V.~S.}\ \bibnamefont {Afrajmovich}}, \bibinfo {author}
  {\bibfnamefont {Y.~S.}\ \bibnamefont {Il'yashenko}}, \ and\ \bibinfo {author}
  {\bibfnamefont {L.~P.}\ \bibnamefont {Shil'nikov}},\ }\href@noop {} {\emph
  {\bibinfo {title} {Dynamical Systems V: Bifurcation Theory and Catastrophe
  Theory}}}\ (\bibinfo  {publisher} {Springer, Berlin},\ \bibinfo {year}
  {1994})\BibitemShut {NoStop}%
\bibitem [{\citenamefont {Alligood}\ \emph {et~al.}(1996)\citenamefont
  {Alligood}, \citenamefont {Sauer},\ and\ \citenamefont
  {Yorke}}]{AlligoodBook1996}%
  \BibitemOpen
  \bibfield  {author} {\bibinfo {author} {\bibfnamefont {K.~T.}\ \bibnamefont
  {Alligood}}, \bibinfo {author} {\bibfnamefont {T.~D.}\ \bibnamefont {Sauer}},
  \ and\ \bibinfo {author} {\bibfnamefont {J.~A.}\ \bibnamefont {Yorke}},\
  }\href@noop {} {\emph {\bibinfo {title} {Chaos: An Introduction to Dynamical
  Systems}}}\ (\bibinfo  {publisher} {Springer, Berlin},\ \bibinfo {year}
  {1996})\BibitemShut {NoStop}%
\bibitem [{\citenamefont {Katok}\ and\ \citenamefont
  {Hasselblatt}(1997)}]{KatokBook1997}%
  \BibitemOpen
  \bibfield  {author} {\bibinfo {author} {\bibfnamefont {A.}~\bibnamefont
  {Katok}}\ and\ \bibinfo {author} {\bibfnamefont {B.}~\bibnamefont
  {Hasselblatt}},\ }\href@noop {} {\emph {\bibinfo {title} {Introduction to the
  Modern Theory of Dynamical Systems}}}\ (\bibinfo  {publisher} {Cambridge
  University Press, Cambridge},\ \bibinfo {year} {1997})\BibitemShut {NoStop}%
\bibitem [{\citenamefont {Kuznetsov}(2004)}]{KuznetsovBook2004}%
  \BibitemOpen
  \bibfield  {author} {\bibinfo {author} {\bibfnamefont {Y.~A.}\ \bibnamefont
  {Kuznetsov}},\ }\href@noop {} {\emph {\bibinfo {title} {Elements of Applied
  Bifurcation Theory}}}\ (\bibinfo  {publisher} {Springer},\ \bibinfo {year}
  {2004})\BibitemShut {NoStop}%
\bibitem [{\citenamefont {Wieczorek}\ \emph {et~al.}(2005)\citenamefont
  {Wieczorek}, \citenamefont {Krauskopf}, \citenamefont {Simpson},\ and\
  \citenamefont {Lenstra}}]{WieczorekPhysRep2005}%
  \BibitemOpen
  \bibfield  {author} {\bibinfo {author} {\bibfnamefont {S.}~\bibnamefont
  {Wieczorek}}, \bibinfo {author} {\bibfnamefont {B.}~\bibnamefont
  {Krauskopf}}, \bibinfo {author} {\bibfnamefont {T.~B.}\ \bibnamefont
  {Simpson}}, \ and\ \bibinfo {author} {\bibfnamefont {D.}~\bibnamefont
  {Lenstra}},\ }\href@noop {} {\bibfield  {journal} {\bibinfo  {journal} {Phys.
  Rep.}\ }\textbf {\bibinfo {volume} {416}},\ \bibinfo {pages} {1} (\bibinfo
  {year} {2005})}\BibitemShut {NoStop}%
\bibitem [{\citenamefont {Stufler}\ \emph {et~al.}(2006)\citenamefont
  {Stufler}, \citenamefont {Machnikowski}, \citenamefont {Ester}, \citenamefont
  {Bichler}, \citenamefont {Axt}, \citenamefont {Kuhn},\ and\ \citenamefont
  {Zrenner}}]{Stufler2006}%
  \BibitemOpen
  \bibfield  {author} {\bibinfo {author} {\bibfnamefont {S.}~\bibnamefont
  {Stufler}}, \bibinfo {author} {\bibfnamefont {P.}~\bibnamefont
  {Machnikowski}}, \bibinfo {author} {\bibfnamefont {P.}~\bibnamefont {Ester}},
  \bibinfo {author} {\bibfnamefont {M.}~\bibnamefont {Bichler}}, \bibinfo
  {author} {\bibfnamefont {V.~M.}\ \bibnamefont {Axt}}, \bibinfo {author}
  {\bibfnamefont {T.}~\bibnamefont {Kuhn}}, \ and\ \bibinfo {author}
  {\bibfnamefont {A.}~\bibnamefont {Zrenner}},\ }\href@noop {} {\bibfield
  {journal} {\bibinfo  {journal} {Phys. Rev. B}\ }\textbf {\bibinfo {volume}
  {73}},\ \bibinfo {pages} {125304} (\bibinfo {year} {2006})}\BibitemShut
  {NoStop}%
\bibitem [{\citenamefont {Jundt}\ \emph {et~al.}(2008)\citenamefont {Jundt},
  \citenamefont {Robledo}, \citenamefont {H\"ogele}, \citenamefont {F\"alt},\
  and\ \citenamefont {Imamo\u{g}lu}}]{Jundt2008}%
  \BibitemOpen
  \bibfield  {author} {\bibinfo {author} {\bibfnamefont {G.}~\bibnamefont
  {Jundt}}, \bibinfo {author} {\bibfnamefont {L.}~\bibnamefont {Robledo}},
  \bibinfo {author} {\bibfnamefont {A.}~\bibnamefont {H\"ogele}}, \bibinfo
  {author} {\bibfnamefont {S.}~\bibnamefont {F\"alt}}, \ and\ \bibinfo {author}
  {\bibfnamefont {A.}~\bibnamefont {Imamo\u{g}lu}},\ }\href@noop {} {\bibfield
  {journal} {\bibinfo  {journal} {Phys. Rev. Lett.}\ }\textbf {\bibinfo
  {volume} {100}},\ \bibinfo {pages} {177401} (\bibinfo {year}
  {2008})}\BibitemShut {NoStop}%
\bibitem [{\citenamefont {Gerardot}\ \emph {et~al.}(2009)\citenamefont
  {Gerardot}, \citenamefont {Brunner}, \citenamefont {Dalgarno}, \citenamefont
  {Karrai}, \citenamefont {Badolato}, \citenamefont {Petroff},\ and\
  \citenamefont {Warburton}}]{Gerardot2009}%
  \BibitemOpen
  \bibfield  {author} {\bibinfo {author} {\bibfnamefont {B.}~\bibnamefont
  {Gerardot}}, \bibinfo {author} {\bibfnamefont {D.}~\bibnamefont {Brunner}},
  \bibinfo {author} {\bibfnamefont {P.}~\bibnamefont {Dalgarno}}, \bibinfo
  {author} {\bibfnamefont {K.}~\bibnamefont {Karrai}}, \bibinfo {author}
  {\bibfnamefont {A.}~\bibnamefont {Badolato}}, \bibinfo {author}
  {\bibfnamefont {P.}~\bibnamefont {Petroff}}, \ and\ \bibinfo {author}
  {\bibfnamefont {R.}~\bibnamefont {Warburton}},\ }\href@noop {} {\bibfield
  {journal} {\bibinfo  {journal} {New J. Phys.}\ }\textbf {\bibinfo {volume}
  {11}},\ \bibinfo {pages} {013028} (\bibinfo {year} {2009})}\BibitemShut
  {NoStop}%
\bibitem [{\citenamefont {Lindblad}(1976)}]{Lindblad1976}%
  \BibitemOpen
  \bibfield  {author} {\bibinfo {author} {\bibfnamefont {G.}~\bibnamefont
  {Lindblad}},\ }\href@noop {} {\bibfield  {journal} {\bibinfo  {journal}
  {Commun. Math. Phys.}\ }\textbf {\bibinfo {volume} {48}},\ \bibinfo {pages}
  {119} (\bibinfo {year} {1976})}\BibitemShut {NoStop}%
\bibitem [{\citenamefont {Blum}(2012)}]{Blum2012}%
  \BibitemOpen
  \bibfield  {author} {\bibinfo {author} {\bibfnamefont {K.}~\bibnamefont
  {Blum}},\ }\href@noop {} {\emph {\bibinfo {title} {Density Matrix: Theory and
  applications (3rd edition)}}}\ (\bibinfo  {publisher} {Springer},\ \bibinfo
  {year} {2012})\BibitemShut {NoStop}%
\bibitem [{\citenamefont {Zaitsev}\ \emph {et~al.}(1983)\citenamefont
  {Zaitsev}, \citenamefont {Malyshev},\ and\ \citenamefont
  {Trifonov}}]{Zaitsev1983}%
  \BibitemOpen
  \bibfield  {author} {\bibinfo {author} {\bibfnamefont {A.~I.}\ \bibnamefont
  {Zaitsev}}, \bibinfo {author} {\bibfnamefont {V.}~\bibnamefont {Malyshev}}, \
  and\ \bibinfo {author} {\bibfnamefont {E.~D.}\ \bibnamefont {Trifonov}},\
  }\href@noop {} {\bibfield  {journal} {\bibinfo  {journal} {Sov. Phys. JETP}\
  }\textbf {\bibinfo {volume} {57}},\ \bibinfo {pages} {285} (\bibinfo {year}
  {1983})}\BibitemShut {NoStop}%
\bibitem [{\citenamefont {Benedict}\ \emph {et~al.}(1996)\citenamefont
  {Benedict}, \citenamefont {Ermolaev}, \citenamefont {Malyshev}, \citenamefont
  {Sokolov},\ and\ \citenamefont {Trifonov}}]{Benedict1996}%
  \BibitemOpen
  \bibfield  {author} {\bibinfo {author} {\bibfnamefont {M.~G.}\ \bibnamefont
  {Benedict}}, \bibinfo {author} {\bibfnamefont {A.~M.}\ \bibnamefont
  {Ermolaev}}, \bibinfo {author} {\bibfnamefont {V.~A.}\ \bibnamefont
  {Malyshev}}, \bibinfo {author} {\bibfnamefont {I.~V.}\ \bibnamefont
  {Sokolov}}, \ and\ \bibinfo {author} {\bibfnamefont {E.~D.}\ \bibnamefont
  {Trifonov}},\ }\href@noop {} {\emph {\bibinfo {title} {Super-radiance:
  Multiatomic Coherent Emission}}}\ (\bibinfo  {publisher} {IOP Publishing
  (Bristol)},\ \bibinfo {year} {1996})\BibitemShut {NoStop}%
\bibitem [{Note2()}]{Note2}%
  \BibitemOpen
  \bibinfo {note} {Strictly speaking, this field should be the field acting
  {\protect \it inside} the SQD, the latter differs from the field acting
  {\protect \it on} the SQD by a screening factor which depends on the system
  geometry and material parameters. In the simplest case of a spherical dot in
  a homogeneous environment this factor can be obtained analytically (see, e.
  g., Ref.~\protect \rev@citealp {Malyshev2011}). A realistic SQD array is a
  considerably more complicated system involving a non-homogeneous host, at
  least three different materials, and a number of geometrical parameters. We
  believe that explicit calculation of the screening factors in this case would
  introduce unnecessary level of detail and obscure further analysis.
  Therefore, for the sake of simplicity, we consider a SQD as a point-like
  system in a homogeneous host; all the fields entering the Lindblad equations
  should be interpreted as those rescaled by appropriate screening
  factors.}\BibitemShut {Stop}%
\bibitem [{\citenamefont {Born}\ and\ \citenamefont
  {Wolf}(1980)}]{BornAndWolf}%
  \BibitemOpen
  \bibfield  {author} {\bibinfo {author} {\bibfnamefont {M.}~\bibnamefont
  {Born}}\ and\ \bibinfo {author} {\bibfnamefont {E.}~\bibnamefont {Wolf}},\
  }\href@noop {} {\emph {\bibinfo {title} {Principles of Optics (6-th
  edition)}}}\ (\bibinfo  {publisher} {Springer},\ \bibinfo {year}
  {1980})\BibitemShut {NoStop}%
\bibitem [{\citenamefont {Friedberg}\ \emph {et~al.}(1973)\citenamefont
  {Friedberg}, \citenamefont {Hartmann},\ and\ \citenamefont
  {Manassah}}]{Friedberg1973}%
  \BibitemOpen
  \bibfield  {author} {\bibinfo {author} {\bibfnamefont {R.}~\bibnamefont
  {Friedberg}}, \bibinfo {author} {\bibfnamefont {S.~R.}\ \bibnamefont
  {Hartmann}}, \ and\ \bibinfo {author} {\bibfnamefont {J.~T.}\ \bibnamefont
  {Manassah}},\ }\href@noop {} {\bibfield  {journal} {\bibinfo  {journal}
  {Phys. Rep. C}\ }\textbf {\bibinfo {volume} {7}},\ \bibinfo {pages} {101}
  (\bibinfo {year} {1973})}\BibitemShut {NoStop}%
\bibitem [{\citenamefont {Hopf}\ \emph {et~al.}(1984)\citenamefont {Hopf},
  \citenamefont {Bowden},\ and\ \citenamefont {Louisell}}]{Hopf1984}%
  \BibitemOpen
  \bibfield  {author} {\bibinfo {author} {\bibfnamefont {F.~A.}\ \bibnamefont
  {Hopf}}, \bibinfo {author} {\bibfnamefont {C.~M.}\ \bibnamefont {Bowden}}, \
  and\ \bibinfo {author} {\bibfnamefont {W.~H.}\ \bibnamefont {Louisell}},\
  }\href@noop {} {\bibfield  {journal} {\bibinfo  {journal} {Phys. Rev. A}\
  }\textbf {\bibinfo {volume} {29}},\ \bibinfo {pages} {2591} (\bibinfo {year}
  {1984})}\BibitemShut {NoStop}%
\bibitem [{\citenamefont {Malyshev}\ and\ \citenamefont
  {Conejero~Jarque}(1997{\natexlab{a}})}]{Malyshev1997a}%
  \BibitemOpen
  \bibfield  {author} {\bibinfo {author} {\bibfnamefont {V.}~\bibnamefont
  {Malyshev}}\ and\ \bibinfo {author} {\bibfnamefont {E.}~\bibnamefont
  {Conejero~Jarque}},\ }\href@noop {} {\bibfield  {journal} {\bibinfo
  {journal} {J. Opt. Soc. Am. B}\ }\textbf {\bibinfo {volume} {16}},\ \bibinfo
  {pages} {1167} (\bibinfo {year} {1997}{\natexlab{a}})}\BibitemShut {NoStop}%
\bibitem [{\citenamefont {Malyshev}\ and\ \citenamefont
  {Conejero~Jarque}(1997{\natexlab{b}})}]{Malyshev1997b}%
  \BibitemOpen
  \bibfield  {author} {\bibinfo {author} {\bibfnamefont {V.}~\bibnamefont
  {Malyshev}}\ and\ \bibinfo {author} {\bibfnamefont {E.}~\bibnamefont
  {Conejero~Jarque}},\ }\href@noop {} {\bibfield  {journal} {\bibinfo
  {journal} {Opt. Spectrosc.}\ }\textbf {\bibinfo {volume} {82}},\ \bibinfo
  {pages} {582} (\bibinfo {year} {1997}{\natexlab{b}})}\BibitemShut {NoStop}%
\bibitem [{Note3()}]{Note3}%
  \BibitemOpen
  \bibinfo {note} {Note that Eqs.~(\ref {rho11})-~(\ref {Omega21}) are
  algebraically equivalent to the equations for a heterodimer comprising a
  metallic nanoparticle and a semiconductor quantum dot subjected to a
  quasi-resonant irradiation (see Refs.~\protect \rev@citealp {Artuso2013}
  and~\protect \rev@citealp {Nugroho2017})}\BibitemShut {NoStop}%
\bibitem [{\citenamefont {Glasser}(1972)}]{Glasser1972}%
  \BibitemOpen
  \bibfield  {author} {\bibinfo {author} {\bibfnamefont {M.~L.}\ \bibnamefont
  {Glasser}},\ }\href@noop {} {\bibfield  {journal} {\bibinfo  {journal} {J.
  Math. Phys.}\ }\textbf {\bibinfo {volume} {14}},\ \bibinfo {pages} {409}
  (\bibinfo {year} {1972})}\BibitemShut {NoStop}%
\bibitem [{\citenamefont {Christiansen}\ \emph {et~al.}(1998)\citenamefont
  {Christiansen}, \citenamefont {Gaididei}, \citenamefont {Johansson},
  \citenamefont {Rasmussen}, \citenamefont {Mezentsev},\ and\ \citenamefont
  {Rasmussen}}]{Christiansen1998}%
  \BibitemOpen
  \bibfield  {author} {\bibinfo {author} {\bibfnamefont {P.~L.}\ \bibnamefont
  {Christiansen}}, \bibinfo {author} {\bibfnamefont {Y.~B.}\ \bibnamefont
  {Gaididei}}, \bibinfo {author} {\bibfnamefont {M.}~\bibnamefont {Johansson}},
  \bibinfo {author} {\bibfnamefont {K.~O.}\ \bibnamefont {Rasmussen}}, \bibinfo
  {author} {\bibfnamefont {V.~K.}\ \bibnamefont {Mezentsev}}, \ and\ \bibinfo
  {author} {\bibfnamefont {J.~J.}\ \bibnamefont {Rasmussen}},\ }\href@noop {}
  {\bibfield  {journal} {\bibinfo  {journal} {Phys. Rev. B}\ }\textbf {\bibinfo
  {volume} {57}},\ \bibinfo {pages} {11303} (\bibinfo {year}
  {1998})}\BibitemShut {NoStop}%
\bibitem [{\citenamefont {Dicke}(1954)}]{Dicke1954}%
  \BibitemOpen
  \bibfield  {author} {\bibinfo {author} {\bibfnamefont {R.~H.}\ \bibnamefont
  {Dicke}},\ }\href@noop {} {\bibfield  {journal} {\bibinfo  {journal} {Phys.
  Rev.}\ }\textbf {\bibinfo {volume} {93}},\ \bibinfo {pages} {99} (\bibinfo
  {year} {1954})}\BibitemShut {NoStop}%
\bibitem [{\citenamefont {Mai}\ \emph {et~al.}(2014)\citenamefont {Mai},
  \citenamefont {Barrette}, \citenamefont {Yu}, \citenamefont {Semenov},
  \citenamefont {Kim}, \citenamefont {Cao},\ and\ \citenamefont
  {Gundogdu}}]{Mai2014}%
  \BibitemOpen
  \bibfield  {author} {\bibinfo {author} {\bibfnamefont {C.}~\bibnamefont
  {Mai}}, \bibinfo {author} {\bibfnamefont {A.}~\bibnamefont {Barrette}},
  \bibinfo {author} {\bibfnamefont {Y.}~\bibnamefont {Yu}}, \bibinfo {author}
  {\bibfnamefont {Y.~G.}\ \bibnamefont {Semenov}}, \bibinfo {author}
  {\bibfnamefont {K.~W.}\ \bibnamefont {Kim}}, \bibinfo {author} {\bibfnamefont
  {L.}~\bibnamefont {Cao}}, \ and\ \bibinfo {author} {\bibfnamefont
  {K.}~\bibnamefont {Gundogdu}},\ }\href@noop {} {\bibfield  {journal}
  {\bibinfo  {journal} {Nano Lett.}\ }\textbf {\bibinfo {volume} {14}},\
  \bibinfo {pages} {202} (\bibinfo {year} {2014})}\BibitemShut {NoStop}%
\bibitem [{\citenamefont {Mak}\ and\ \citenamefont {Shan}(2016)}]{Mak2016}%
  \BibitemOpen
  \bibfield  {author} {\bibinfo {author} {\bibfnamefont {K.~F.}\ \bibnamefont
  {Mak}}\ and\ \bibinfo {author} {\bibfnamefont {J.}~\bibnamefont {Shan}},\
  }\href@noop {} {\bibfield  {journal} {\bibinfo  {journal} {Nat. Photonics}\
  }\textbf {\bibinfo {volume} {10}},\ \bibinfo {pages} {216} (\bibinfo {year}
  {2016})}\BibitemShut {NoStop}%
\bibitem [{\citenamefont {Lorenz}(1963)}]{Lorenz1963}%
  \BibitemOpen
  \bibfield  {author} {\bibinfo {author} {\bibfnamefont {E.~N.}\ \bibnamefont
  {Lorenz}},\ }\href@noop {} {\bibfield  {journal} {\bibinfo  {journal} {J.
  Atmos. Sci.}\ }\textbf {\bibinfo {volume} {20}},\ \bibinfo {pages} {130}
  (\bibinfo {year} {1963})}\BibitemShut {NoStop}%
\bibitem [{Note4()}]{Note4}%
  \BibitemOpen
  \bibinfo {note} {Additionally, one can apply the same criteria to the average
  $A_n=(\protect \qopname \relax m{min}_{\Delta T}{|\Omega (t)|}+\protect
  \qopname \relax m{max}_{\Delta T}{|\Omega (t)|})/2$ to account for and get
  rid of the drift of the average towards an attractor}\BibitemShut {NoStop}%
\bibitem [{\citenamefont {Shimada}\ and\ \citenamefont
  {Nagashima}(1979)}]{Shimada1979}%
  \BibitemOpen
  \bibfield  {author} {\bibinfo {author} {\bibfnamefont {I.}~\bibnamefont
  {Shimada}}\ and\ \bibinfo {author} {\bibfnamefont {T.}~\bibnamefont
  {Nagashima}},\ }\href {\doibase 10.1143/PTP.61.1605} {\bibfield  {journal}
  {\bibinfo  {journal} {Progress of Theoretical Physics}\ }\textbf {\bibinfo
  {volume} {61}},\ \bibinfo {pages} {1605} (\bibinfo {year}
  {1979})}\BibitemShut {NoStop}%
\bibitem [{\citenamefont {Benettin}\ \emph
  {et~al.}(1980{\natexlab{a}})\citenamefont {Benettin}, \citenamefont
  {Galgani}, \citenamefont {Giorgilli},\ and\ \citenamefont
  {Strelcyn}}]{Benettin1980a}%
  \BibitemOpen
  \bibfield  {author} {\bibinfo {author} {\bibfnamefont {G.}~\bibnamefont
  {Benettin}}, \bibinfo {author} {\bibfnamefont {L.}~\bibnamefont {Galgani}},
  \bibinfo {author} {\bibfnamefont {A.}~\bibnamefont {Giorgilli}}, \ and\
  \bibinfo {author} {\bibfnamefont {M.}~\bibnamefont {Strelcyn}},\ }\href@noop
  {} {\bibfield  {journal} {\bibinfo  {journal} {Meccanica}\ }\textbf {\bibinfo
  {volume} {15}},\ \bibinfo {pages} {9} (\bibinfo {year}
  {1980}{\natexlab{a}})}\BibitemShut {NoStop}%
\bibitem [{\citenamefont {Benettin}\ \emph
  {et~al.}(1980{\natexlab{b}})\citenamefont {Benettin}, \citenamefont
  {Galgani}, \citenamefont {Giorgilli},\ and\ \citenamefont
  {Strelcyn}}]{Benettin1980b}%
  \BibitemOpen
  \bibfield  {author} {\bibinfo {author} {\bibfnamefont {G.}~\bibnamefont
  {Benettin}}, \bibinfo {author} {\bibfnamefont {L.}~\bibnamefont {Galgani}},
  \bibinfo {author} {\bibfnamefont {A.}~\bibnamefont {Giorgilli}}, \ and\
  \bibinfo {author} {\bibfnamefont {M.}~\bibnamefont {Strelcyn}},\ }\href
  {\doibase 10.1007/BF02128237} {\bibfield  {journal} {\bibinfo  {journal}
  {Meccanica}\ }\textbf {\bibinfo {volume} {15}},\ \bibinfo {pages} {21}
  (\bibinfo {year} {1980}{\natexlab{b}})}\BibitemShut {NoStop}%
\bibitem [{\citenamefont {Wolf}\ \emph {et~al.}(1985)\citenamefont {Wolf},
  \citenamefont {B.~Swift}, \citenamefont {Swinney},\ and\ \citenamefont
  {A.~Vastano}}]{Wolf1985}%
  \BibitemOpen
  \bibfield  {author} {\bibinfo {author} {\bibfnamefont {A.}~\bibnamefont
  {Wolf}}, \bibinfo {author} {\bibfnamefont {J.}~\bibnamefont {B.~Swift}},
  \bibinfo {author} {\bibfnamefont {H.}~\bibnamefont {Swinney}}, \ and\
  \bibinfo {author} {\bibfnamefont {J.}~\bibnamefont {A.~Vastano}},\ }\href
  {\doibase 10.1016/0167-2789(85)90011-9} {\bibfield  {journal} {\bibinfo
  {journal} {Physica D: Nonlinear Phenomena}\ }\textbf {\bibinfo {volume}
  {16}},\ \bibinfo {pages} {285} (\bibinfo {year} {1985})}\BibitemShut
  {NoStop}%
\bibitem [{\citenamefont {Dieci}\ and\ \citenamefont
  {Vleck}(1995)}]{Dieci1994}%
  \BibitemOpen
  \bibfield  {author} {\bibinfo {author} {\bibfnamefont {L.}~\bibnamefont
  {Dieci}}\ and\ \bibinfo {author} {\bibfnamefont {E.~S.~V.}\ \bibnamefont
  {Vleck}},\ }\href {\doibase https://doi.org/10.1016/0168-9274(95)00033-Q}
  {\bibfield  {journal} {\bibinfo  {journal} {Applied Numerical Mathematics}\
  }\textbf {\bibinfo {volume} {17}},\ \bibinfo {pages} {275 } (\bibinfo {year}
  {1995})},\ \bibinfo {note} {special Issue on Numerical Methods for Ordinary
  Differential Equations}\BibitemShut {NoStop}%
\bibitem [{\citenamefont {Dieci}\ \emph {et~al.}(1997)\citenamefont {Dieci},
  \citenamefont {Russell},\ and\ \citenamefont {S.~Van~Vleck}}]{Dieci1997}%
  \BibitemOpen
  \bibfield  {author} {\bibinfo {author} {\bibfnamefont {L.}~\bibnamefont
  {Dieci}}, \bibinfo {author} {\bibfnamefont {R.}~\bibnamefont {Russell}}, \
  and\ \bibinfo {author} {\bibfnamefont {E.}~\bibnamefont {S.~Van~Vleck}},\
  }\href@noop {} {\bibfield  {journal} {\bibinfo  {journal} {Siam Journal on
  Numerical Analysis}\ }\textbf {\bibinfo {volume} {34}} (\bibinfo {year}
  {1997})}\BibitemShut {NoStop}%
\bibitem [{\citenamefont {Scales}\ and\ \citenamefont
  {S.~Van~Vleck}(1997)}]{Scales1997}%
  \BibitemOpen
  \bibfield  {author} {\bibinfo {author} {\bibfnamefont {J.}~\bibnamefont
  {Scales}}\ and\ \bibinfo {author} {\bibfnamefont {E.}~\bibnamefont
  {S.~Van~Vleck}},\ }\href {\doibase 10.1006/jcph.1997.5643} {\bibfield
  {journal} {\bibinfo  {journal} {Journal of Computational Physics}\ }\textbf
  {\bibinfo {volume} {133}},\ \bibinfo {pages} {27} (\bibinfo {year}
  {1997})}\BibitemShut {NoStop}%
\bibitem [{\citenamefont {Frederickson}\ \emph {et~al.}(1983)\citenamefont
  {Frederickson}, \citenamefont {L~Kaplan}, \citenamefont {D~Yorke},\ and\
  \citenamefont {A~Yorke}}]{Frederickson1983}%
  \BibitemOpen
  \bibfield  {author} {\bibinfo {author} {\bibfnamefont {P.}~\bibnamefont
  {Frederickson}}, \bibinfo {author} {\bibfnamefont {J.}~\bibnamefont
  {L~Kaplan}}, \bibinfo {author} {\bibfnamefont {E.}~\bibnamefont {D~Yorke}}, \
  and\ \bibinfo {author} {\bibfnamefont {J.}~\bibnamefont {A~Yorke}},\ }\href
  {\doibase 10.1016/0022-0396(83)90011-6} {\bibfield  {journal} {\bibinfo
  {journal} {Journal of Differential Equations}\ }\textbf {\bibinfo {volume}
  {49}},\ \bibinfo {pages} {185} (\bibinfo {year} {1983})}\BibitemShut
  {NoStop}%
\bibitem [{\citenamefont {L.~Kaplan}\ and\ \citenamefont
  {A.~Yorke}(1979)}]{Kaplan2006}%
  \BibitemOpen
  \bibfield  {author} {\bibinfo {author} {\bibfnamefont {J.}~\bibnamefont
  {L.~Kaplan}}\ and\ \bibinfo {author} {\bibfnamefont {J.}~\bibnamefont
  {A.~Yorke}},\ }\enquote {\bibinfo {title} {Chaotic behavior of
  multidimensional difference equations},}\ in\ \href {\doibase
  10.1007/BFb0064319} {\emph {\bibinfo {booktitle} {Functional Differential
  Equations and Approximation of Fixed Points. Lecture Notes in
  Mathematics}}},\ Vol.\ \bibinfo {volume} {730},\ \bibinfo {editor} {edited
  by\ \bibinfo {editor} {\bibfnamefont {H.-O. P.-O.}\ \bibnamefont {Walther}}}\
  (\bibinfo  {publisher} {Springer, Berlin, Heidelberg},\ \bibinfo {year}
  {1979})\ pp.\ \bibinfo {pages} {204--227}\BibitemShut {NoStop}%
\bibitem [{\citenamefont {Lai}\ and\ \citenamefont {T\'{e}l}(2011)}]{Lai2011}%
  \BibitemOpen
  \bibfield  {author} {\bibinfo {author} {\bibfnamefont {Y.-C.}\ \bibnamefont
  {Lai}}\ and\ \bibinfo {author} {\bibfnamefont {T.}~\bibnamefont {T\'{e}l}},\
  }\href@noop {} {\emph {\bibinfo {title} {Transient Chaos. Complex dynamics in
  finite-time scales}}}\ (\bibinfo  {publisher} {Springer, Berlin},\ \bibinfo
  {year} {2011})\BibitemShut {NoStop}%
\bibitem [{\citenamefont {T\'{e}l}(2015)}]{Tel2015}%
  \BibitemOpen
  \bibfield  {author} {\bibinfo {author} {\bibfnamefont {T.}~\bibnamefont
  {T\'{e}l}},\ }\href {\doibase 10.1063/1.4917287} {\bibfield  {journal}
  {\bibinfo  {journal} {Chaos}\ }\textbf {\bibinfo {volume} {25}} (\bibinfo
  {year} {2015}),\ 10.1063/1.4917287}\BibitemShut {NoStop}%
\bibitem [{Note5()}]{Note5}%
  \BibitemOpen
  \bibinfo {note} {When the system is exactly in a stationary state it remains
  there forever. However, due to the final precision of numerical methods and
  the initial state itself, the system is in very small vicinity of the exact
  stationary state. Therefore it is either attracted to the steady state (if it
  is a stable fixed point) or drifts away from it if the stationary state is
  unstable.}\BibitemShut {Stop}%
\bibitem [{\citenamefont {Nugroho}\ \emph {et~al.}(2017)\citenamefont
  {Nugroho}, \citenamefont {Iskandar}, \citenamefont {Malyshev},\ and\
  \citenamefont {Knoester}}]{Nugroho2017}%
  \BibitemOpen
  \bibfield  {author} {\bibinfo {author} {\bibfnamefont {B.~S.}\ \bibnamefont
  {Nugroho}}, \bibinfo {author} {\bibfnamefont {A.~A.}\ \bibnamefont
  {Iskandar}}, \bibinfo {author} {\bibfnamefont {V.~A.}\ \bibnamefont
  {Malyshev}}, \ and\ \bibinfo {author} {\bibfnamefont {J.}~\bibnamefont
  {Knoester}},\ }\href@noop {} {\bibfield  {journal} {\bibinfo  {journal} {J.
  Opt.}\ }\textbf {\bibinfo {volume} {19}},\ \bibinfo {pages} {015004}
  (\bibinfo {year} {2017})}\BibitemShut {NoStop}%
\bibitem [{\citenamefont {Friedberg}\ \emph {et~al.}(1989)\citenamefont
  {Friedberg}, \citenamefont {Hartmann},\ and\ \citenamefont
  {Manassah}}]{Friedberg1989}%
  \BibitemOpen
  \bibfield  {author} {\bibinfo {author} {\bibfnamefont {R.}~\bibnamefont
  {Friedberg}}, \bibinfo {author} {\bibfnamefont {S.~R.}\ \bibnamefont
  {Hartmann}}, \ and\ \bibinfo {author} {\bibfnamefont {J.~T.}\ \bibnamefont
  {Manassah}},\ }\href@noop {} {\bibfield  {journal} {\bibinfo  {journal}
  {Phys. Rev. A}\ }\textbf {\bibinfo {volume} {39}},\ \bibinfo {pages} {3444}
  (\bibinfo {year} {1989})}\BibitemShut {NoStop}%
\bibitem [{\citenamefont {Malyshev}(2012)}]{Malyshev2012}%
  \BibitemOpen
  \bibfield  {author} {\bibinfo {author} {\bibfnamefont {A.~V.}\ \bibnamefont
  {Malyshev}},\ }\href@noop {} {\bibfield  {journal} {\bibinfo  {journal}
  {Phys. Rev. A}\ }\textbf {\bibinfo {volume} {86}},\ \bibinfo {pages} {065804}
  (\bibinfo {year} {2012})}\BibitemShut {NoStop}%
\bibitem [{\citenamefont {Back}\ \emph {et~al.}(2018)\citenamefont {Back},
  \citenamefont {Zeytinoglu}, \citenamefont {Ijaz}, \citenamefont {Kroner},\
  and\ \citenamefont {Imamo\ifmmode~\breve{g}\else \u{g}\fi{}lu}}]{Back2018}%
  \BibitemOpen
  \bibfield  {author} {\bibinfo {author} {\bibfnamefont {P.}~\bibnamefont
  {Back}}, \bibinfo {author} {\bibfnamefont {S.}~\bibnamefont {Zeytinoglu}},
  \bibinfo {author} {\bibfnamefont {A.}~\bibnamefont {Ijaz}}, \bibinfo {author}
  {\bibfnamefont {M.}~\bibnamefont {Kroner}}, \ and\ \bibinfo {author}
  {\bibfnamefont {A.}~\bibnamefont {Imamo\ifmmode~\breve{g}\else
  \u{g}\fi{}lu}},\ }\href {\doibase 10.1103/PhysRevLett.120.037401} {\bibfield
  {journal} {\bibinfo  {journal} {Phys. Rev. Lett.}\ }\textbf {\bibinfo
  {volume} {120}},\ \bibinfo {pages} {037401} (\bibinfo {year}
  {2018})}\BibitemShut {NoStop}%
\bibitem [{\citenamefont {Scuri}\ \emph {et~al.}(2018)\citenamefont {Scuri},
  \citenamefont {Zhou}, \citenamefont {High}, \citenamefont {Wild},
  \citenamefont {Shu}, \citenamefont {De~Greve}, \citenamefont {Jauregui},
  \citenamefont {Taniguchi}, \citenamefont {Watanabe}, \citenamefont {Kim},
  \citenamefont {Lukin},\ and\ \citenamefont {Park}}]{Scuri2018}%
  \BibitemOpen
  \bibfield  {author} {\bibinfo {author} {\bibfnamefont {G.}~\bibnamefont
  {Scuri}}, \bibinfo {author} {\bibfnamefont {Y.}~\bibnamefont {Zhou}},
  \bibinfo {author} {\bibfnamefont {A.~A.}\ \bibnamefont {High}}, \bibinfo
  {author} {\bibfnamefont {D.~S.}\ \bibnamefont {Wild}}, \bibinfo {author}
  {\bibfnamefont {C.}~\bibnamefont {Shu}}, \bibinfo {author} {\bibfnamefont
  {K.}~\bibnamefont {De~Greve}}, \bibinfo {author} {\bibfnamefont {L.~A.}\
  \bibnamefont {Jauregui}}, \bibinfo {author} {\bibfnamefont {T.}~\bibnamefont
  {Taniguchi}}, \bibinfo {author} {\bibfnamefont {K.}~\bibnamefont {Watanabe}},
  \bibinfo {author} {\bibfnamefont {P.}~\bibnamefont {Kim}}, \bibinfo {author}
  {\bibfnamefont {M.~D.}\ \bibnamefont {Lukin}}, \ and\ \bibinfo {author}
  {\bibfnamefont {H.}~\bibnamefont {Park}},\ }\href {\doibase
  10.1103/PhysRevLett.120.037402} {\bibfield  {journal} {\bibinfo  {journal}
  {Phys. Rev. Lett.}\ }\textbf {\bibinfo {volume} {120}},\ \bibinfo {pages}
  {037402} (\bibinfo {year} {2018})}\BibitemShut {NoStop}%
\bibitem [{\citenamefont {Gao}\ and\ \citenamefont {Chen}(2008)}]{Gao2008}%
  \BibitemOpen
  \bibfield  {author} {\bibinfo {author} {\bibfnamefont {T.}~\bibnamefont
  {Gao}}\ and\ \bibinfo {author} {\bibfnamefont {Z.}~\bibnamefont {Chen}},\
  }\href@noop {} {\bibfield  {journal} {\bibinfo  {journal} {Phys. Lett. A 372
  and 394}\ } (\bibinfo {year} {2008})}\BibitemShut {NoStop}%
\bibitem [{\citenamefont {Vlasov}\ \emph
  {et~al.}(2013{\natexlab{a}})\citenamefont {Vlasov}, \citenamefont {Lemeza},\
  and\ \citenamefont {Gladush}}]{Vlasov2013a}%
  \BibitemOpen
  \bibfield  {author} {\bibinfo {author} {\bibfnamefont {R.~A.}\ \bibnamefont
  {Vlasov}}, \bibinfo {author} {\bibfnamefont {A.~M.}\ \bibnamefont {Lemeza}},
  \ and\ \bibinfo {author} {\bibfnamefont {M.~G.}\ \bibnamefont {Gladush}},\
  }\href@noop {} {\bibfield  {journal} {\bibinfo  {journal} {J. Appl.
  Spectrosc.}\ }\textbf {\bibinfo {volume} {80}},\ \bibinfo {pages} {698}
  (\bibinfo {year} {2013}{\natexlab{a}})}\BibitemShut {NoStop}%
\bibitem [{\citenamefont {Vlasov}\ \emph
  {et~al.}(2013{\natexlab{b}})\citenamefont {Vlasov}, \citenamefont {Lemeza},\
  and\ \citenamefont {Gladush}}]{Vlasov2013b}%
  \BibitemOpen
  \bibfield  {author} {\bibinfo {author} {\bibfnamefont {R.~A.}\ \bibnamefont
  {Vlasov}}, \bibinfo {author} {\bibfnamefont {A.~M.}\ \bibnamefont {Lemeza}},
  \ and\ \bibinfo {author} {\bibfnamefont {M.~G.}\ \bibnamefont {Gladush}},\
  }\href@noop {} {\bibfield  {journal} {\bibinfo  {journal} {Las. Phys. Lett.}\
  }\textbf {\bibinfo {volume} {10}},\ \bibinfo {pages} {045401} (\bibinfo
  {year} {2013}{\natexlab{b}})}\BibitemShut {NoStop}%
\bibitem [{\citenamefont {Malikov}\ \emph {et~al.}(2017)\citenamefont
  {Malikov}, \citenamefont {Ryzhov},\ and\ \citenamefont
  {Malyshev}}]{MalikovEPJWebConf2017}%
  \BibitemOpen
  \bibfield  {author} {\bibinfo {author} {\bibfnamefont {R.~F.}\ \bibnamefont
  {Malikov}}, \bibinfo {author} {\bibfnamefont {I.~V.}\ \bibnamefont {Ryzhov}},
  \ and\ \bibinfo {author} {\bibfnamefont {V.~A.}\ \bibnamefont {Malyshev}},\
  }\href@noop {} {\bibfield  {journal} {\bibinfo  {journal} {EPJ Web of
  Conference}\ }\textbf {\bibinfo {volume} {161}},\ \bibinfo {pages} {02014}
  (\bibinfo {year} {2017})}\BibitemShut {NoStop}%
\bibitem [{\citenamefont {Malyshev}\ \emph {et~al.}(2019)\citenamefont
  {Malyshev}, \citenamefont {Zapatero}, \citenamefont {Malyshev}, \citenamefont
  {Malikov},\ and\ \citenamefont {Ryzhov}}]{MalyshevJPhysConfSer2019}%
  \BibitemOpen
  \bibfield  {author} {\bibinfo {author} {\bibfnamefont {V.~A.}\ \bibnamefont
  {Malyshev}}, \bibinfo {author} {\bibfnamefont {P.~A.}\ \bibnamefont
  {Zapatero}}, \bibinfo {author} {\bibfnamefont {A.~V.}\ \bibnamefont
  {Malyshev}}, \bibinfo {author} {\bibfnamefont {R.~F.}\ \bibnamefont
  {Malikov}}, \ and\ \bibinfo {author} {\bibfnamefont {I.~V.}\ \bibnamefont
  {Ryzhov}},\ }\href@noop {} {\bibfield  {journal} {\bibinfo  {journal} {J.
  Phys.: Conf. Ser.}\ }\textbf {\bibinfo {volume} {1220}},\ \bibinfo {pages}
  {012006} (\bibinfo {year} {2019})}\BibitemShut {NoStop}%
\bibitem [{\citenamefont {Malyshev}\ and\ \citenamefont
  {Malyshev}(2011)}]{Malyshev2011}%
  \BibitemOpen
  \bibfield  {author} {\bibinfo {author} {\bibfnamefont {A.~V.}\ \bibnamefont
  {Malyshev}}\ and\ \bibinfo {author} {\bibfnamefont {V.~A.}\ \bibnamefont
  {Malyshev}},\ }\href@noop {} {\bibfield  {journal} {\bibinfo  {journal}
  {Phys. Rev. B}\ }\textbf {\bibinfo {volume} {84}},\ \bibinfo {pages} {035314}
  (\bibinfo {year} {2011})}\BibitemShut {NoStop}%
\bibitem [{\citenamefont {Artuso}\ and\ \citenamefont
  {Bryant}(2013)}]{Artuso2013}%
  \BibitemOpen
  \bibfield  {author} {\bibinfo {author} {\bibfnamefont {R.~D.}\ \bibnamefont
  {Artuso}}\ and\ \bibinfo {author} {\bibfnamefont {G.~W.}\ \bibnamefont
  {Bryant}},\ }\href@noop {} {\bibfield  {journal} {\bibinfo  {journal} {Phys.
  Rev. B}\ }\textbf {\bibinfo {volume} {87}},\ \bibinfo {pages} {125423}
  (\bibinfo {year} {2013})}\BibitemShut {NoStop}%
\end{thebibliography}%

\end{document}